\documentclass{aa}  

\usepackage{graphicx}
\usepackage{txfonts}
\usepackage[colorlinks=true, linktocpage, linkcolor={blue!60!black}, citecolor={blue!60!black}, urlcolor={blue!60!black}]{hyperref}

\usepackage[dvipsnames,svgnames,x11names]{xcolor}

\begin{document}

   \title{The strongest cool core in REXCESS:\\ Missing X-ray cavities in RXC~J2014.8-2430\thanks{Images and datacubes are available in electronic format at the CDS via anonymous ftp to cdsarc.u-strasbg.fr (130.79.128.5) or via \url{http://cdsweb.u-strasbg.fr/cgi-bin/qcat?J/A+A/}.}}

   \titlerunning{Missing X-ray cavities in RXC~J2014.8-2430}
   \authorrunning{Mroczkowski et al.}

   \author{Tony Mroczkowski\inst{1}\and
    Megan Donahue\inst{2}\and
    Joshiwa van~Marrewijk\inst{1}\and
    Tracy E.\ Clarke\inst{3}\and
    Aaron Hoffer\inst{2}\and
    Huib Intema\inst{4,5}\and
    Luca Di~Mascolo\inst{6,7,8}\and
    Gerg\"o Popping\inst{1}\and
    Gabriel W.\ Pratt\inst{9}\and
    Ming Sun\inst{10}\and
    Mark Voit\inst{2}
    }

   \institute{European Southern Observatory (ESO), Karl-Schwarzschild-Strasse 2, Garching 85748, Germany\\
    \email{tony.mroczkowski@eso.org}
    \and
        Michigan State University, Physics \& Astronomy Dept., East Lansing, MI 48824-2320
    \and
        Naval Research Laboratory, Code 7213, 4555 Overlook Ave SW, Washington, DC 20375 USA
    \and
        International Centre for Radio Astronomy Research (ICRAR), Curtin University, Bentley, WA 6102, Australia
    \and
        Leiden Observatory, Leiden University, Niels Bohrweg 2, 2333 CA, Leiden, The Netherlands
    \and
        Astronomy Unit, Department of Physics, University of Trieste, via Tiepolo 11, Trieste 34131, Italy
    \and 
        INAF – Osservatorio Astronomico di Trieste, via Tiepolo 11, Trieste 34131, Italy
    \and 
        IFPU - Institute for Fundamental Physics of the Universe, Via Beirut 2, 34014 Trieste, Italy
    \and
        Université Paris-Saclay, Université Paris Cité, CEA, CNRS, AIM, 91191, Gif-sur-Yvette, France
    \and
        University of Alabama in Huntsville, Department of Physics, Huntsville, AL, 35899
        }

   \date{Received April 6, 2022; accepted July 18, 2022}

 
  \abstract
  {We present a broad, multi-wavelength study of RXC~J2014.8-2430, the most extreme cool-core cluster in the Representative {\it XMM-Newton} Cluster Structure Survey (REXCESS), using {\it Chandra} X-ray, Southern Astrophysical Research (SOAR) Telescope spectroscopic and narrow-band imaging, Atacama Large Millimeter/submillimeter Array (ALMA), Very Large Array, and Giant Metrewave Radio Telescope observations. 
  While feedback from an active galactic nucleus (AGN) is thought to be the dominant mechanism by which a cooling flow is suppressed, the {\it Chandra} imaging observations surprisingly do not reveal the bi-lateral X-ray cavities one might expect to see in the intracluster medium (ICM) of an extreme cool core hosting a powerful radio source, though cavities commonly appear in many similar sources. We discuss the limits on the properties of putative radio bubbles associated with any undetected X-ray cavities. We place upper limits on any significant X-ray AGN in the brightest cluster galaxy (BCG) and show that the X-ray peak is offset from the central radio source, which exhibits a steep low-frequency radio spectrum indicative of electron ageing. 
  The imaging and spectroscopy provided by SOAR reveal an extended, luminous optical emission-line source. From our narrow-band H$\alpha$ imaging of the BCG, the central H$\alpha$ peak is coincident with the radio observations, yet offset from the X-ray peak, consistent with sloshing found previously in this cluster. ALMA observations of the CO(1-0) emission reveal a large reservoir of molecular gas that traces the extended H$\alpha$ emission in the direction of the cool core.
  We conclude either that the radio source and its cavities in the X-ray gas are nearly aligned along the line of sight, or that ram pressure induced by sloshing has significantly displaced the cool molecular gas feeding it, perhaps preempting the AGN feedback cycle.
  We argue that the sloshing near the core is likely subsonic, as expected, given the co-location of the H$\alpha$, CO(1-0), radio continuum, and stellar emission peaks and their proximity to the X-ray peak. Further, the X-ray emission from the core is strongly concentrated, as is the distribution of metals, indicating the cool core remains largely intact.    
  Deeper {\it Chandra} observations will be crucial for definitively establishing the presence or lack of X-ray cavities, while X-ray micro-calorimetric observations from {\it Athena} could establish if the motion of the cold and warm gas is dominated by large-scale motions of the surrounding ICM.}

   
   \maketitle
%

\section{Introduction}

The first evidence that a radio-loud active galactic nucleus (AGN) could disturb the X-ray atmosphere of a cluster of galaxies was seen with the High Resolution Imager on board the Roentgen Satellite (ROSAT) \citep{1993MNRAS.264L..25B}, in a historic observation of the Perseus cluster. This image showed two depressions in the X-ray surface brightness map, bracketing the central AGN. Early {\it Chandra} observations of nearby clusters revealed similar patterns of cavities around radio sources in brightest cluster galaxies \citep[BCGs; see e.g.][]{2000ApJ...534L.135M, 2001ApJ...562L.149M}. These cavities have generally been found to be filled with radio emission, though often it is too faint to be detected in shallow radio surveys and other short observations at 1.4~GHz \citep{2012MNRAS.427.3468B}. Long-wavelength radio observations and deeper observations of the Perseus cluster by \citet{2000MNRAS.318L..65F} revealed older and larger cavities farther out in the cluster, filled by low-frequency radio emission presumably originating from an ageing population of relativistic electrons. More recent work by \cite{Fabian2022} has shown that the outer cavities in Perseus could be suppressed by the passage of cold fronts, though the inner cavities remain intact.

Before the launches of {\it Chandra} and {\it XMM-Newton}, astronomers had assumed that the rate of kinetic energy emerging from radio AGNs would be similar to their radiative luminosities (i.e.\ low). The assumption that there was no source of energy to counterbalance the prodigious release of radiative energy from the gas led to the notion of a `cooling flow', where the entire gas atmosphere slowly compresses and cools to temperatures $\lesssim 1$~keV \citep{1977MNRAS.180..479F, 1977ApJ...215..723C, 1994ARAA..32..277F}. The lack of huge reservoirs of cold gas and the constraints on star formation rates at least an order of magnitude lower than the cooling flow rate call into question the simple cooling flow model. The door was apparently closed on this model by high resolution spectroscopy made by the Reflection Grating Spectrograph on board the {\it Newton} X-ray Multi Mirror (XMM) telescope \citep{2003ApJ...590..207P}. These observations showed that in most relaxed clusters the strong X-ray emission lines from gas around $10^7$ K ($\sim$0.9~keV) predicted by the simple cooling flow model were generally not present at their predicted strengths. However, there has been growing evidence for molecular gas cooling directly onto the central galaxy \citep{Hamer2012}, and notable exceptions supporting the existence of true cooling flows have been found over the last decade; examples include the Phoenix Cluster \citep{McDonald2012}, SpARCS~J104922.6+564032.5 \citep{Hlavacek-Larrondo2020}, and eMACS~J0252.4-2100 \citep{Ebeling2021}, which all have star formation rates of several hundred $\rm M_\odot~yr^{-1}$ in their BCGs.  In the case of the Phoenix Cluster, \cite{Russell2017} showed through Atacama Large Millimeter/submillimeter Array (ALMA)  observations that the mass in molecular gas is similar to that in other cool-core clusters, indicating that the salient distinction of a cooling flow may be the star formation rate rather than the amount of cold molecular gas.

Despite the exceptions, the question remains as to how it is possible that approximately half of all X-ray luminous clusters have X-ray luminous cool cores, while so few exhibit true cooling flows. The answer, for isolated systems at least, may lie in episodic feedback by AGNs \citep{Li2015}. 
The discovery of cavities commonly associated with AGNs in the atmospheres of cool cores posed a `smoking-gun' answer. This scenario was bolstered by the discovery that the size of the cavity and the X-ray gas pressure confining the cavity were consistent with the energy lost by the cluster cooling core, using the buoyant rise time as the relevant timescale \citep{2005Natur.433...45M, 2007ARAA..45..117M, 2008ApJ...686..859B}. To prevent the intracluster medium (ICM) in a cool core from catastrophically cooling, it is possible for the central supermassive black hole of the BCG in its AGN phase to quench cooling \citep[e.g.][]{1995MNRAS.276..663B,2001ApJ...554..261C}. 
The AGN could accomplish this heating through the creation of X-ray cavities (`bubbles') that buoyantly rise from the cluster centre \citep[e.g.][]{2003ApJ...592..839B, 2002Natur.418..301B}.  More recently, the structures of cooling gas have been shown in several studies to correspond to entrained filaments behind rising radio bubbles \citep[e.g.][]{McNamara2016, Russell2016, Tremblay2018, Olivares2019, Vantyghem2021}.
In the radio, \citet{2004ApJ...607..800B} found a strong correlation of AGN jet power with the X-ray cooling rate in clusters and groups, with only a small amount of scatter.  They inferred that the mechanical energies from X-ray cavities must be in the range 1$pV$ to 16$pV$ per cavity in order to quench the cooling suggested by the central entropy of the cluster. Following this, \citet{2012MNRAS.421.1360H, 2013MNRAS.431.1638H, 2015ApJ...805...35H} demonstrated that this correlation extends to clusters up to $z=1.2$.  The correlation between AGN jet power and X-ray luminosity therefore suggests that radio bubble formation scales strongly with the amount of cooling. 

Over the past decade, a number of studies have established the presence of cold molecular and warm atomic gas surrounding the BCGs in cool cores in filamentary structures.
The emission-line nebulae associated with cool cores provide another diagnostic for the physical processes occurring there. 
The origin of the warm gas in these dusty, optically luminous filaments is not entirely understood \citep[][]{2011ApJ...738L..24V}, and the processes that excite the optical emission are similarly mysterious. For some filaments, photoionisation by hot stars may well be the dominant source of excitation, but some additional source of heating might be required to explain both the brightest forbidden line emission and the lack of He II recombination lines \citep{1997ApJ...486..242V, 2004cgpc.symp..143D, 2009MNRAS.392.1475F, 2009ApJ...704L..20S, 2013ApJ...767..153W}. Nevertheless, emission-line nebulae are providing significant clues to unlocking the cool core mystery: they only appear in clusters with cool cores, that is, clusters with low central gas entropy and short cooling times \citep{2009ApJS..182...12C}. 
Cool-core cluster nebulae extend up to $\sim$70~kpc from the cluster core \citep[e.g.][]{1996ApJ...466L...9M}.
Typically, the morphology of the BCG's H$\alpha$ correlates well with that seen in the soft ($<$ 1 keV) X-ray emission \citep[e.g.][]{2004ApJ...607..294S, 2006MNRAS.366..417F, 2010MNRAS.407.2063W}.

In this paper we present observations taken with the {\it Chandra} X-ray Observatory, the Southern Astrophysical Research (SOAR) Telescope operated by the National Optical-Infrared Astronomy Research Laboratory (NOIRLab), ALMA, the Very Large Array (VLA),  the VLA Low-band Ionosphere and Transient Experiment (VLITE),
and the Giant Metrewave Radio Telescope (GMRT) of the galaxy cluster \object{RXC J2014.8-2430}, which is renowned as the strongest cool-core cluster in the Representative {\it XMM-Newton} Cluster Structure Survey (REXCESS). The REXCESS sample \citep{2007AA...469..363B} is a representative $z\sim0.1$ sample of 31 clusters spanning a wide range in luminosity, mass, and temperature. It was designed to avoid bias in X-ray morphology or central surface brightness. These observations were followed up with a short (20 ks) {\it Chandra} observation to complement the {\it XMM-Newton} data, in which the cluster core was studied with arcsecond resolution. From the {\it XMM-Newton} data, \citet{2008AA...487..431C} found that RXC~J2014.8-2430 is strongly peaked and has the shortest central cooling time of any member in the sample at 0.550 $\pm$ 0.026 Gyr, which is 15\% shorter than the central cooling time for the next coolest cluster. \citet{2010ApJ...715..881D} determined an H$\alpha$ luminosity of 6.4$\times 10^{41}$ h$^{-2}_{70}$ erg s$^{-1}$, which is twice as large as that of any other cluster in REXCESS. Additionally, in relation to other REXCESS clusters hosting radio sources, it has a moderate strength radio source in the centre \citep{1998AJ....115.1693C}, and its minimum star formation rate, based on UV {\it XMM-Newton} optical monitor data uncorrected for dust extinction, is 8-14 M$_\odot$ yr$^{-1}$ \citep{2010ApJ...715..881D}.

For all calculations, the assumed cosmology is $H_0 = 70~\rm{km~s^{-1}~Mpc^{-1}}$, $\Omega_M=0.3$, and $\Omega_{\Lambda}= 0.7$. \citet{2009AA...498..361P} used the redshift obtained from the optical spectroscopic measurements by \cite{reflex}, 
$z = 0.1538$, with no uncertainty reported. We use a more recent optical spectroscopic redshift, $z = 0.1555 \pm 0.0003$, obtained by \citet{2010ApJ...715..881D}, who fit for the H$\alpha$ emission from the BCG.  
Using that updated redshift, the angular scale is 2.694$~\rm{kpc/\arcsec}$ and the luminosity distance is 741.8~Mpc \citep{2006PASP..118.1711W}. 


\section{Observations and data reduction}

\begin{table*}
    \centering
    \caption{\label{tab:2014obs} Observation summary.}
    \begin{tabular}{lllllc}
    \hline\hline\noalign{\smallskip}
    Telescope & Filter & Exposure & Date & ID & PI\\
    \hline
    \noalign{\medskip}
    SOAR/SOI & CTIO 7580/85 & 3$\times$1200s & 2010-Sep-06 &  & Donahue \\
     & CTIO 7384/84 & 3$\times$720s &  &  &  \\
    SOAR/Goodman & 600 l/mm grating & 3$\times$1200s & 2012-Jul-25 &  & Donahue \\
     &  & 3$\times$1200s &  &  &  \\
    Chandra/ACIS-S &  & 20 ks & 2009-Aug-25 & 11757 &  Donahue \\
    XMM/MOS1+MOS2 &  & 26.7 ks & 2004-Oct-08 & 201902201 &  Boehringer \\
    GMRT & 610~MHz & 3.1 hr & 2008-Jun-21 & 14JHC01 &  Croston\\
    VLA & L-band (1--2~GHz) & 71 min & 2014-May-29 & 14A-280 &  Edge\\
    ALMA & Band 3 (86-90, 94-98~GHz) & 20.6 min & 2018-Dec-24 & 2018.1.00940.S & Mroczkowski \\
    \noalign{\smallskip}
    \hline
    \end{tabular}
    \tablefoot{Summary of observations used in this work.}
\end{table*}

\subsection{Chandra}\label{sec:xray}

\subsubsection{Chandra X-ray observation}\label{sec:xray_obs}

\begin{figure*}
\begin{center}
 \includegraphics[clip,trim=00mm 0mm 0mm 0mm,height=0.33\textwidth]{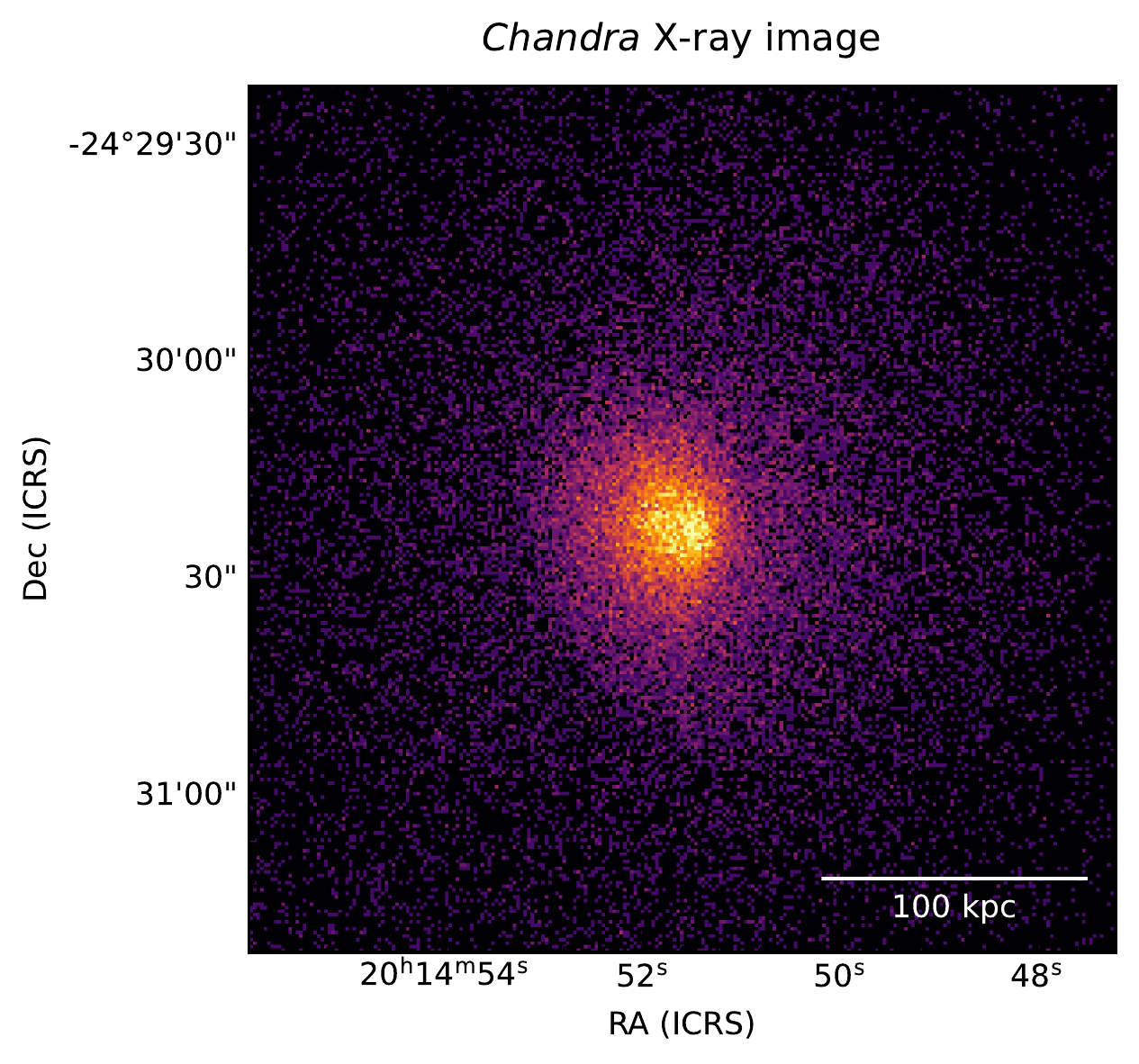}
 \includegraphics[clip,trim=27mm 0mm 0mm 0mm,height=0.33\textwidth]{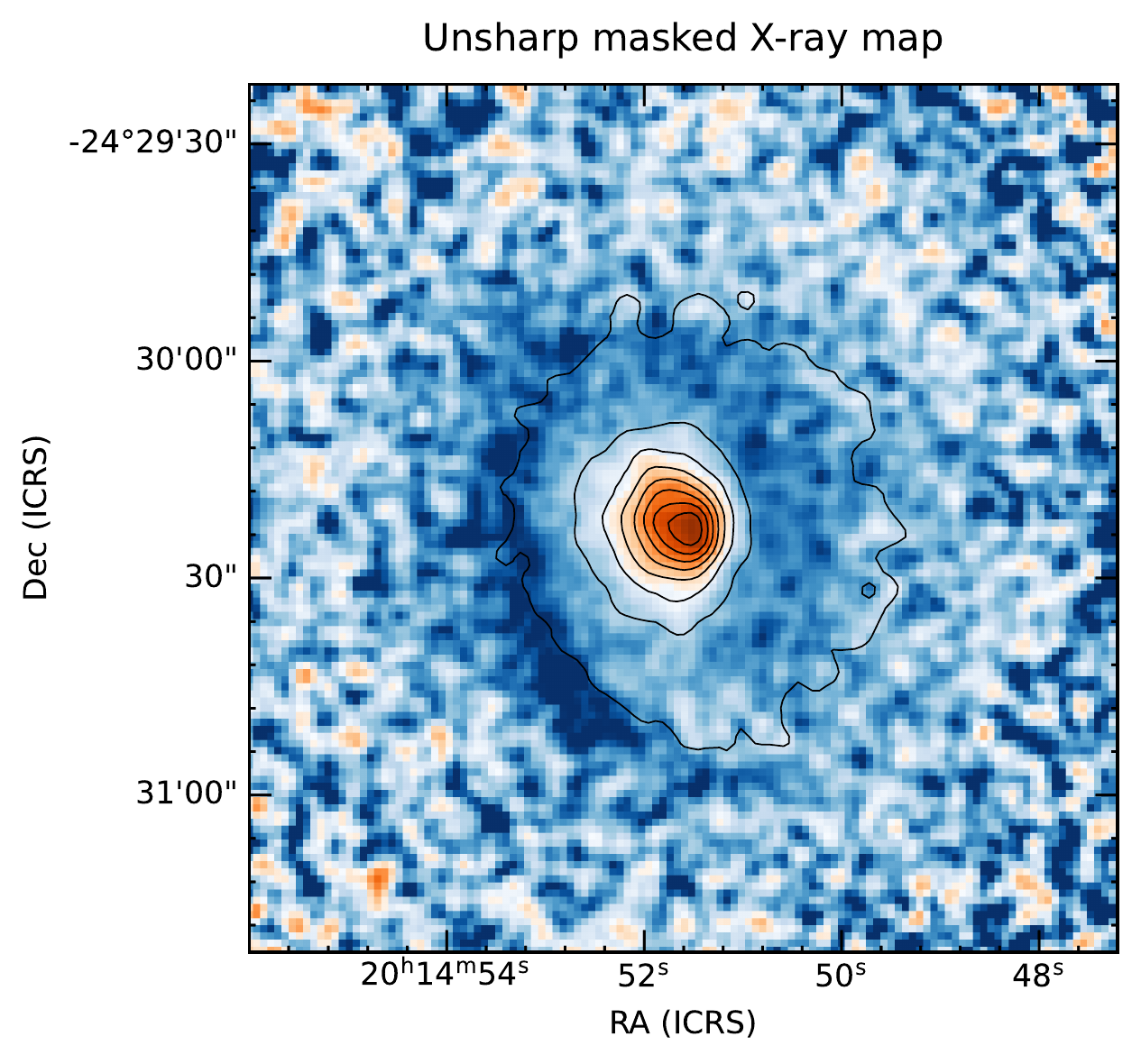}
 \includegraphics[clip,trim=27mm 0mm 0mm 0mm,height=0.33\textwidth]{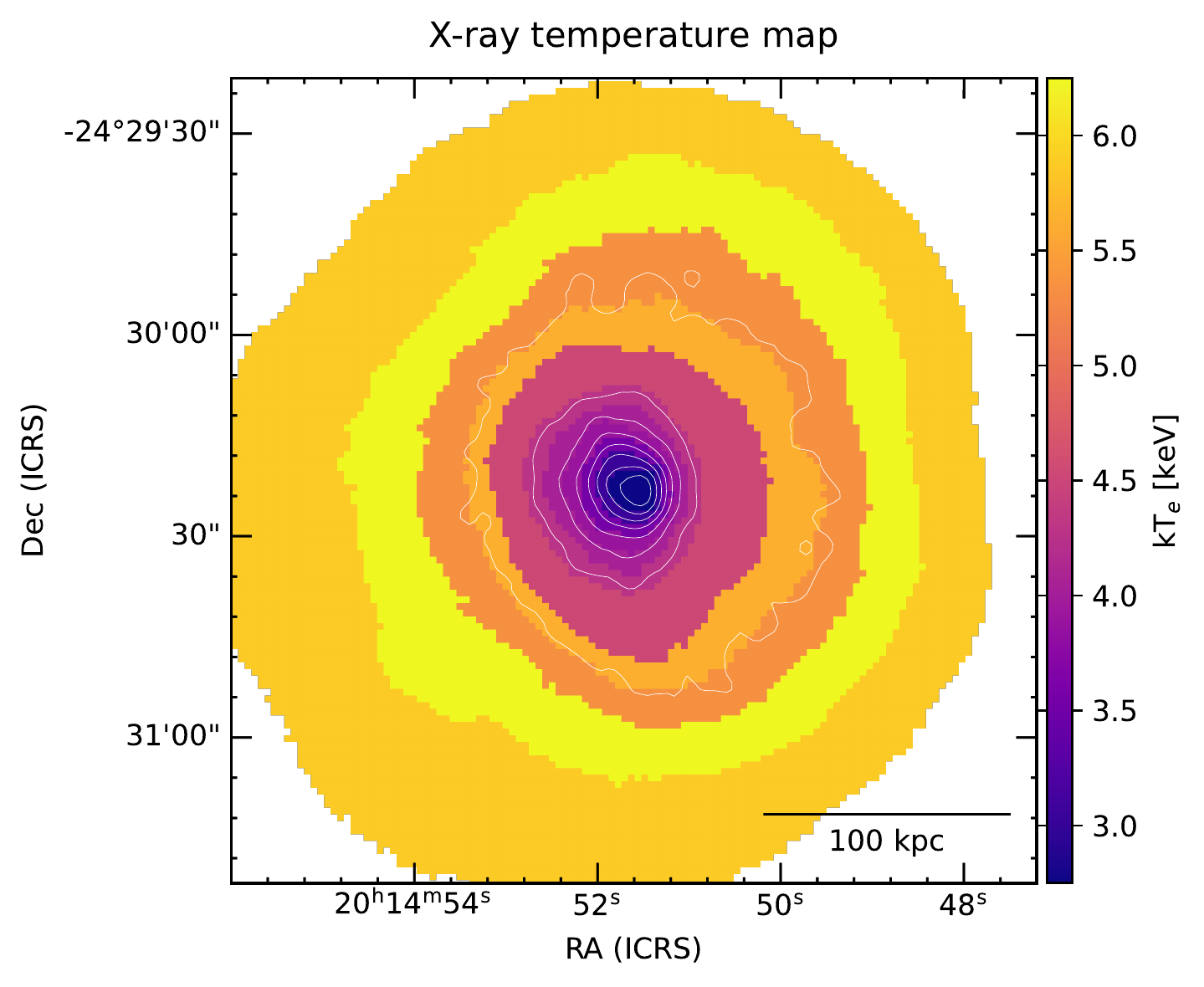}
\end{center}
\caption{\label{fig:xray}
         {\it Chandra} X-ray imaging and derived temperature of RXC~J2014.8-2430. {Left:} Broadband {\it Chandra} X-ray image of RXC~J2014.8-2430 in the 0.5-7.0~keV range, shown on a square root scale, roughly following projected density.  We note that no point sources were removed from the image. Right is west, and north is up. 
         {\bf Middle:} Unsharp masked {\it Chandra} X-ray surface brightness shown on a logarithmic scale, with X-ray contours from the left panel. The image was produced to highlight surface brightness edges. The image reveals a spiral decrement characteristic of cold fronts induced by sloshing, running anticlockwise from the western portion of the core outwards.
         {\bf Right:} Temperature map derived by applying the contour binning technique of \cite{2006MNRAS.371..829S} to the {\it Chandra} X-ray data for the core region of RXC~J2014.8-2430. Contours are from the left panel.}
\end{figure*}

A 20 ksec {\it Chandra} observation of RXC~J2014.8-2430 was taken on August 25, 2009 (ObsID 11757) as part of the {\it Chandra} Guest Observers programme. Observation details are listed in Table~\ref{tab:2014obs}. Observations were taken with the {\it Chandra} Advanced Charge-coupled device (CCD) Imaging Spectrometer (ACIS-S) centred on the back-illuminated ACIS-S3 chip in very faint (VFAINT) mode. The data were reprocessed using version 4.3 of the {\it Chandra} Interactive Analysis of Observations (CIAO)\footnote{\url{https://cxc.cfa.harvard.edu/ciao/}} tools with CALibration DataBase (CALDB) version 4.4.6. We used deep background files, located using the CIAO {\tt acis\_bkgrnd\_lookup} command.  The background events files are re-projected to match the observation pointing and roll angle, and scaled to match the high energy (9.5-12.0 keV) background count rate measured in a large source-free region in the main observation data \citep[as recommended by][]{2006ApJ...645...95H}. 
A more detailed discussion of the reduction and analysis methods is provided in \cite{donahue2014clash}.
Point sources around the cluster were identified primarily using the CIAO {\tt wavdetect} routine, which identified 13 sources within $r_{500}$. We note that {\tt wavdetect} identified the core as a compact X-ray source. We did not exclude the central region of the cluster from our analyses of the X-ray data, as we found no evidence for a central X-ray point source (\S3.1).  Several other compact sources not flagged by {\tt wavdetect} were identified by eye and manually excluded from the analysis. These additional sources were at large radii ($>0.5 r_{500}$) and do not impact the fit near the centre of the cluster.
To estimate the location of the centroid, we used the procedure from \citet{2008ApJ...682..821C}.  For a relaxed cool-core cluster with a slight asymmetry in the cluster core, this method determined an X-ray peak that is coincident with the peak pixel (0.492$\arcsec \times$0.492$\arcsec$) in the clean and point source-subtracted image. The X-ray centre is RA 20$^h$14$^m$51.65$^s$, Dec -24$^d$30$^m$21.1$^s$. 

\subsubsection{Chandra X-ray analysis}\label{sec:xray_analysis}

We fitted a 2$\times$2 binned 0.5-7.0 keV exposure-corrected flux map (Figure~\ref{fig:xray}, left), created using the \emph{fluximage} script,\footnote{\url{http://cxc.harvard.edu/ciao/ahelp/fluximage.html}} with elliptical profiles using the {\it ellipse} \citep{1987MNRAS.226..747J} function from the Image Reduction and Analysis Facility (IRAF) \citep{1993ASPC...52..173T}. Following the method described in, for example, \cite{Cassano2010}, we estimated the centroid shift $w = \frac{1}{R_{max}}\times \sqrt{\frac{\Sigma(\Delta_i - <\Delta>)^2}{N-1}}$ where N is the total number of apertures considered, $\Delta_{i}$ is the separation between centroid and that found for each isophote computed within $R_{max}$ and within the $i^{th}$ aperture, and $<\Delta>$ is the average value of the for all apertures. Using our elliptical profile fit, our centroid shift parameter $w$ = 0.018 computed within 500~kpc. 
For comparison to the literature, \citet{2012MNRAS.421.1583M} chose $w$ = 0.006 defined with respect to $R_{500}$ to separate relaxed from disturbed clusters in their sample of 114 clusters with observations using the Advanced CCD Imaging Spectrometer imaging camera (ACIS-I) on {\it Chandra}. This centroid shift cut was found to distinguish well the cool-core (CC) from non-CC clusters, with only 3 CC clusters in their sample having $w > 0.006$.  
In order to approximate the centroid shift we would obtain for $R_{max} = R_{500}$, we use the mass $M_{500}=(6.0\substack{+1.6\\-1.3}) \times 10^{14}~\rm M_\odot$ reported by \cite{Hilton2021} for this cluster using Sunyaev-Zeldovich (SZ; \citealt{1972CoASP...4..173S}) measurements from the Atacama Cosmology Telescope (ACT).\footnote{Here $M_{500}$ is the mass computed using the radius $R_{500}$ within which the average density is $500\times$ the critical density $\rho_{c}(z)$ of the Universe at the cluster redshift (i.e. $M_{500} = (4 \pi / 3) \, 500 \rho_{c}(z) \, R_{500}^3$).}  
This mass corresponds to $R_{500} = 1.2\pm0.1$~Mpc, in good agreement with the value of $R_{500} = 1155.29$~kpc previously found by \cite{2008AA...487..431C} using \emph{XMM-Newton} observations.  Conservatively assuming the centroid shift is dominated by scales $\lesssim 0.5~\rm Mpc$, we estimate a corrected $w\approx 0.0075$ with respect to $R_{500}$, implying RXC~J2014.8-2430 has a similarly high eccentricity as the three most eccentric cool-core clusters in \cite{2012MNRAS.421.1583M}. However, as \cite{2014MNRAS.441L..31W} demonstrated that the sloshing persists out to $R_{500}$, the centroid shift on these scales could in fact be larger.

\begin{figure}
\begin{center}
  \includegraphics[height=4.2cm]{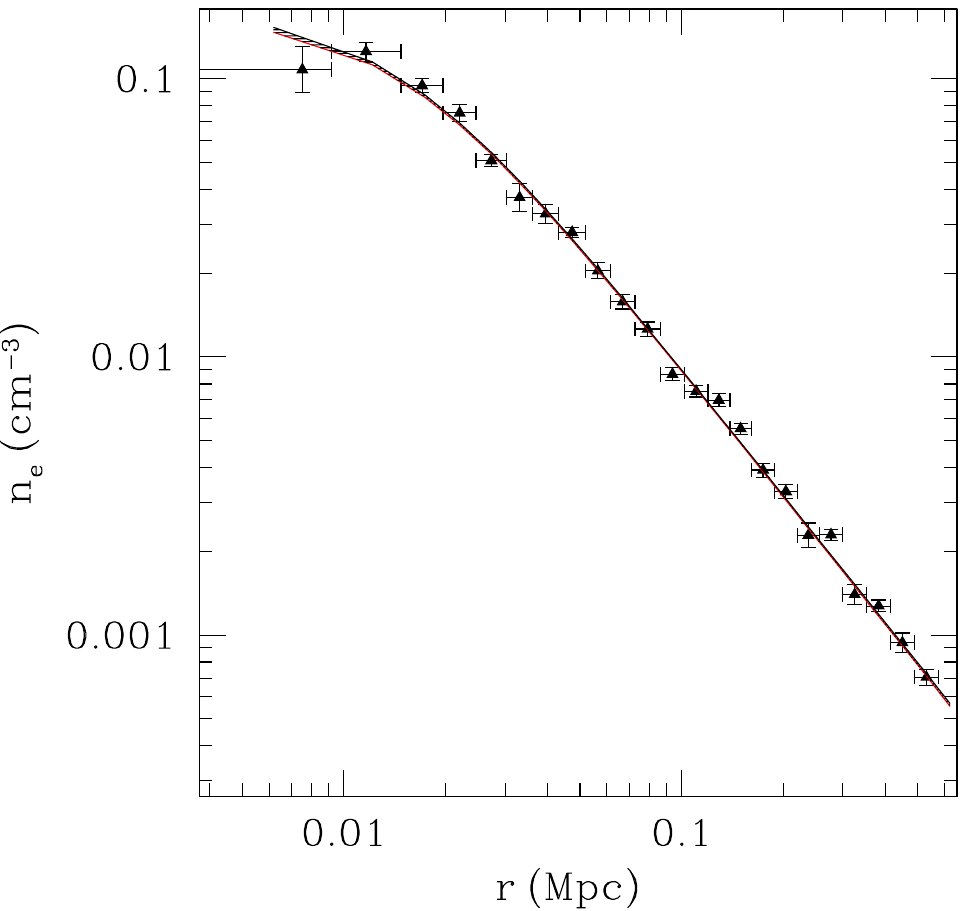}
  \includegraphics[height=4.2cm]{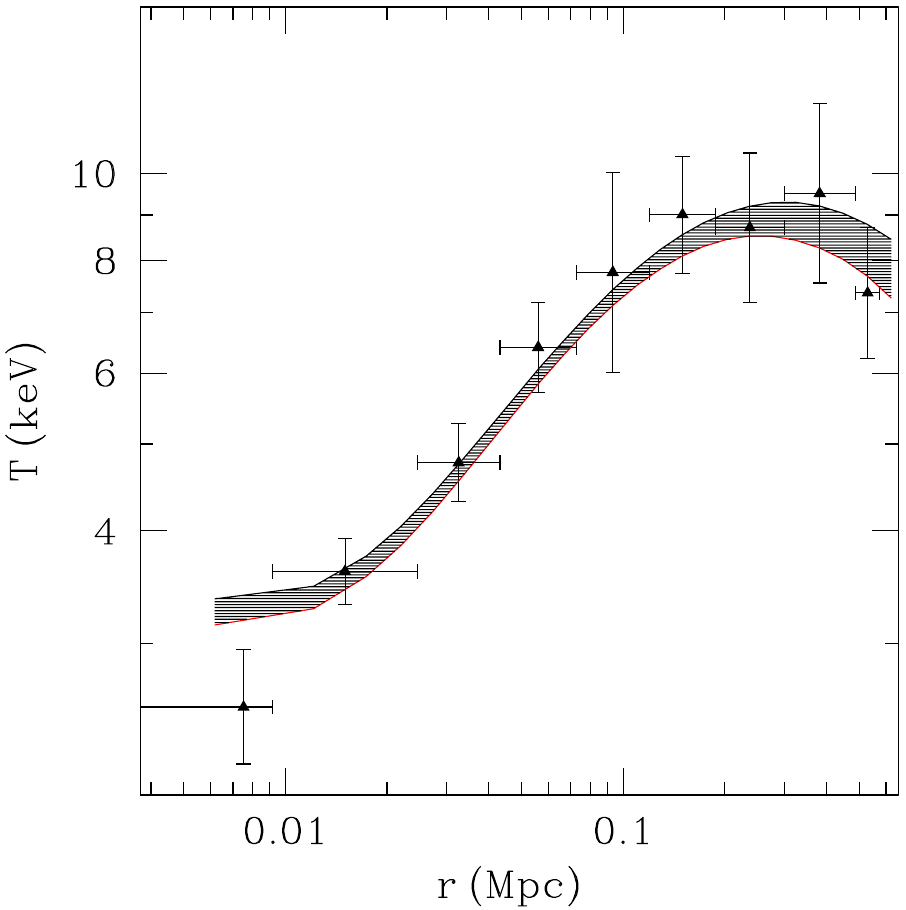}\\
  \includegraphics[height=4.2cm]{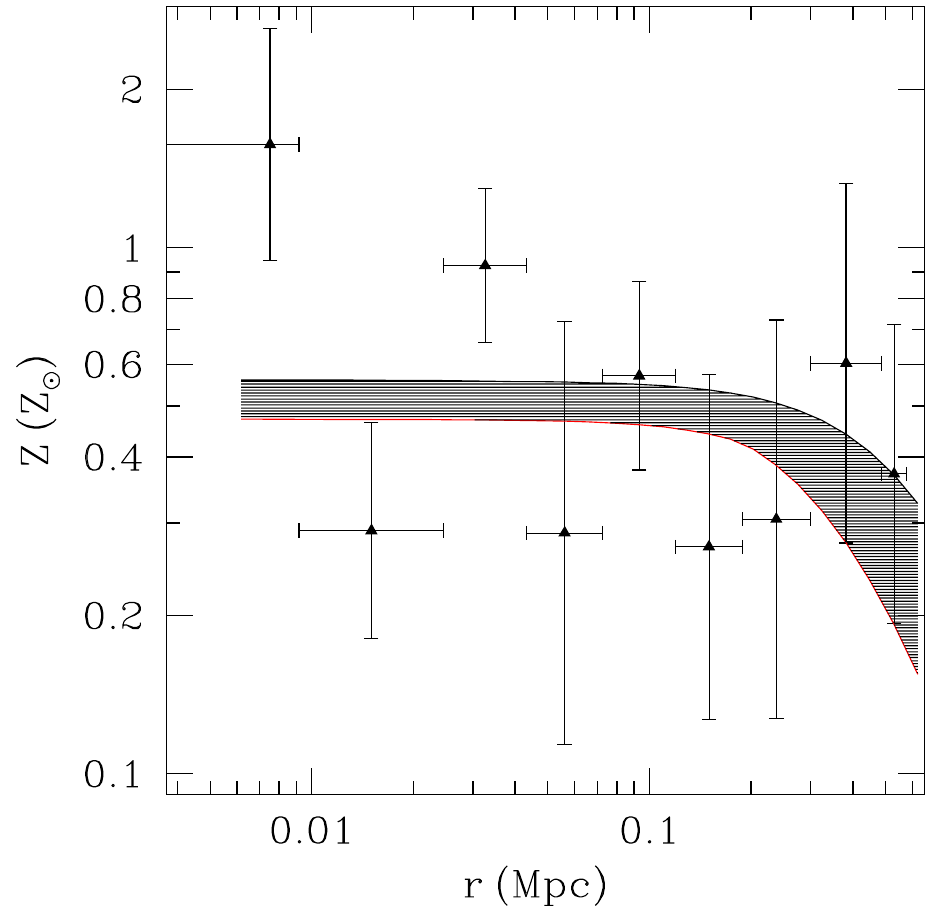}
  \includegraphics[height=4.2cm]{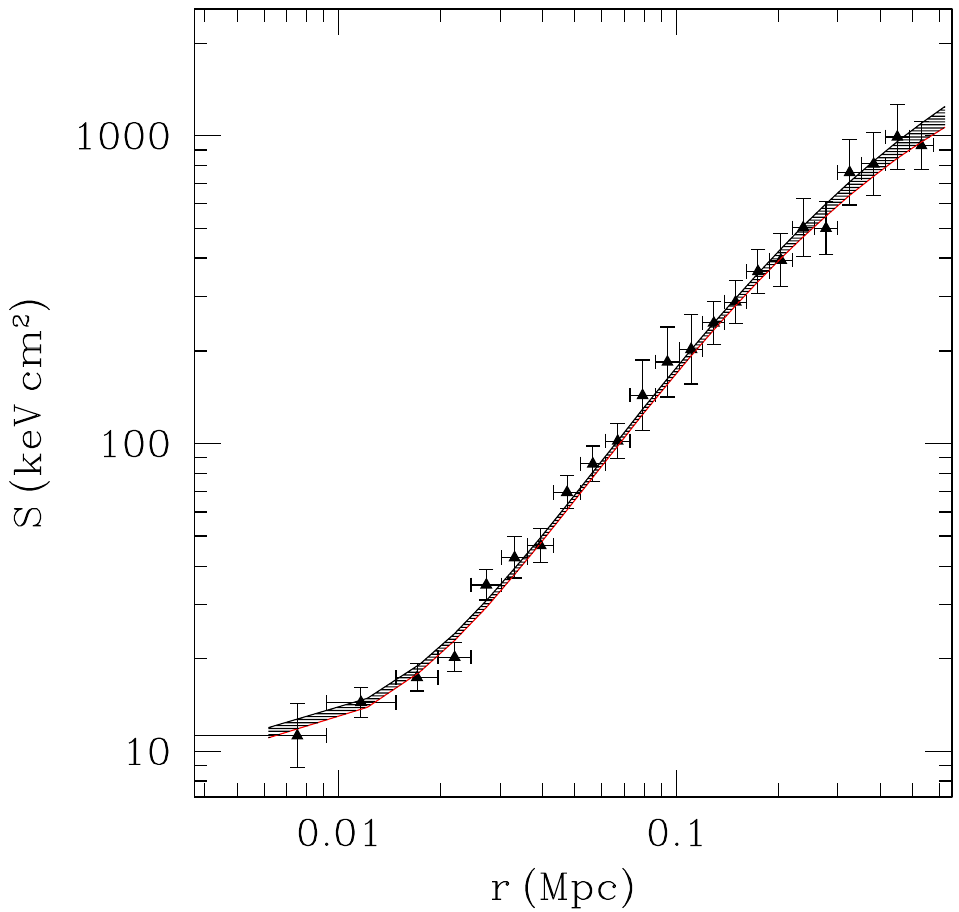}
\end{center}
\caption{Electron density, temperature, metallicity, and entropy parameter for the cluster as a function of physical distance from the cluster centre. The horizontal error bar describes the width of the annulus where the XSPEC model was fit. The error bars depict the 68\% statistical uncertainties on the data points, while the shaded regions show those on the fit profiles.}
\label{fig:xray_profiles}
\end{figure}

To search for structure in the X-ray image, we constructed an unsharp masked image using the same process as \citet{2009ApJ...700.1404R} where we took the exposure-corrected 2$\times$2 binned broadband image of the cluster, smoothed it by a one pixel (0.984$\arcsec$) Gaussian, and divided that by the same image smoothed by a ten pixel (9.84$\arcsec$) Gaussian. The unsharp masked image, shown in the middle panel of Figure~\ref{fig:xray}, clearly exhibits a spiralling decrement structure around cluster centre and an excess at the cluster centre. The cluster does not exhibit any obvious X-ray cavities.

We fitted a 2D elliptical $\beta$ model profile,  which has the form 
\begin{equation}
I(b) = I_0\left[1+\left(\frac{b}{r_\mathrm{c}}\right)^2\right]^{-3\beta + 0.5}
\end{equation}
\citep[see][]{Cavaliere1976,1988xrec.book.....S}, where $I_0$ is the normalisation, $b$ is the projected angular distance from the centre, the coordinate $b$ is transformed such that $b \longrightarrow \sqrt{x^2 + (y/R)^2}$ for $x$ and $y$ rotated to align respectively with the major and minor axes, $R$ is the minor to major axis ratio, $r_\mathrm{c}$ is the core radius in units of angle, and a classic $\beta$ defines the power law. We fitted a more general elliptical model to the 2$\times$2 binned image described previously, using Sherpa \citep{2001SPIE.4477...76F}. We used the Cash statistic \citep{1979ApJ...228..939C} in the fitting, and allowed all the parameters to vary (including the centre and orientation angle). The parameters for the 2D $\beta$ model fit (with 1$\sigma$ errors) are: $r_\mathrm{c}$=5.53$\arcsec$ $\pm$ 0.08$\arcsec$, I$_0$=83.5 $\pm$ 1.6 counts pixel$^{-2}$, ellipticity $\epsilon = 1-R = 0.098 \pm 0.006$, $\beta=0.998\pm 0.002$. The position angle of the major axis of the ellipse points directly north at $-0\fdg8 \pm 1\fdg7$. The major axis angle is equivalent to a 0\degr\ position angle for the IRAF ellipse fit, which agrees over most of the cluster. We use the residual image to look for any additional deviations from a smooth $\beta$ model profile. 

In the right panel of Fig.~\ref{fig:xray}, we show a temperature map derived by applying the contour binning technique {\tt contbin} of \cite{2006MNRAS.371..829S}, which selects regions of similar X-ray surface brightness. For both the contour bins and for radial bins used later in the is work,
X-ray spectra were extracted using {\it specextract} in the energy range 0.3-11.0 keV over circular annuli centred at the X-ray peak with at least 2500 net counts.  Analysis is restricted to the ACIS-S2 and ACIS-S3 chips. We used the Mewe-Kaastra-Liedahl (MEKAL) model \citep{1985AAS...62..197M,1986AAS...65..511M,1993AAS...97..443K,1995ApJ...438L.115L} to find the temperature and metal abundance in XSPEC 12.6.0 \citep{1996ASPC..101...17A}. The metallicity parameter fit to the central bin, where the signal to noise is highest, is treated as independent. Outside the central bin, the metallicity parameters are determined using pairs of consecutive annuli. For all spectral fits, the Galactic foreground extinction is fixed at N$_H$ = 7.4$\times 10^{20}$ cm$^{-2}$, the value reported in \citet{1990ARAA..28..215D} for this region of the sky. The Galactic foreground column density as well as the \citet{1998SSRv...85..161G} relative solar abundances are fixed parameters for the photoelectric absorption (PHABS) model used for Galactic extinction. We fit the unbinned spectral data with the Cash statistics \citep{1979ApJ...228..939C} implementation in XSPEC (modified c-stat). 

We fit the projected temperature and metallicity of the annuli and the resultant temperature, deprojected electron density profile, and entropy parameter profile using the joint analysis of cluster observations (JACO) code \citep{2007ApJ...664..162M}, following the procedures in \citet{donahue2014clash}. 
We adopt the common definition for the entropy parameter $K \equiv k_B T_e n_e^{-2/3}$ as in, for example, \cite{2001ApJ...559L..71B}.
The resulting density, temperature, metallicity, and entropy parameter profiles are shown in Figure~\ref{fig:xray_profiles}. In this procedure we have made the approximation that the projected temperature is approximately equal to the deprojected temperature. We fit the entropy profile using the functional form from \citet{2006ApJ...643..730D}, $K(r) = K_0 + K_{100}(r/100~\rm{kpc})^{\alpha}$. We find a central entropy $K_0 = 11.6 \pm 3.9 \rm{~keV~cm^2}$, with fit parameters $K_{100} = 159.0 \pm 57.2$, and $\alpha = 1.284 \pm 0.002$ for the radial dependence ($\chi^2_{red}$ = 0.725).  For comparison, \citet{2009ApJS..182...12C} found that all clusters with K$_0 \lesssim 50~\rm{keV~cm^2}$ to be cool cores, with an average of  K$_0 \sim 15~\rm{keV~cm^2}$.  The low central entropy and temperature gradient both firmly indicate this is a strong cool-core cluster. Additionally, hints of a break in the entropy radial profile at a radius of 30~kpc, which is due to both a drop in density and rise in temperature, may be due to the inner sloshing cold front. Such discontinuities are expected as the cold front separates low entropy from high-entropy gas \citep{Roediger2011}.

\subsection{Southern Astrophysical Research (SOAR) Telescope}\label{sec:soar}

\subsubsection{SOAR H$\alpha$ imaging and spectral observations}\label{sec:soar_obs}

\begin{figure*}
\begin{center}
 \includegraphics[clip,trim=0mm 3mm 20mm 12mm,width=\textwidth]{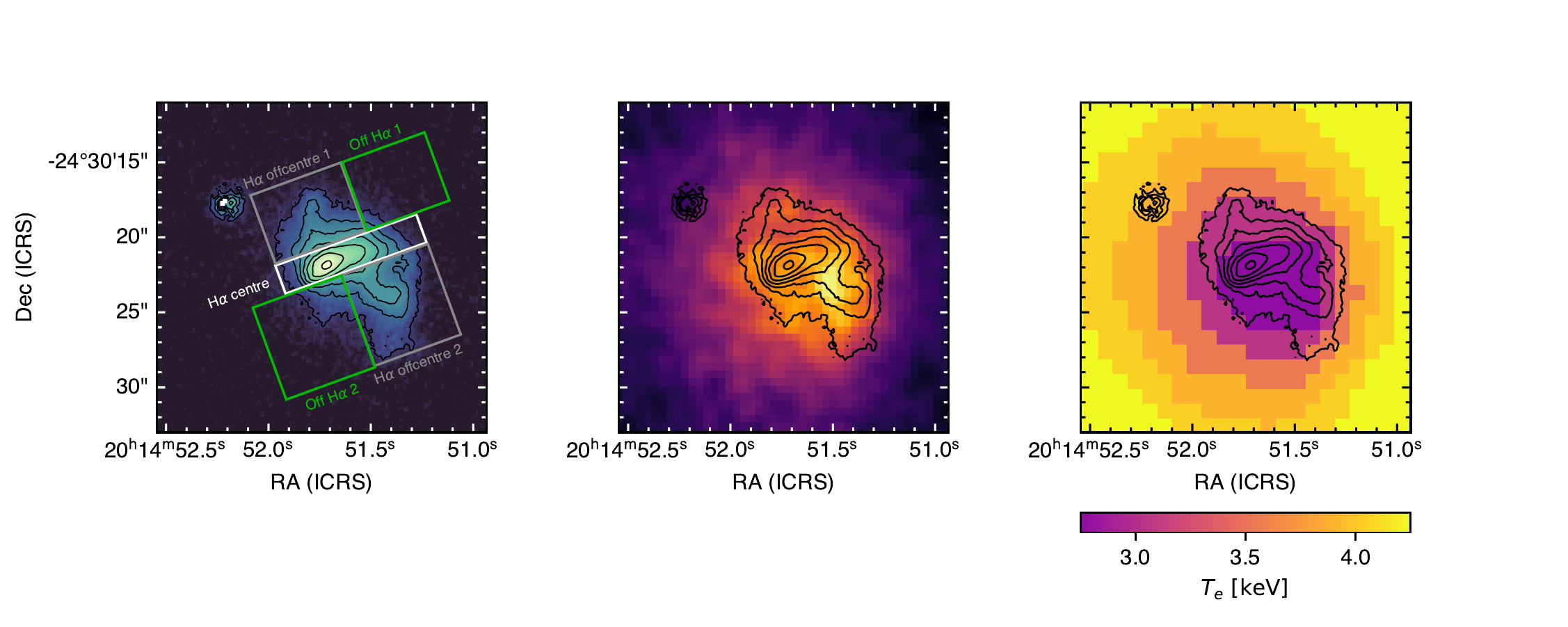}
\end{center}
\caption{\label{fig:core}
        SOAR H$\alpha$ imaging in comparison to the X-ray properties of RXC~J2014.8-2430. {\bf Left:} SOAR net H$\alpha$ image. The regions used for spectral extractions reported in Table~\ref{tab:xrayhalpha} are shown here. The white box labelled `H-$\alpha$ centre' corresponds to the region used for spectroscopy. We note that the bright source to the north-east of the H$\alpha$ peak is a foreground star.
        {\bf Middle:} {\it Chandra} X-ray surface brightness from Fig.~\ref{fig:xray} (left) with SOAR contours from the left panel. 
        {\bf Right:} Temperature map from Fig.~\ref{fig:xray} (right).  The peak in the H$\alpha$ emission is offset from the peak in the X-ray surface brightness.  However, the south-western spur in the H$\alpha$ emission appears to trace the coolest, brightest gas seen in X-ray.
    The images are 22\arcsec\ wide, corresponding to 59.3~kpc.}
\end{figure*}

\begin{table*}
    \centering
    \caption{ICM environment local to the H$\alpha$-emitting regions}
    \begin{tabular}{lcccccc}
    \hline\hline\noalign{\smallskip}    
    Region & RA & Dec & Box Size & Net Counts & T$_\textsc{x}$ & $Z$ \\ 
    (---) & (---) & (---) &  (arcsec$^2$) & (---) & (keV) & ($Z_\odot$)  \\ \noalign{\smallskip}
    \hline
    \noalign{\medskip}
    H$\alpha$ centre & 20$^h$14$^m$51.610$^s$ & -24$^d$30$^m$21.24$^s$ & 20.00 & 1204 & 3.51$^{+0.45}_{-0.37}$ & 0.91$^{+0.57}_{-0.41}$ \\
H$\alpha$ offcentre 1 & 20$^h$14$^m$51.820$^s$ & -24$^d$30$^m$18.51$^s$ & 30.96 & 1433 & 4.15$^{+0.64}_{-0.49}$ & 0.37$^{+0.39}_{-0.29}$ \\
H$\alpha$ offcentre 2 & 20$^h$14$^m$51.364$^s$ & -24$^d$30$^m$24.63$^s$ & 39.26 & 1962 & 3.21$^{+0.29}_{-0.26}$ & 1.10$^{+0.46}_{-0.35}$ \\
Off H$\alpha$ 1 & 20$^h$14$^m$51.391$^s$ & -24$^d$30$^m$16.39$^s$ & 28.26 & 747 & 4.04$^{+0.87}_{-0.64}$ & 0.36$^{+0.49}_{-0.31}$ \\
Off H$\alpha$ 2 & 20$^h$14$^m$51.791$^s$ & -24$^d$30$^m$26.81$^s$ & 41.10 & 1861 & 4.81$^{+0.73}_{-0.57}$ & 0.50$^{+0.42}_{-0.33}$ \\
Off H$\alpha$ 1+2 tied  & --- & --- & 69.37 & 2608 & 4.45$^{+0.48}_{-0.41}$ & 0.47$^{+0.27}_{-0.21}$ \\
Offcentre H$\alpha$ 1+2 tied & --- & --- & 70.22 & 3395 & 3.63$^{+0.26}_{-0.25}$ & 0.98$^{+0.32}_{-0.28}$ \\
Offcentre H$\alpha$ 1+2+centre tied & --- & --- & 90.22 & 4599 & 3.65$^{+0.23}_{-0.20}$ & 1.13$^{+0.29}_{-0.25}$\\
    \noalign{\smallskip}
    \hline
    \end{tabular}
    \tablefoot{The regions used for the X-ray spectral extractions are displayed in the left-hand panel of Fig.~\ref{fig:core}.}
    \label{tab:xrayhalpha}
\end{table*}

Narrow-band optical imaging was taken on the SOAR Telescope with the SOAR Optical Imager (SOI) \citep{2003SPIE.4841..286W} on September 6, 2010. A series of three exposures each with an exposure time of 1200 seconds was taken with the narrow-band filter (7580/85) that was centred on the redshifted H$\alpha$ (7572~\AA). A second set of three exposures of 720 seconds each was taken with a narrow-band continuum filter (7384/84) to determine the contribution from the continuum emission to the `on band' image. We flux calibrated the combined image with the spectrophotometric star LTT 7379 \citep{1994PASP..106..566H}. 
The observing conditions for both the optical imaging and spectra from SOAR were excellent, with sub-arcsecond seeing.
The images were aligned to the world coordinate system (WCS) using stars from the Two Micron All Sky Survey (2MASS). The resulting image, shown in the left panel of Figure~\ref{fig:core}, is aligned to within 0.\!$\arcsec$2 astrometric tolerance in right ascension and 0.\!$\arcsec$3 tolerance in declination. To correct the image from foreground Galactic extinction we used an E(B-V) = 0.1491 and assumed A$_V$/E(B-V) = 3.1 \citep{1998ApJ...500..525S}. We calculated a total H$\alpha$+[NII] luminosity of $189\pm6\times10^{40}$~erg s$^{-1}$ for the cluster inside a circular aperture with a radius of 8$\arcsec$ (21.5~kpc) and centred it on 20$^h$14$^m$51.57$^s$ -24$^d$30$^m$22.3$^s$ to avoid a poorly subtracted star near the galaxy. We took a background from an 8$\arcsec$ radius circle of blank sky centred on 20$^h$14$^m$54.29$^s$ -24$^d$30$^m$13.8$^s$. In the middle and right panels of Figure~\ref{fig:core}, we overlay contours tracing the SOAR emission on the inner regions of the broadband X-ray surface brightness image and contour binned temperature maps from Fig.~\ref{fig:xray}.

Optical spectra of the BCG were taken with the Goodman spectrograph \citep{2004SPIE.5492..331C} on July 25, 2012. The Goodman spectra were taken with the 600 $\ell$/mm grating ($\sim$2600~\AA~coverage) centred on 6500~\AA\ with the 1\farcs68 wide slit, which corresponds to an approximate rest wavelength range of 4510-6760~\AA. We observed the BCG at a position angle $110\degr$ east of north, aligned with the elongation of the central H$\alpha$ region, and centred on the brightest pixel of the BCG. 

\subsubsection{SOAR analysis}\label{sec:soar_analysis}

We reduced the 2D spectra using the standard IRAF spectral reduction routines in the NOAO {\it onedspec} and {\it twodspec} packages. FeAr and quartz lamps were observed before and after each observation. The CCD on Goodman exhibits severe spectroscopic (multiplicative) fringing from the interference patterns of the monochromatic light. The fringing is approximately 20\% peak-to-peak in wavelengths beyond 7000~\AA. To make a fringe correction frame we normalised the overall response in the quartz flat to a third order spline. We did not detect variations in the fringe pattern between the before and after quartzes: normalised fringe frame variations were $<$ 0.5\%. We were able to reduce 20\% peak-to-peak fringing down to 2\%. For wavelength calibration we identified lines in the FeAr lamp (which is contaminated with helium) spectra. We verified the centres of the night sky lines in the object frames ($\lambda\lambda$5577, 5889, 6300, 6363, 6863, 6923, 7276, 7316, and 7340~\AA) were within 1~\AA\ of our wavelength solution. We flux calibrated the spectra using the {\it APALL} super-task with observations of the spectrophotometric standard star LTT 9491. To examine emission features in the spectra we extracted 3 pixel (0\farcs45) wide 1D spectra. The results of these spectra are in Table~\ref{tab:goodmanlines}. We note that pixel \#812 corresponds to the BCG peak and that the pixel number increases from west to east (positive in RA).

All fluxes and equivalent widths, computed using {\it splot}, are calculated in the observer's frame such that we can also estimate redshifts in each of the lines to track variation of the velocities of different elements in the cluster. We use a Gaussian profile to fit each of the lines and the {\it splot} bootstrap resampling (100 realisations) method to compute errors on the Gaussian profile fits.  We estimated our background and continuum subtraction from a linear fit of two continuum ranges $\approx$ 20~\AA\ from the outer edges of each measured emission line. The error on the continuum was from the root mean square from an emission-free region. We fix this at $\sigma_0$ = 4.196$\times$10$^{-18}$ erg cm$^{-2}$ s$^{-1}$ \AA$^{-1}$ over the range 5600-7600~\AA. Based on the width of single unsaturated
FeAr lines in the calibration spectra, the instrumental line velocity width of the SOAR/Goodman setup used was 294 km~s$^{-1}$, consistent with expectations. We subtracted the instrumental velocity from the observed velocity width in quadrature to estimate residual observed velocity widths. 

We estimated the redshift of the stellar emission by Fourier cross-correlating the continuum emission in RXC~J2014.8-2430 to the spectrum of an elliptical galaxy SDSS J120028.87-000724.8 ($z = 0.0813\pm0.0002$). Using the {\it fxcor} task in IRAF, we shifted the Sloan Digital Sky Survey (SDSS) spectrum to the baseline estimated redshift (0.1555) of RXC~J2014.8-2430. We binned the SDSS spectrum to match the lower spectral resolution SOAR spectrum. We extracted a 40 pixel wide spectrum for RXC~J2014.8-2430 centred on our nominal centre. The correlation result was based on an emission-line-free range between 6100-7100~\AA\ and the error is based on results from 1000 randomly selected sections 200 \AA\ wide. We find a velocity shift to the baseline estimate of $(-10.57\pm29.10)~\rm km~s^{-1}$, which is statistically consistent with zero shift from the nominal emission-line redshift. 

In order to probe the local environment of the H$\alpha$ filaments, we extracted and fit X-ray spectra of regions defined to be on and off of the most emissive H$\alpha$ region. The results are presented in Table~\ref{tab:xrayhalpha}. We find the coolest and most metal-rich region, `H$\alpha$ centre', corresponds to the brightest H$\alpha$ emission.

\subsection{ALMA Band 3}\label{sec:alma}

\subsubsection{ALMA Band 3 observations}\label{sec:almaobs}

We use millimetre-wave observations performed by ALMA on December 24, 2018, to probe the location and dynamics of the molecular gas traced by the lowest transition of the dominant form of carbon monoxide, CO(1-0), which has a rest frequency of 115.271~GHz.  

Through the ALMA Cycle 7 project 2018.1.00940.S, the central region of RXC~J2014.8-2430 was observed in Band 3 with the 12-metre `main' array of ALMA.
The data were reduced by the calibrated measurement set (CalMS) service of the European ALMA Regional Centre using the Common Astronomy Software Applications (CASA) version 5.4.0, with the methodology closely following the ALMA pipeline.
The ALMA observations span four spectral windows -- 84.16-86.15, 86.12-88.10, 96.16-98.14, and 98.21-100.09~GHz, where the last window was chosen to include the redshifted emission from CO(1-0).
At the redshift of the cluster, the maximum recoverable scale of 26\farcs8 in the observation corresponds to 72.2~kpc.
The ALMA observations reach a continuum sensitivity of $\approx 12~\mu$Jy RMS, and a line sensitivity of $\approx 6$~mJy per $10~\rm km~s^{-1}$ channel.

\begin{figure}
  \includegraphics[width=80.2mm]{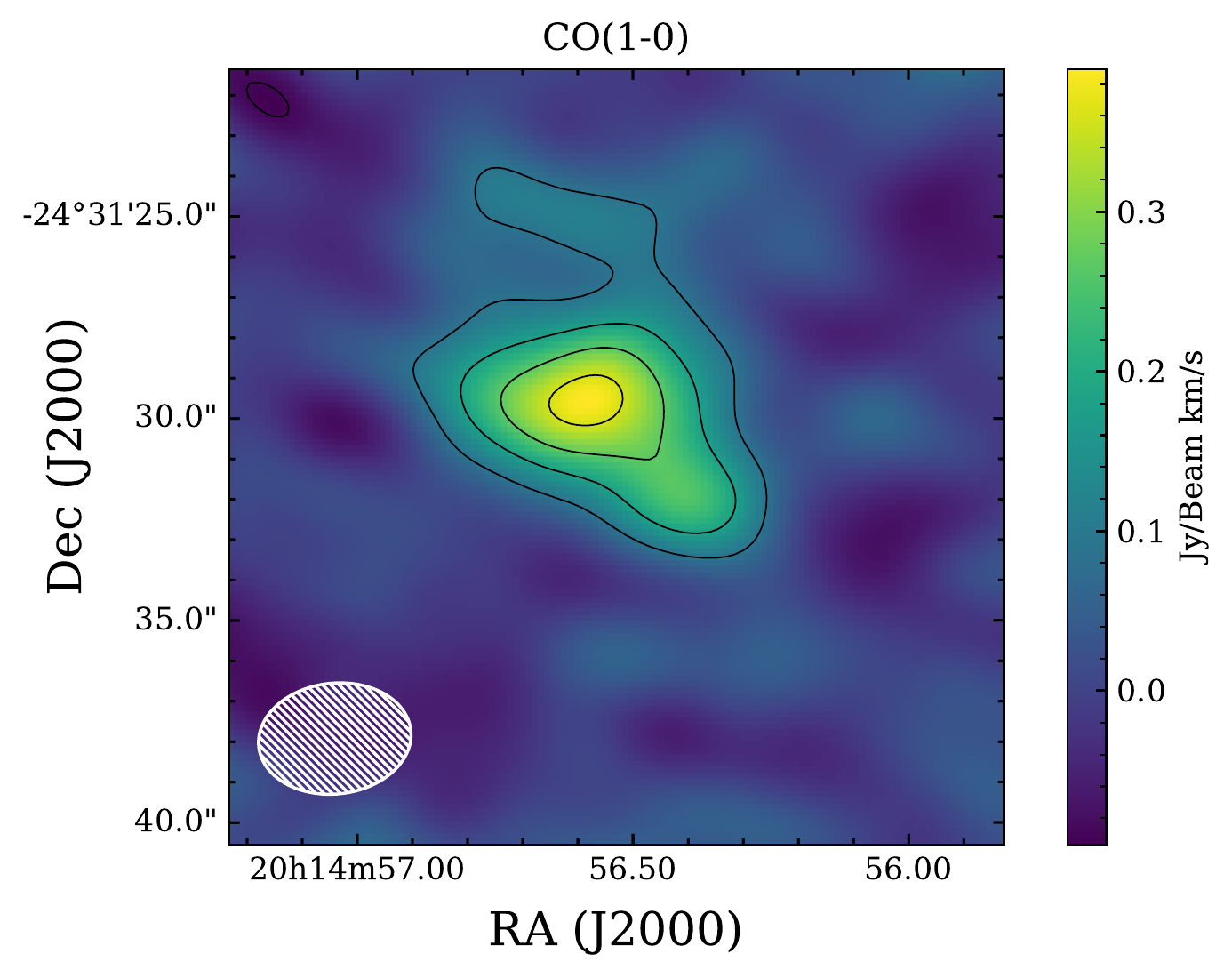}\\
  \includegraphics[width=83mm]{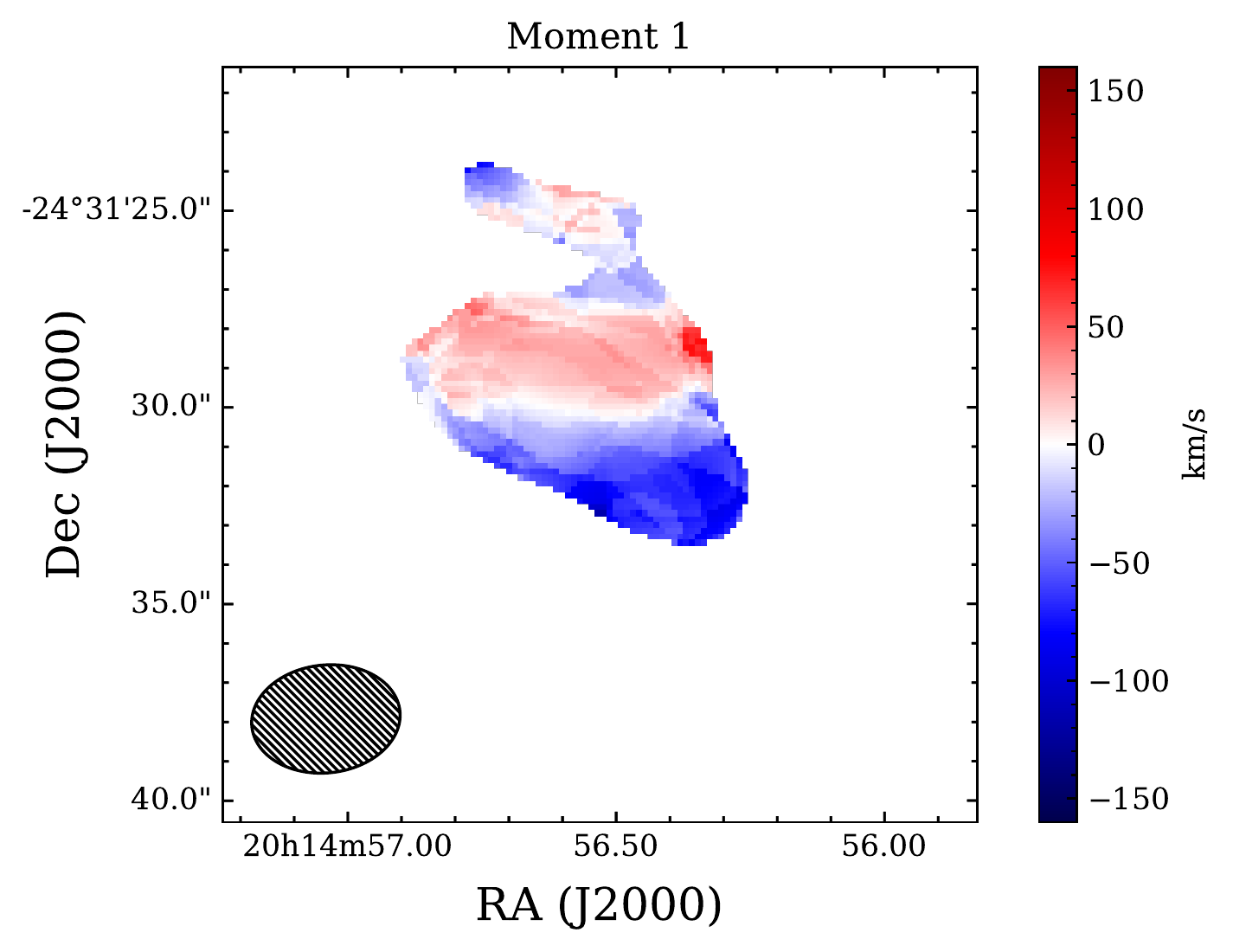}\\
  \includegraphics[width=81.5mm]{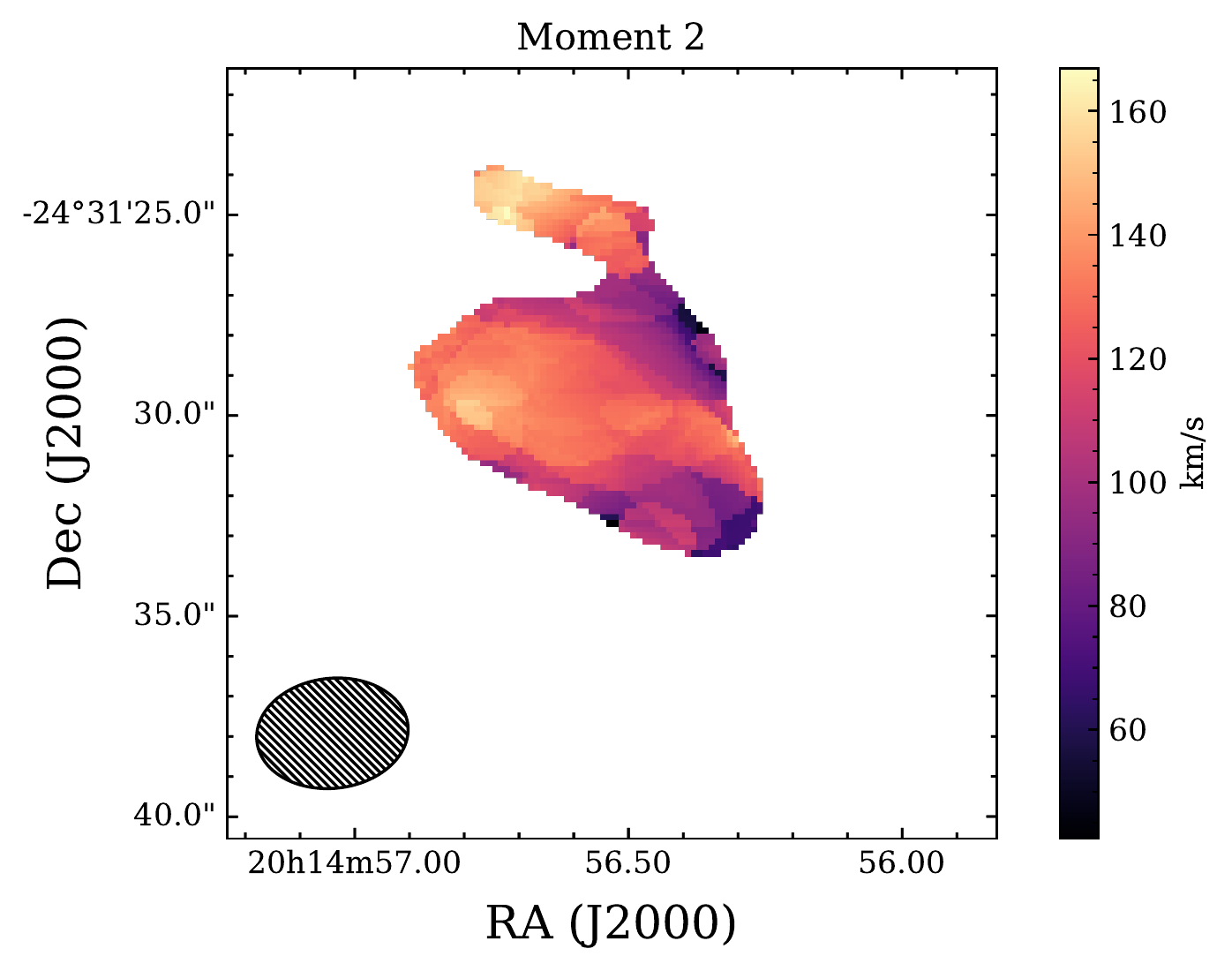}
\caption{\label{fig:CO_moments}
        Moment maps of the CO(1-0) emission.  Contours in the moment=0 map (first panel, showing CO(1-0) line intensity) start at 2$\sigma$ and are incremented in 2$\sigma$ steps. 
        The moment=1 and 2 maps, tracing velocity and velocity dispersion, are masked for values where the intensity of the CO(1-0) is less than 2$\sigma$.
        The synthesised beam of ALMA is depicted in the lower-left corners (hatched ellipses). We note the moment maps are not corrected for beam smearing effects.}
\end{figure}

\subsubsection{ALMA CO(1-0) analysis}\label{sec:alma_CO}

The calibrated ALMA data were reimaged using the latest version of CASA at the time of writing (CASA 6.3), using the \texttt{tclean} task to produce a spectral datacube for the CO(1-0) at 30~km~s$^{-1}$ resolution, which is used in the following sections.  For all imaging, we use natural weighting of the visibilities.
For completeness, we include individual CO(1-0) channel maps binned to 80~km~s$^{-1}$ spectral resolution in Fig.~\ref{fig:alma_co_channelmaps} in Appendix~\ref{sec:appendix:channelmaps}.

\begin{figure*}
\centerline{
 \includegraphics[clip,trim=00mm 0mm 0mm 0mm,height=56mm]{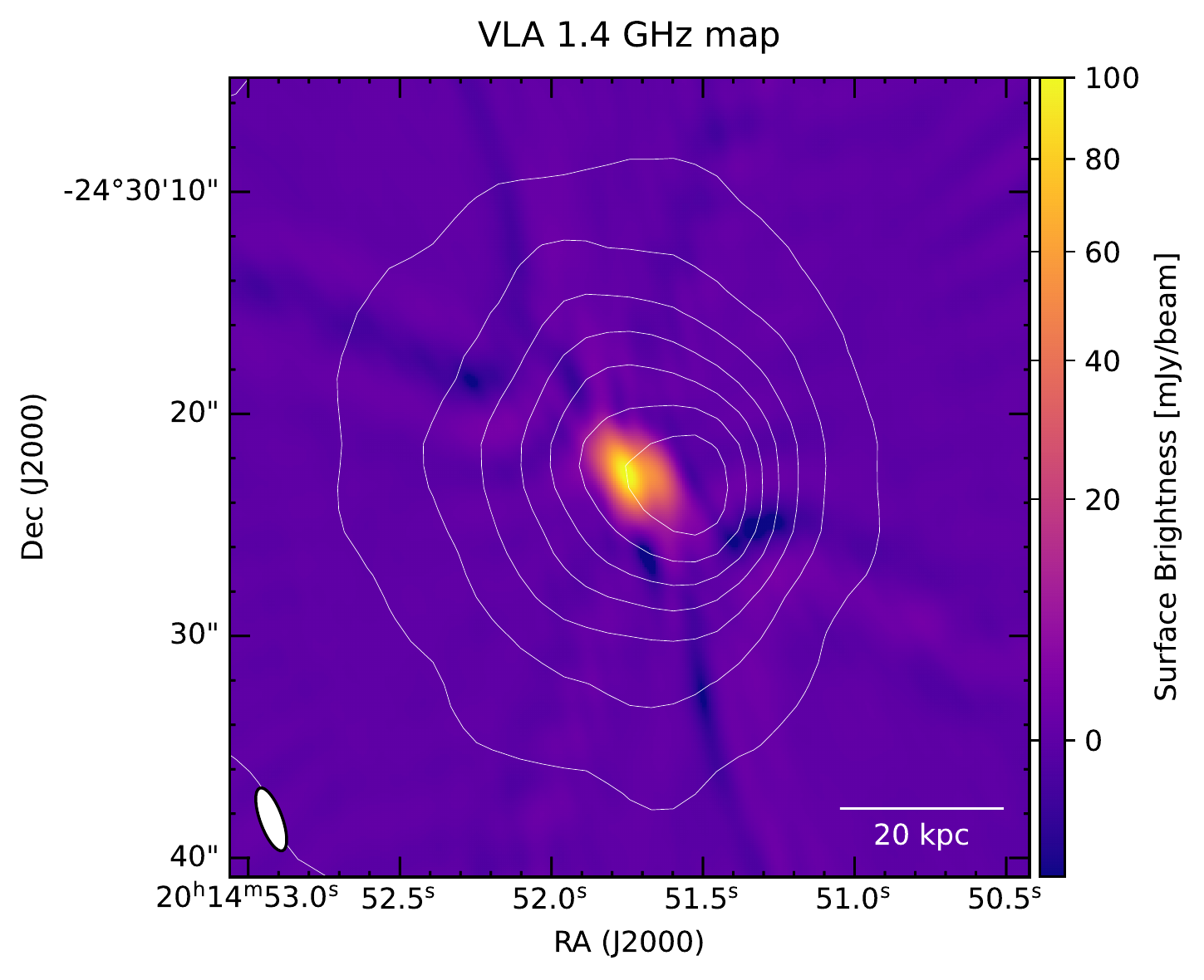}
 \includegraphics[clip,trim=27mm 0mm 0mm 0mm,height=56mm]{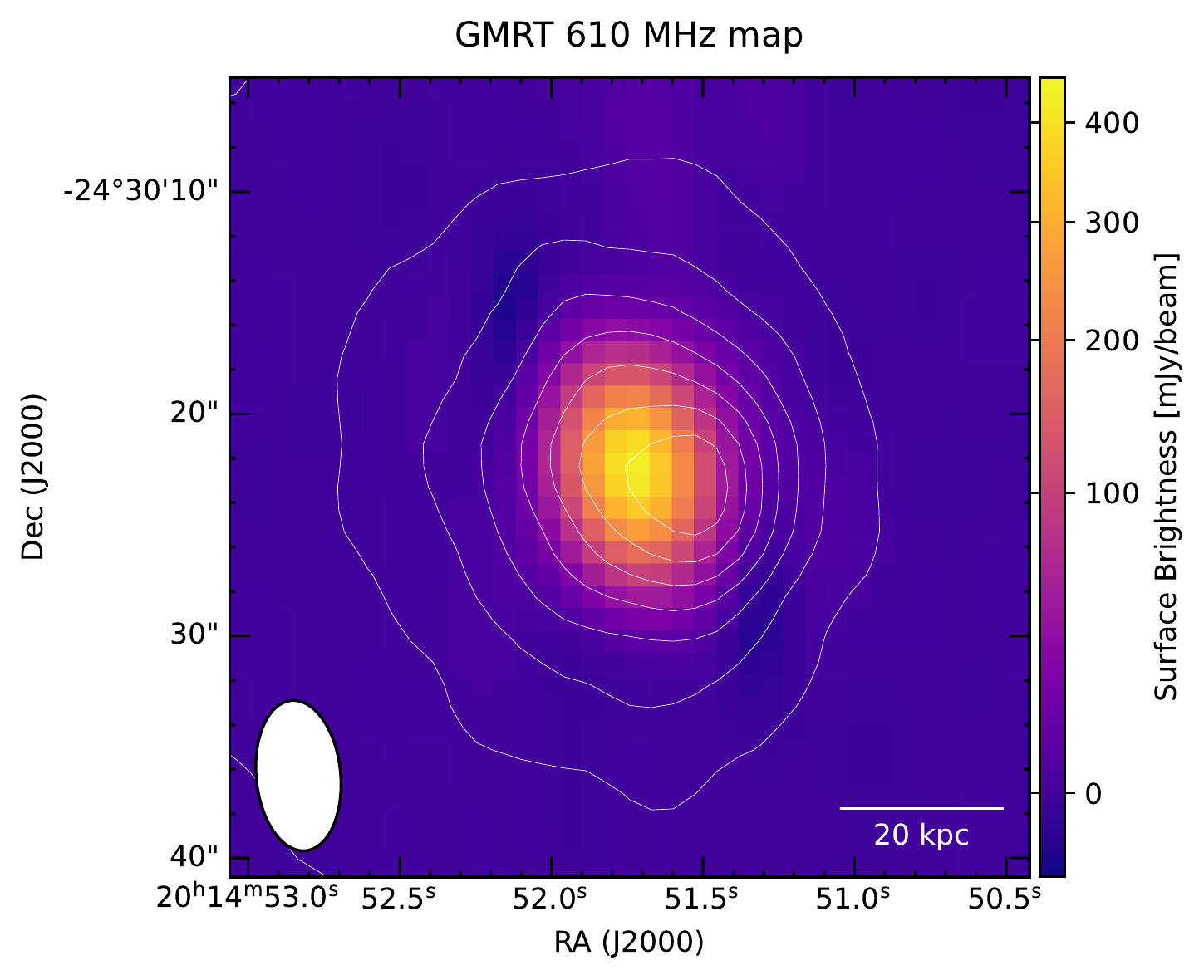}
 \includegraphics[clip,trim=27mm 0mm 0mm 0mm,height=56mm]{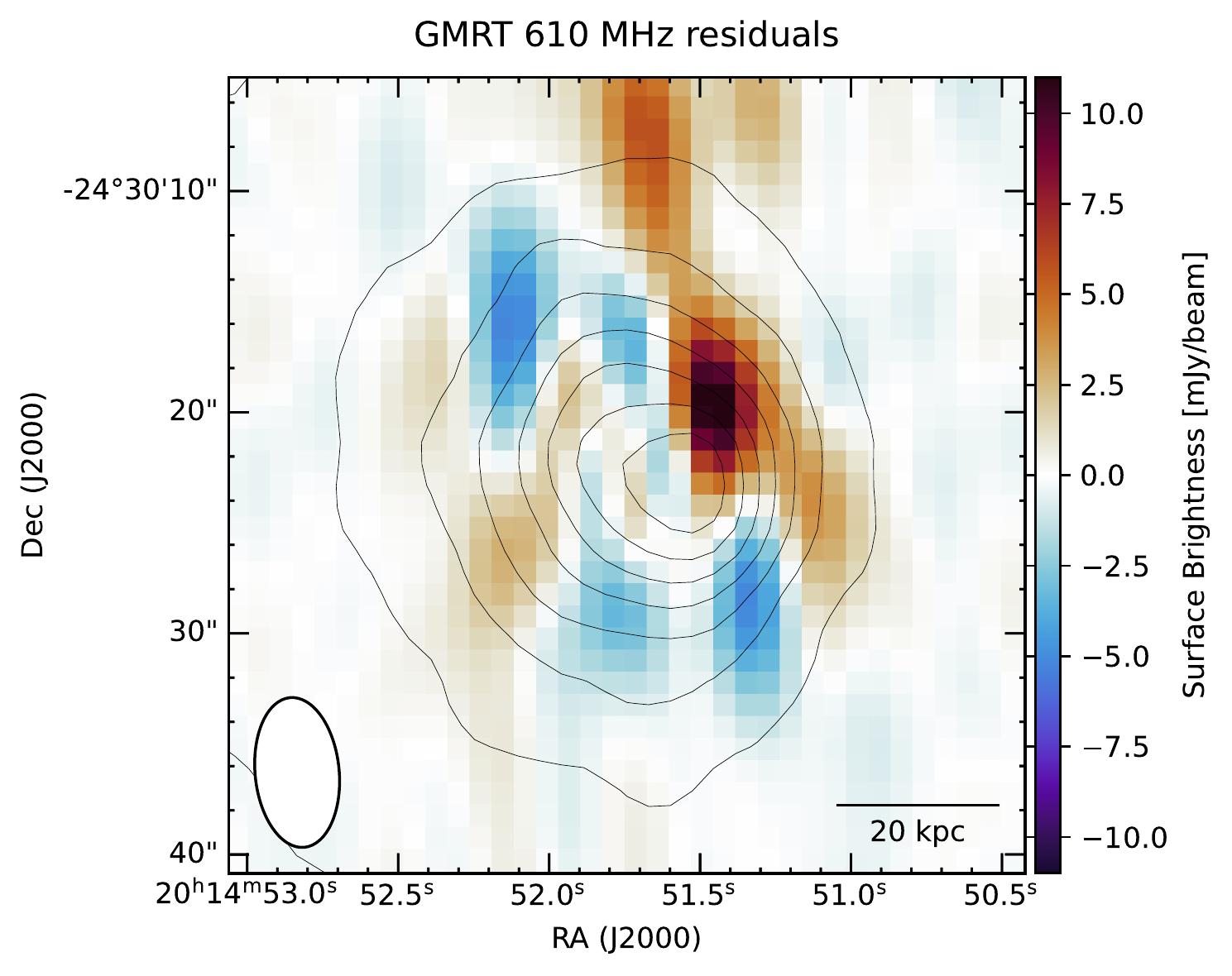}
}
\caption{\label{fig:radio}
        Low-frequency radio imaging and residuals for RXC~J2014.8-2430. {\bf Left:} VLA 1.4~GHz image of RXC~J2014.8-2430, shown on a square-root scale, with contours from the X-ray image in Fig.~\ref{fig:xray} overlaid. The beam size and position angle are depicted in the lower-left corner.
        {\bf Middle:} GMRT 610~MHz image of RXC~J2014.8-2430, shown on a square-root scale, with contours from the X-ray image in Fig.~\ref{fig:xray} overlaid. The beam size and position angle are depicted in the lower-left corner.
        {\bf Right:} Residuals in the GMRT imaging after subtraction of a Gaussian source model convolved with the synthesised beam, revealing a large residual feature to the west of the AGN. Contours from the X-ray image in Fig.~\ref{fig:xray} are overlaid. The beam size and position angle are depicted in the lower-left corner, and the scale is linear.}
\end{figure*}

The reference frequency chosen to make the moment maps was 99.759~GHz, corresponding to the CO(1-0) line being redshifted to $z=0.1555$. Fig.~\ref{fig:CO_moments} shows the moment=0, 1, and 2 maps of the CO(1-0) transition, respectively tracing the total CO(1-0) intensity, the spatial dependence of the velocity, and the velocity dispersion of the line.
The moment=1 map (second panel of Fig.~\ref{fig:CO_moments}), corresponding to velocity, shows a largely linear gradient from south to north, with the south-western portion more blueshifted.  
The moment=2 map (third panel of Fig.~\ref{fig:CO_moments}), corresponding to velocity dispersion, shows that the dispersion is higher towards the core of the BCG, but lower towards the south-western spur.
We note that the moment maps are not corrected for beam smearing effects, as they are largely being used to gauge the phenomenological behaviour of the cold gas.  
Using the channel maps (Fig.~\ref{fig:alma_co_channelmaps}), we chose several regions of interest for more detailed spectral analysis, presented in Section~\ref{sec:velocity_struct}.

\subsection{Low-frequency radio observations}\label{sec:radio}

\begin{table}
    \centering
    \caption{Archival radio observations.}
    \begin{tabular}{cccc}
    \hline\hline\noalign{\smallskip}    
    Survey & $\nu$ & Flux & $\sigma_{\rm flux}$ \\
    Name & (MHz) & (Jy) &  (Jy) \\ \noalign{\smallskip}
    \hline
    \noalign{\medskip}
    GLEAM\tablefootmark{a} & 80 & 3.7 & 0.16  \\
    GLEAM & 88 & 3.45 & 0.12  \\
    GLEAM & 95 & 2.92 & 0.1  \\
    GLEAM & 103 & 2.93 & 0.09  \\
    GLEAM & 111 & 2.58 & 0.07  \\
    GLEAM & 118 & 2.53 & 0.06  \\
    GLEAM & 126 & 2.22 & 0.06  \\
    GLEAM & 134 & 2.23 & 0.06  \\
    GLEAM & 147 & 1.93 & 0.05  \\
    TGSS ADR1\tablefootmark{b} & 150 & 1.77 & 0.18 \\
    GLEAM & 154 & 1.93 & 0.04\\
    GLEAM & 162 & 1.75 & 0.04 \\
    GLEAM & 170 & 1.61 & 0.04 \\
    GLEAM & 177 & 1.64 & 0.06 \\
    GLEAM & 185 & 1.58 & 0.04 \\
    GLEAM & 193 & 1.38 & 0.04 \\
    GLEAM & 200 & 1.45 & 0.04 \\
    GLEAM & 208 & 1.37 & 0.07 \\
    GLEAM & 223 & 1.36 & 0.07 \\
    GLEAM & 231 & 1.27 & 0.03 \\
    WISH\tablefootmark{c} & 352 & 0.842 & 0.006  \\
    Texas\tablefootmark{d} & 365 & 0.849 & 0.03 \\
    RACS\tablefootmark{e} & 887 & 0.376 & 0.027 \\
    NVSS\tablefootmark{f} & 1400 & 0.229 & 0.007 \\
    VLASSQL\tablefootmark{g} & 3000 & 0.0964 & 0.0006 \\
    \noalign{\smallskip}
    \hline
    \end{tabular}
    \tablefoot{
        \tablefoottext{a}{GLEAM data can be found in \citet{2017MNRAS.464.1146H}.} 
        \tablefoottext{b}{TGSS ADR1 data can be found in \citet{TGSS}.}
        \tablefoottext{c}{WISH data can be found in \citet{WISH}.}
        \tablefoottext{d}{Texas data can be found in \citet{1996AJ....111.1945D}.}
        \tablefoottext{e}{RACS data can be found in \citet{2021PASA...38...58H}.}
        \tablefoottext{f}{NVSS data can be found in \citet{1998AJ....115.1693C}.}
        \tablefoottext{g}{VLASSQL data can be found in \citet{2021ApJS..255...30G}.}
    }
    \label{tbl:flux}
\end{table}

We performed an extensive cross-check of our target in archival radio observations and public catalogues. Radio emission associated with the BCG was detected from 80~MHz to 4~GHz. We summarise in Table~\ref{tbl:flux} the total flux measurements of the archival catalogues, including data from the GaLactic and Extragalactic All-sky Murchison Widefield Array (MWA) Survey \citet[GLEAM;][]{2017MNRAS.464.1146H}, The  GMRT 150 MHz All-sky Radio Survey: First Alternative Data Release \citep[TGSS ADR1;][]{TGSS}, the Westerbork in the Southern Hemisphere survey \citep[WISH;][]{WISH}, the Texas Survey \citep{1996AJ....111.1945D}, the Rapid Australia Square Kilometer Array Pathfinder (ASKAP) Continuum Survey \citep[RACS;][]{2021PASA...38...58H}, the National Radio Astronomical Observatory (NRAO) VLA Sky Survey  \citep[NVSS;][]{1998AJ....115.1693C}, and the Karl G. Jansky Very Large Array Sky Survey \citep[VLASS;][]{2020PASP..132c5001L,2021ApJS..255...30G}.

Further, we examined in more detail the archival images from NVSS and VLASS. 
NVSS detected the radio AGN as a compact source, but the 45$\arcsec$ resolution is not sufficient to probe the radio bubbles. VLASS on the other hand, provides a resolution of 2\farcs5, though the image may be dynamic range limited by the bright central AGN.  Neither one reveals any obvious radio bubble emission.
In order to search for radio bubbles in more detail, we reduced and analysed two low-frequency ($\nu < 2~\rm GHz$) archival radio observations of RXC~J2014.8-2430, described below.

\subsubsection{VLA radio observations}\label{sec:vla}

The VLA performed L-band (1--2~GHz) observations on May 29, 2014, of the continuum radio emission from RXC~J2014.8-2430 as part of Project code 14A-280.  The total on source time was 71 minutes. The data were processed in CASA 4.1.0.24668 using Expanded VLA (EVLA) Pipeline 1.2.0. After flagging, $85\%$ of the on-source data remain unflagged.  Additionally, 25\% of the bands, corresponding to spectral windows 1, 3, 8, and 9, were excluded by hand due to the presence of radio frequency interference (RFI). Several iterations of the \texttt{clean} algorithm and self-calibration of the phases were applied to produce a model image with symmetric sidelobes in the residuals. The resulting VLA image is shown in the left panel of Fig.~\ref{fig:radio}. The observations were taken in A Configuration (the most extended configuration) and the synthesised beam in our VLA imaging is $1\arcsec \times 3\arcsec$ at a position angle of $20\degr$. 

\subsubsection{GMRT radio observations}\label{sec:gmrt}

As part of a larger observing programme (project code 14JHC01), RXC~J2014.8-2430 was observed at 610~MHz by the GMRT for 3.1~hours in 6~equal blocks during the night of June 21, 2008. The data were recorded every 16.8~seconds in 256~channels covering 32~MHz in one polarisation (RR). The flux calibrator 3C48, also used for bandpass calibration, was observed for 16~minutes directly following the last block. We processed the archival radio observations of RXC~J2014.8-2430 using the Source Peeling and Atmospheric Modelling (SPAM) package \citep{2009A&A...501.1185I} for the data reduction, following a standardised GMRT data reduction recipe \citep[e.g. see][]{2014ApJ...785....1B}.
About 40\% of the data were lost due to various non-active and malfunctioning telescopes, and another 30\% due to RFI and peculiarities in the data. The latter is unfortunately not uncommon for older, hardware-correlator data. The final image, made with robust weighting, has an RMS background noise of $\sim 0.17$~mJy/beam, and a (synthesised beam) resolution of $6.\!\arcsec8 \times 3.\!\arcsec8$ at a position angle (PA) of $5\fdg5$. The resulting GMRT image is shown in the middle panel of Figure~\ref{fig:radio}. Due to dynamic range limitations ($\sim 2500$ in this image), residual sidelobes from the strong central radio source in RXC~J2014.8-2430 are a dominant contribution to the central background noise, raising the local noise by a factor of two.

\subsubsection{VLITE observations}\label{sec:VLITE}
\begin{figure}
\centerline{
  \includegraphics[width=0.495\textwidth]{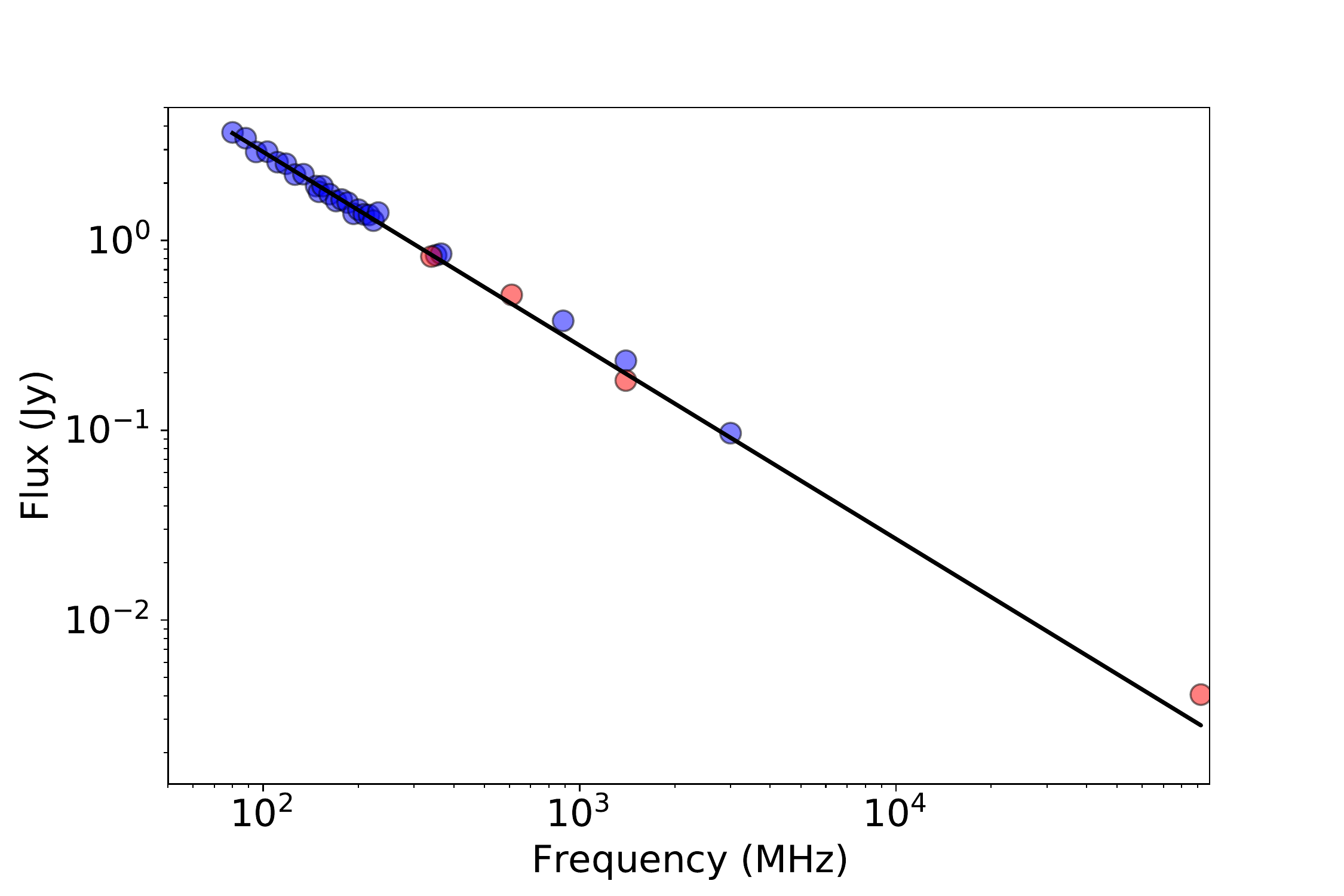}
}
\caption{\label{fig:radio_spec} Low-frequency radio spectral index of the central compact source. Blue points show flux measurements from the literature (Table~\ref{tbl:flux}), and red points show, in order of frequency, the new VLITE, GMRT, VLA, and ALMA measurements presented here. Flux measurements (excluding the 92~GHz measurement from ALMA) were fitted with a singular power law spectrum, shown in black ($\alpha=-1.02$). We note that the uncertainties are too small to be seen on the scale of the plot.}
\end{figure}

The VLITE is a commensal 338 MHz system that operates on the NRAO VLA in parallel with
standard observing programmes that use the 1-50 GHz feeds \citep{2016SPIE.9906E..5BC}. VLITE's field of view is significantly larger than that of the higher frequencies that it operates in parallel with, thus making it a powerful resource for detecting sources both within and beyond the field of view of the higher-frequency primary observing programme. RXC J2014.8-2430 was detected in six separate VLITE observations at distances between $0\fdg5$ and $1\fdg4$ from the phase centre between December 10, 2017, and November 22, 2020. All observations were calibrated, imaged, and catalogued following standard VLITE processing as described in \citet{2015ApJ...812..153O}. The central radio source is unresolved in all VLITE observations and the catalogued fluxes obtained from the VLITE database pipeline \citep{2019ASPC..523..441P} were averaged together. We conservatively adopt a 20\% flux density error that includes the local image noise as well as uncertainties related to source fitting and the flux scale. The new VLITE flux measurement of $S_{\mbox{\tiny338~\rm MHz}}=838.8 \pm 167.8$~mJy is shown in red in Figure~\ref{fig:radio_spec}.

\subsubsection{Low-frequency radio analysis}\label{sec:radio_analysis}

The dominant central 610~MHz radio source (Figure~\ref{fig:radio}, middle) can be accurately modelled by a single 2D Gaussian of dimensions $6\farcs8 \times4\farcs6$ and PA $8\fdg2$, with uncertainties on these dimensions being less than 0.1~\%. The peak and integrated flux for this Gaussian are $430.4 \pm 0.3$~mJy/beam and $S_{\mbox{\tiny610~\rm MHz}}=515.6 \pm 0.5$~mJy, respectively. The minor axis is slightly larger than the synthesised beam minor axis, which indicates that the radio source is somewhat resolved in approximately the east-west direction. This indicates that something beyond the dimensions of the central AGN is producing radio emission. Given the well-studied nature of central BCGs, this is most likely some radio jet activity within the boundaries of the BCG in RXC~J2014.8-2430.  Alternatively, the extended central emission could be due to the presence of a mini-halo, often found in sloshing cool-core clusters \citep[e.g.][]{2014ApJ...781....9G}.

\begin{figure}
\begin{center}
 \includegraphics[clip,trim=0mm 0mm 30mm 10mm,width=0.45\textwidth]{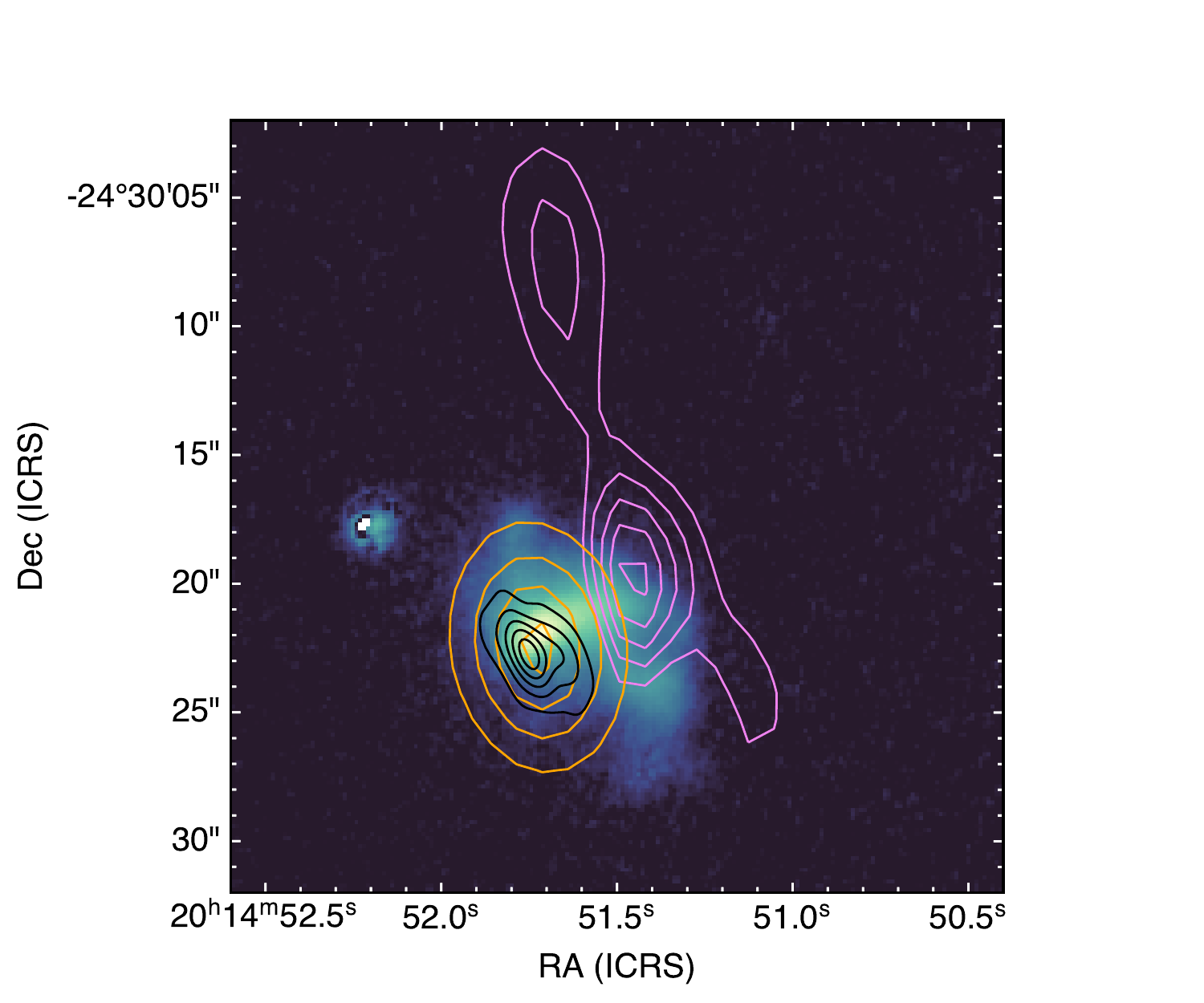}
\end{center}
\caption{\label{fig:radio_overlays}
        SOAR net H$\alpha$ image with contours based on the images in Fig.~\ref{fig:radio}.
    VLA 1.4~GHz contours are overlaid at (10, 30, 50, 70, 90) mJy/beam in black,  GMRT 610~MHz contours are overlaid at (100, 200, 300, 400) mJy/beam in orange, and GMRT 610~MHz residuals at (3, 5, 7, 9, 11) mJy/beam are overlaid in violet.
    The image is 30\arcsec\ wide, corresponding to 80.8~kpc.}
\end{figure}

The peak radio emission at 1.4~GHz agrees within 1$\arcsec$ with that seen in GMRT data at 610~MHz, and with the peak of the SOAR optical imaging (Figure~\ref{fig:radio_overlays}). 
The central radio source has an integrated 1.4~GHz flux density of $S_{\mbox{\tiny 1.4~\rm GHz}}=182.6 \pm 0.4~\rm mJy$, where the reported error is statistical only. 
Variations due to sidelobe structure and aperture choice clearly dominate the errors in this measurement, and are $\approx$1.2~mJy.
From the first epoch VLASS catalogue \citep{Gordon2020}, we find the continuum integrated flux in the 2-4~GHz band, $S_{\mbox{\tiny 2-4~\rm GHz}} = 96.4 \pm 0.6$~mJy.
Taken together with the archival low-frequency radio observations, we find a median spectral index $\alpha=-1.02$ (where $S \propto \nu^\alpha$ (see Figure~\ref{fig:radio_spec}), indicative of ageing in the electron population.  We note that we have excluded the 92~GHz continuum data point provided by ALMA, which at 4.05~mJy $\pm 0.01$~mJy lies above the power law extrapolation.
For a luminosity distance of 741.8~Mpc, the integrated flux density at 1.4~GHz corresponds to a radio power of 
$P_{1400~\rm MHz}=1.2\times10^{32}~\rm erg~s^{-1}~Hz^{-1}$.  
This would position it among the most powerful known AGN hosted by a cool-core cluster
 when compared to the sample of \citet{2012MNRAS.427.3468B}.

Notably, the radio, millimetre-wave, and optical data all locate the AGN 3\farcs5 (9.4~kpc projected separation) to the east of the X-ray peak (see Figure~\ref{fig:radio} \& \ref{fig:radio_overlays}).
This offset is likely due to sloshing, which has displaced the coolest, densest
ICM gas away from the central AGN. At 1.4~GHz, the residual extending along 
the H$\alpha$ tail towards the X-ray peak is smaller than that at 610~MHz.

Subtracting the fit 2D Gaussian from the GMRT 610~MHz image provides a clear view of the fainter substructures directly surrounding the dominant central source. Before interpreting these substructures, it is important to realise that a dynamic-range limited image will have (positive and negative) residual sidelobes around bright sources. The strongest sidelobes are typically found in mirrored pairs, closest to sources with the highest peak fluxes. In the residual image in Figure~\ref{fig:radio} (right panel), it is straightforward to identify two nearby pairs of negative sidelobes. This indicates that any nearby positive residuals of similar amplitude are very likely to be residual sidelobes as well. There are several nearby positive residuals, for which all but one have comparable amplitudes as the negative sidelobes. 

The positive residual, west-northwest of the central source (Figure~\ref{fig:radio}, right panel) has more than twice the amplitude of the largest residual sidelobes. Furthermore, this excess does not have a similar positive counterpart at the mirror location (there is a small positive residual there, but it is much fainter). The strength of the excess and the lack of a counterpart strongly suggest that the brightest positive residual could be genuine radio emission. The excess appears to extend to the north and south-west into additional positive residuals. Although without mirror counterparts, these extensions are of similar amplitude as the residual sidelobes, and we therefore identify these as being possible spurious structure. The total flux in the excess region is approximately $8.5 \pm 1.0$~mJy.


\section{Discussion}

\subsection{Limits on the X-ray luminosity of the AGN}

We place an upper limit on the X-ray flux of an AGN point source using the tool {\tt celldetect} from the CIAO\footnote{\url{https://cxc.cfa.harvard.edu/ciao/}} software package. The events were restricted to the 0.5 - 7.0 keV energy range and the search was limited to fixed cell sizes of 1 pixel and 3 pixels near the cluster centre. The local cluster background was estimated (13.67$\pm$4.54 counts pixels$^{-2}$) from a 5$\arcsec$ circle near the centre of the cluster. We did not find any $>$3$\sigma$ detections of compact sources within 10$\arcsec$ of the cluster centre. Using the CIAO {\it aprates} task we calculated a 3$\sigma$ upper limit on the counts expected on an AGN point source. The algorithm\footnote{The details are found in \url{http://cxc.harvard.edu/csc/memos/files/Kashyap\_xraysrc.pdf}.} determines confidence intervals based on a Bayesian background-marginalised posterior probability distribution function of possible source counts. We assume the possible point source will nearly have all of its flux in a single cell and the background is the mean value computed within 5$\arcsec$ of the centre. Our 3$\sigma$ upper limit of 17 counts for a point source above the local background of extended cluster emission in the core of the cluster, corresponding to a limit of $<$ 1.55$\times$10$^{-14}$ erg~cm$^{-2}$~s$^{-1}$ (at $\bar{\rm{E}}$=2 keV), corresponding to a  0.5-7.0 keV X-ray luminosity of $<$ 5.39$\times$10$^{42}$ erg~s$^{-1}$ for a power-law ($\alpha$=1) point source at the cluster redshift. 

\subsection{Radio bubble limits}

Given the central entropy of the cluster and the existence of a strong central radio source, the apparent lack of a radio bubble may be surprising. The additional lack of radio lobes would suggest that any potential cavities may be along the line of sight. In order to estimate the size of a bubble expected in such a source, we used the relation in \citet{2006ApJ...652..216R}. As in \citet{2006ApJ...652..216R}, we define the X-ray cooling radius to be that within which the gas has a cooling time of less than 7.7$\times$10$^9$ years, which is the cosmic time elapsed since $z = 1$ to the present epoch for the cosmology adopted in this paper.
\citet{2006ApJ...652..216R} consider this cooling time representative of the time it has taken the cluster to relax and establish a cooling flow. We use the same XSPEC model fits in Section 2 and include a fit with the MKCFLOW model added and fix the low temperature to 0.1 keV. Similar to the modelling for the MEKAL model fits, the metallicity is tied across pairs of annuli, outside of the central region. The high temperature and  metallicity components of the MKCFLOW model are tied to the simultaneously fit MEKAL model. However, outside of the cooling radius the MKCFLOW normalisation is set to zero. We calculated the luminosity of each annulus after fitting using {\it lumin} in XSPEC in the extrapolated range of 0.1 to 100 keV to estimate the bolometric luminosity. We estimate the bolometric luminosity L$_{bol}=19.7\times10^{44}$ erg s$^{-1}$ in the range 0.11555 - 100.0 keV rest frame over the total MEKAL model inside all annuli (675~kpc). This luminosity is reasonably consistent with the bolometric luminosity calculated from {\it XMM-Newton} ($L_\textsc{x} = 21.06 \pm 0.07 \times 10^{44}$~erg s$^{-1}$ within $R_{500}$ = 1155.3 $\pm$ 4~kpc; \citealt{2009AA...498..361P}), which was over a slightly larger area. Using the technique from \citet{2004ApJ...607..800B} we define the cooling time for each of the annuli using $t_{cool} = 3nkT_\textsc{x}/2n_e n_H \Lambda(T,Z)$, where $\Lambda(T,Z)$ is the X-ray emissivity as a function of temperature and metallicity. We compute for $\Lambda(T,Z)$ using the normalisation from the MEKAL model and the bolometric luminosity found above. We assume a fully ionised plasma such that the total number density n = $2.3n_H$. We compute the cooling radius at a distance from the cluster centroid such that the cooling time is less than 7.7$\times$10$^9$ years. For this cluster the cooling radius is 105~kpc (39$\arcsec$). Inside this aperture we have a luminosity $L_\textsc{x}(< r_{cool}) = 4.71 \times 10^{44}$ erg s$^{-1}$.  It is worth noting that for cool-core luminosities $L_\textsc{x}(< r_{cool}) \gtrsim 10^{43}$ erg s$^{-1}$, all of the clusters in the samples of \citet{2006ApJ...652..216R}, \citet{2012MNRAS.421.1360H}, \citet{2013MNRAS.431.1638H}, and \citet{2015ApJ...805...35H} show strong evidence for X-ray cavities. With a much larger core X-ray luminosity, RXC~J2014.8-2430 thus presents a strong exception.

We use the relation in Figure 6 of \citet{2006ApJ...652..216R} to estimate the cavity radio power P$_{\rm cav} \sim 5\times 10^{44}$ erg s$^{-1}$ (assuming a $\gamma$=4/3 for relativistic particles, which gives a 4pV enthalpy) for a possible AGN source given our X-ray luminosity inside the cooling radius. With the bubble in pressure equilibrium with the X-ray gas ($\sim 5\times 10^{-10}$ erg cm$^{-3}$ near the core) and a bubble age $t_{\rm cav} 
\sim 10^7$ years (\citet{2004ApJ...607..800B} estimate ages for bubbles seen ranging from 0.5-15 $\times 10^7$ years) we would expect a pair of spherical bubbles to each have a radius of (3P$_{\rm cav} t_{\rm cav} \, 4\pi P_{\rm X})^{1/3}$ $\sim 5\arcsec$. Assuming the typical range of enthalpies from 1-16~pV gives a factor of five in bubble power in either direction. A lower power assuming a lower enthalpy of 1pV would give a bubble pair with radii $\sim 3\arcsec$. The largest bubbles expected would have radii $\sim 9\arcsec$ in size.  
\citet{2004ApJ...607..800B} indicate that this relation between radio power and X-ray luminosity is a limit on how much work the radio source could contribute to the system to compensate for the cooling, it is not necessarily required. Their figure assumed total enthalpy levels of the cluster from 1-16~pV, which is the observed range that clusters fall in if heating and cooling are balanced.

The range of plausible bubble sizes is consistent with the scales resolved by the VLA imaging (Section \ref{sec:vla}), while the smaller end of this range would be unresolved in the GMRT observations (Section \ref{sec:gmrt}).  However, the bright central AGN in both of the radio images leads to dynamic range issues, so that much of the structure left after modelling the sources could be dominated by residuals due to imperfect phase calibration.  Further, the age estimate of 10$^7$ years is at the boundary between the ages of ghost cavity systems, from which no 1.4~GHz emission is detected, and those with radio-filled lobes \citep{2008ApJ...686..859B}, while the spectral break could also lie below 610 MHz.  We would therefore not necessarily expect the radio luminosity of the lobes to be measurable in the available VLA or GMRT images.

\subsection{X-ray cavity toy model}\label{sec:toymodel}

It is possible that bubbles have formed along our line of sight, superposed on the projected centre of the cluster. Such bubbles might be difficult to detect in projection. We created a toy model to help determine potential sizes (3$\arcsec$-9$\arcsec$) of cavities that could not have been missed along our line of sight in {\it Chandra} data. As in Section \ref{sec:xray}, we use a standard $\beta$ model with the gas density, $\rho(r)$, in the form $\rho(r)\propto[1+(r/r_0)^2]^{-3\beta/2}$ \citep[e.g.][]{1988xrec.book.....S} and use Sherpa to fit the data, binned 2$\times$2 pixels. We create a 3D grid with 512 pixels (1$\arcsec$/pixel) per side and apply the $\beta$ model from the centre of the grid with added Poisson noise. We compute the emission measure (EM$\equiv \int n_en_p~dl$) from the 3D model by taking the square of each element and summing along an axis. We expect the projected form $f(b)\propto[1+(b/r_0)^2]^{-3\beta + 1/2}$  \citep{1988xrec.book.....S}. We scale the amplitude of the projected image such that the projected image has the same amplitude as the fit from the 2D $\beta$ model in Section \ref{sec:xray}. Before projecting the images, a pair of simulated bubbles can be added to the density profile along the line of sight by introducing empty spheres of different radii and distances from the cluster centre. 

\begin{figure}
\centerline{
  \includegraphics[trim=5mm 0mm 30mm 10mm, width=0.495\textwidth]{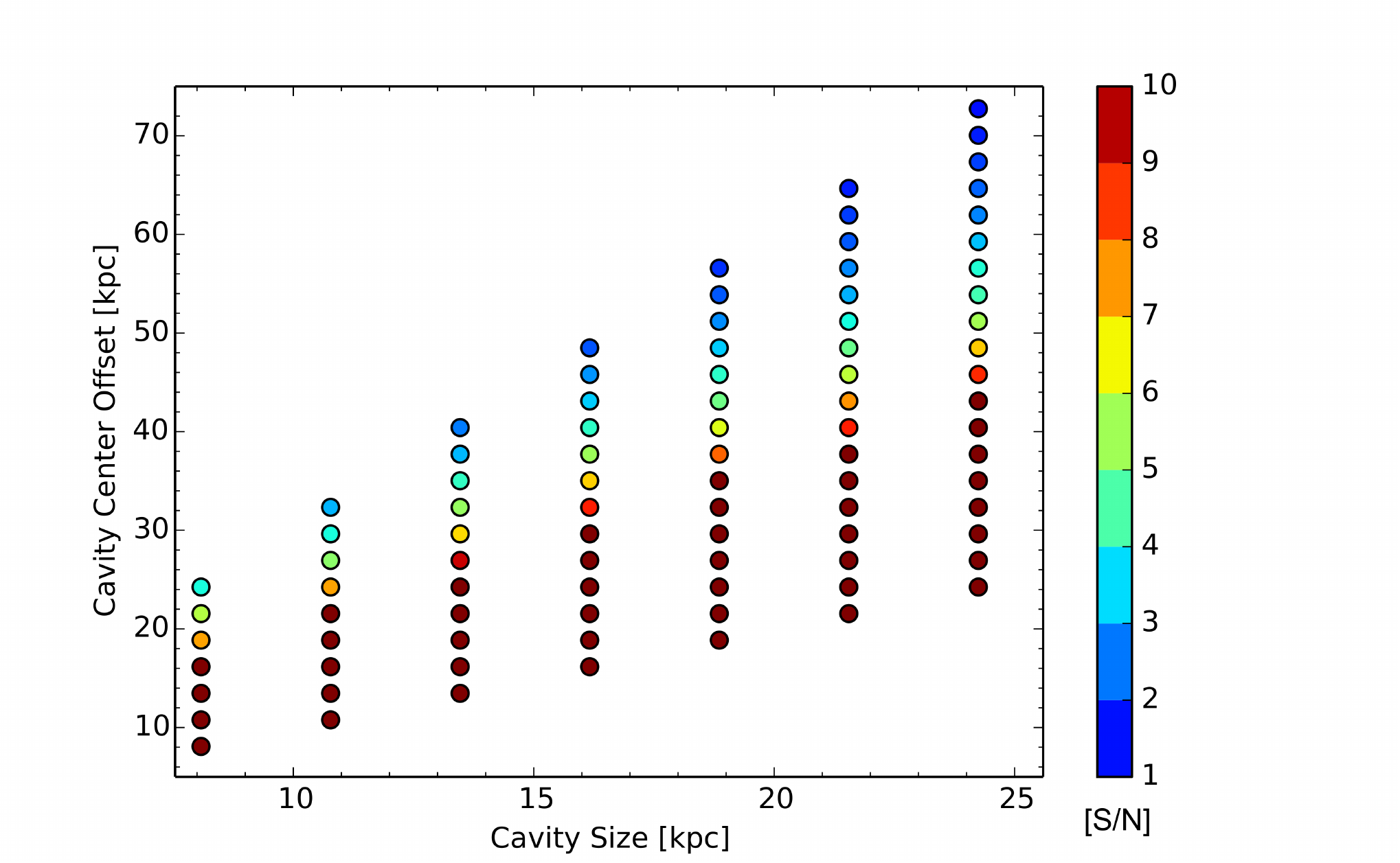}
}
\caption{Signal-to-noise ratio (S/N) at which depressions in the X-ray surface brightness would be detected determined for a range of cavity model configurations The significance level of each cavity configuration is indicated by the colour scale, with each value representing the number of standard deviations from a model without cavities. The cavity centre offset is the distance of the inner edge of the cavity from the centre of the cluster.}\label{fig:toymodel}
\end{figure}

To determine if a cavity created along the line of sight is significant, we compare the total counts in the flattened cluster image with a pair of cavities to the flattened image of the same cluster without cavities. The region of comparison is restricted to a circular aperture equal to the cavity size. In Figure~\ref{fig:toymodel} we show the significance of detection between the two given a Poisson count error. For a pair of cavities to be at a detection threshold of 5$\sigma$ or less, the cavities would have to be at least 20~kpc from the cluster core, with larger cavities needing to be at farther distances. Using the cluster sound speed in the cluster core,  $c_{\mathrm{s},1}=\sqrt{\gamma_{\textsc{p}}{k_{\mathrm{\textsc{b}}}T}/{\mu m_{\textsc{p}}}}$, with $\gamma_{\textsc{p}}$ = 5/3 and $\mu$ = 0.62, it would take a cavity $\sim 45$ million years to reach 20~kpc. With an expected cavity lifetime of less than 100 million years, it is possible for cavities on the smaller end ($<$ 6$\arcsec$) of the expected size to exist in this cluster along our line of sight without being detected in the X-ray image.

\begin{figure}
\begin{center}
  \includegraphics[width=0.24\textwidth]{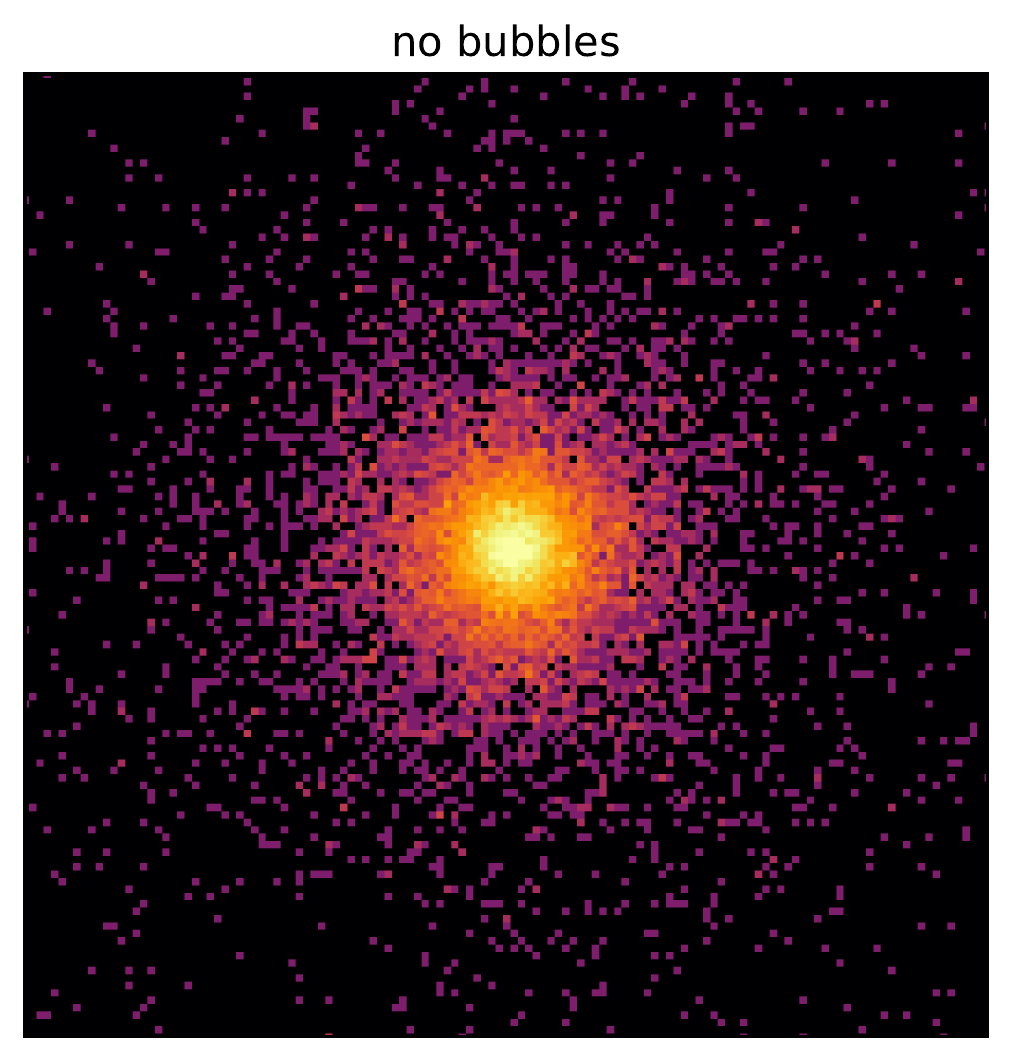} 
  \includegraphics[width=0.24\textwidth]{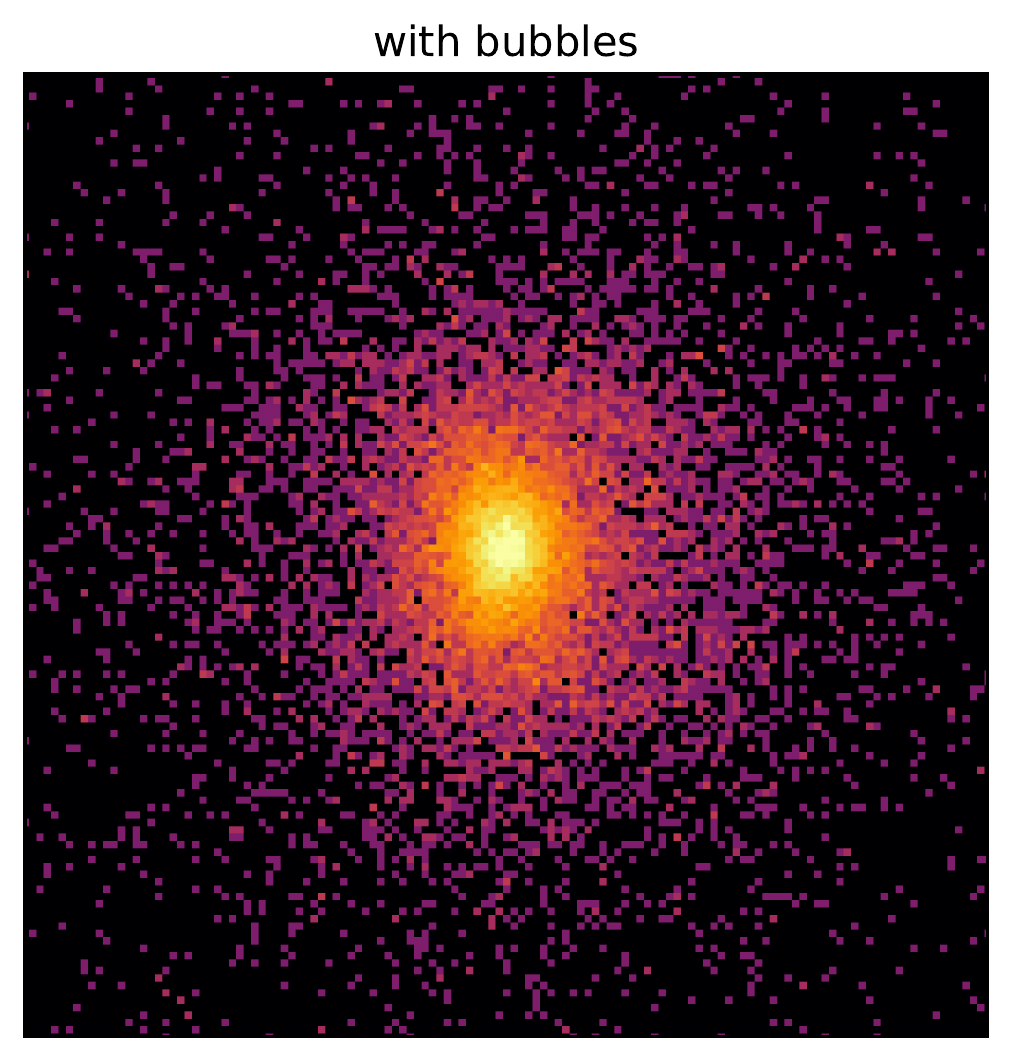}\\
  \includegraphics[width=0.24\textwidth]{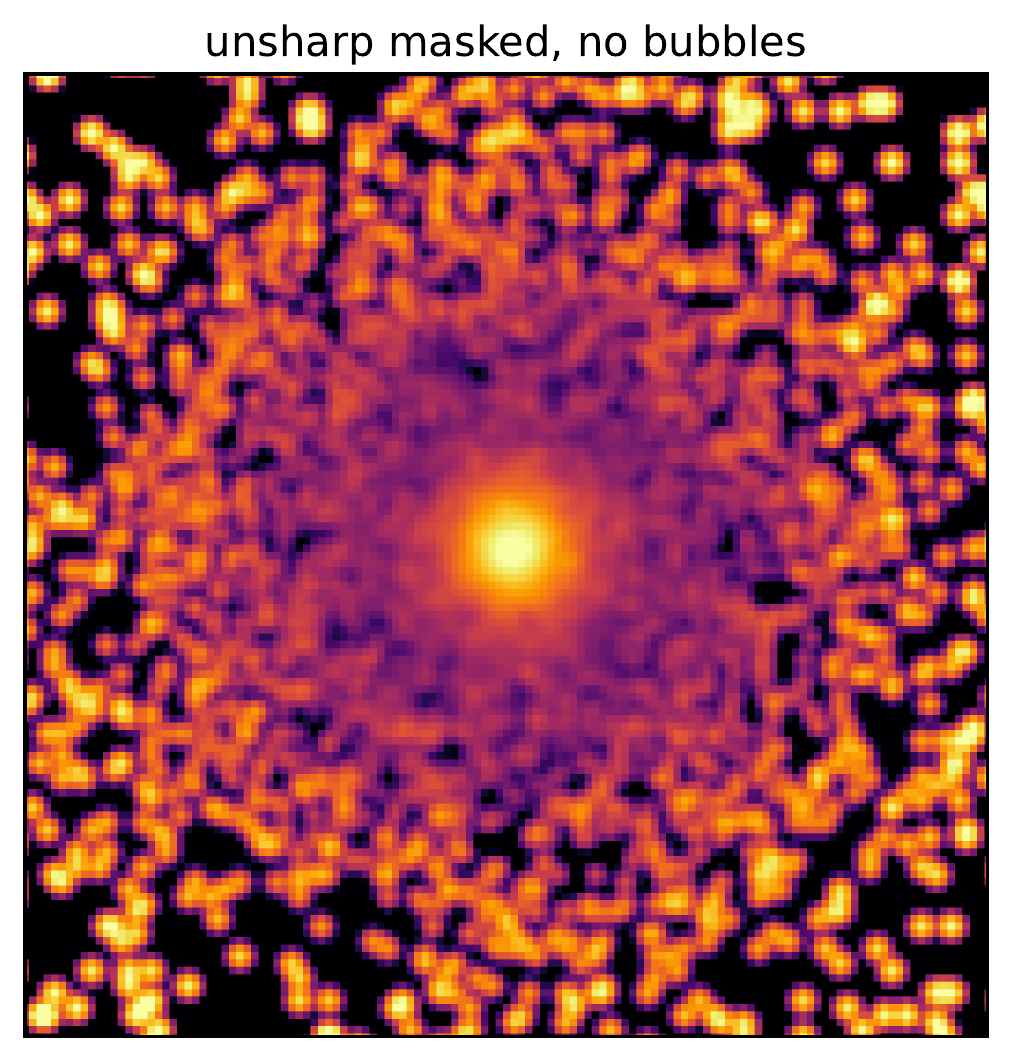} 
  \includegraphics[width=0.24\textwidth]{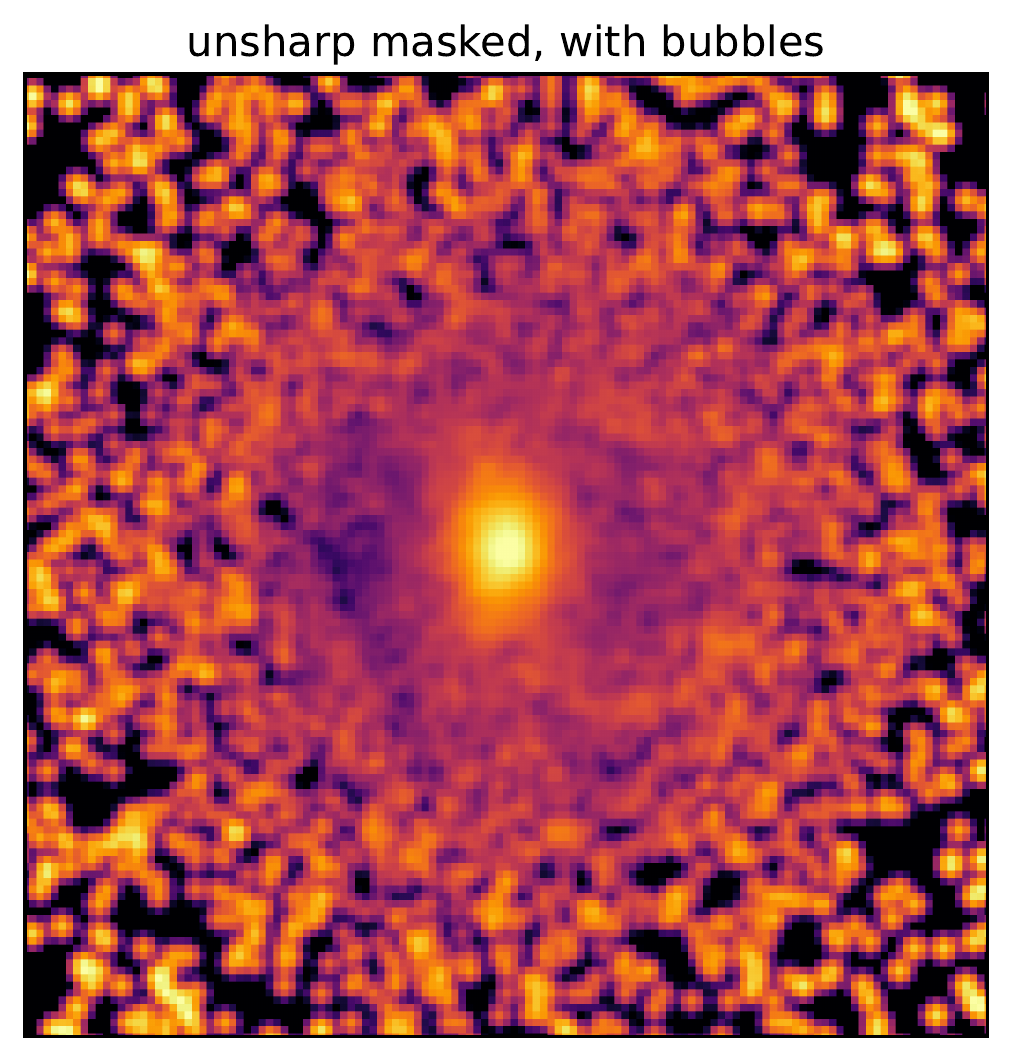} 
\end{center}
\caption{Toy models for possible X-ray cavities in a model cluster. The upper panels show the cluster toy $\beta$ model without (left) and with (right) bubbles added at an angle of 66\degr\ from the plane of the sky.  The lower panels show the corresponding unsharp masked images for the panels in the upper row.}\label{fig:bestmodel}
\end{figure}

In Figure~\ref{fig:bestmodel} we show an example toy $\beta$ model cluster with and without a pair of X-ray cavities. The upper left panel shows the reference toy model without bubbles. In the upper-right panel, a model bubble 10\arcsec\ in radius (26.94~kpc, near the upper limit for the size we expect based on the radio power), 66\degr\ from the plane of the sky, has been subtracted from the reference toy model.  The corresponding unsharp masked images are shown in the lower panels.
Both sloshing, discussed next in Section~\ref{sec:sloshing}, and highly inclined bubbles, such as those in Figure~\ref{fig:bestmodel}, would produce surface brightness edges as well as decrements in the unsharp masked imaging.
It may therefore be possible that a strong sloshing motion over a region similar in size to the predicted bubble could make cavity detection somewhat ambiguous.  In this case, we might miss a cavity that would otherwise have been more obvious. 

\subsection{Sloshing in the cluster core}\label{sec:sloshing}

\citet{2014MNRAS.441L..31W} showed that both the {\it XMM-Newton} and {\it Chandra} data demonstrate structure in the surface brightness as well as cold fronts that are qualitatively similar to simulations of sloshing in cluster cores due to minor mergers. From the {\it Chandra} data, they identified the same inner cold fronts we see here, while the {\it XMM-Newton} data reveal an excess `swirl' pattern reaching out to 800~kpc (approximately half the virial radius) from the centre of the cluster. 
For reference, \citet{2014MNRAS.441L..31W} refer to the inner cold front as `CF1', and the outer as `CF2',  concluding that the two must be two independent sloshing events separated in time.

From the {\it Chandra} unsharp masked image presented here (middle panel, Figure~\ref{fig:xray}), it is clear that the inner cold front permeates to within tens of kiloparsecs of the cool core. 
There are, however, a few salient differences between this cluster and other clusters shown to have sloshing near their cores. For instance, observations of sloshing clusters have been shown to distribute metals from the cluster centre and to flatten their metallicity profile \citep{2010MNRAS.405...91S,2010AA...523A..81D}, while RXC~J2014.8-2430 likely has a peaked metallicity profile (see Figure~\ref{fig:xray_profiles}, third panel).
However, as seen in Figure~\ref{fig:core}, the position of the X-ray peak is offset $\lesssim 10$~kpc in projection from the location of the BCG as the X-ray gas sloshes around the cluster core. 

We compare the morphology of RXC~J2014.8-2430 to the cluster simulations in \citet{2010ApJ...717..908Z}, who find that as the subclusters fall into the cluster core there is a brief period where the X-ray emission is compressed and the luminosity increases (their Figure 17). From Figure 10 in \citet{2010ApJ...717..908Z}, the gas peak is within 10~kpc of the galaxy cluster potential minimum for the first 1 Gyr. After the gas peak shifts away due to sloshing, the peak returns to within $< 10$~kpc from the minimum in the galaxy cluster potential approximately every 500 Myrs. Also, they find that the presence of viscosity as well as the addition of a large BCG potential (their Figure 26) in the cluster core will decrease the core heating expected from sloshing, in agreement with what we observe in RXC~J2014.8-2430.

We consider here the impact strong sloshing near the core may have had in producing the south-western spur of cold molecular gas and warmer atomic gas seen respectively in the ALMA CO(1-0) line imaging and SOAR H$\alpha$ narrow-band imaging.  
While formally the entrainment of cold molecular and warm atomic gas by the sloshing ICM may not be considered ram pressure stripping, as often such filamentary gas is observed to be nearly in hydrostatic equilibrium with the surrounding \citep[e.g.][]{Russell2016}, we use estimates from ram pressure to gauge the feasibility that the cooler gas could be offset by sloshing, a scenario discussed recently in \cite{Inoue2022}.

Following \cite{GunnGott1972}, the effect of ram pressure scales as $P_{\rm ram} = \rho_{\rm ext} \varv^2$, where $\rho_{\rm ext}$ is the external gas density acting on the gas in question, and $\varv$ is the velocity.
From \cite{2010ApJ...717..908Z, 2019SSRv..215...24S}, the typical velocities involved in sloshing have Mach number $\mathcal{M} \sim 0.5$, which for the cool ICM near the core ($3.21\pm0.3$~keV; Table \ref{tab:xrayhalpha}) corresponds to $c_{\mathrm{s},1} \approx 460~\rm km~s^{-1}$.  
Here we have assumed the ideal monatomic gas sound speed, as in Section \ref{sec:toymodel}.
Using this value and our estimate for the ICM density near the core ($\rho_{\rm ext} = m_p \mu_e n_e$, where $n_e \geq 0.1~\rm cm^{-3}$; see Figure~\ref{fig:xray_profiles}) and assuming $\mu_e=1.17$, typical for ICM gas with metallicity reported in Figure~\ref{fig:xray_profiles}), the effect of ram pressure is $\geq 4.1 \times 10^{-10}~\rm erg~cm^{-3}$, comparable to the peak total thermal pressure of the system ($\geq 0.1~\rm cm^{-3} (1+\mu_e) \times 3.2~keV = 10.4 \times 10^{-10}~\rm erg~cm^{-3}$).
The comparable levels of thermal pressure and ram pressure $P_{\rm ram}$ support the idea that the sloshing motion could be a factor influencing the observed spatial distribution of the cold molecular and warm atomic gas.

\subsection{Goodman line ratios}\label{sec:bpt}

\begin{figure}
\centerline{
  \includegraphics[width=0.45\textwidth]{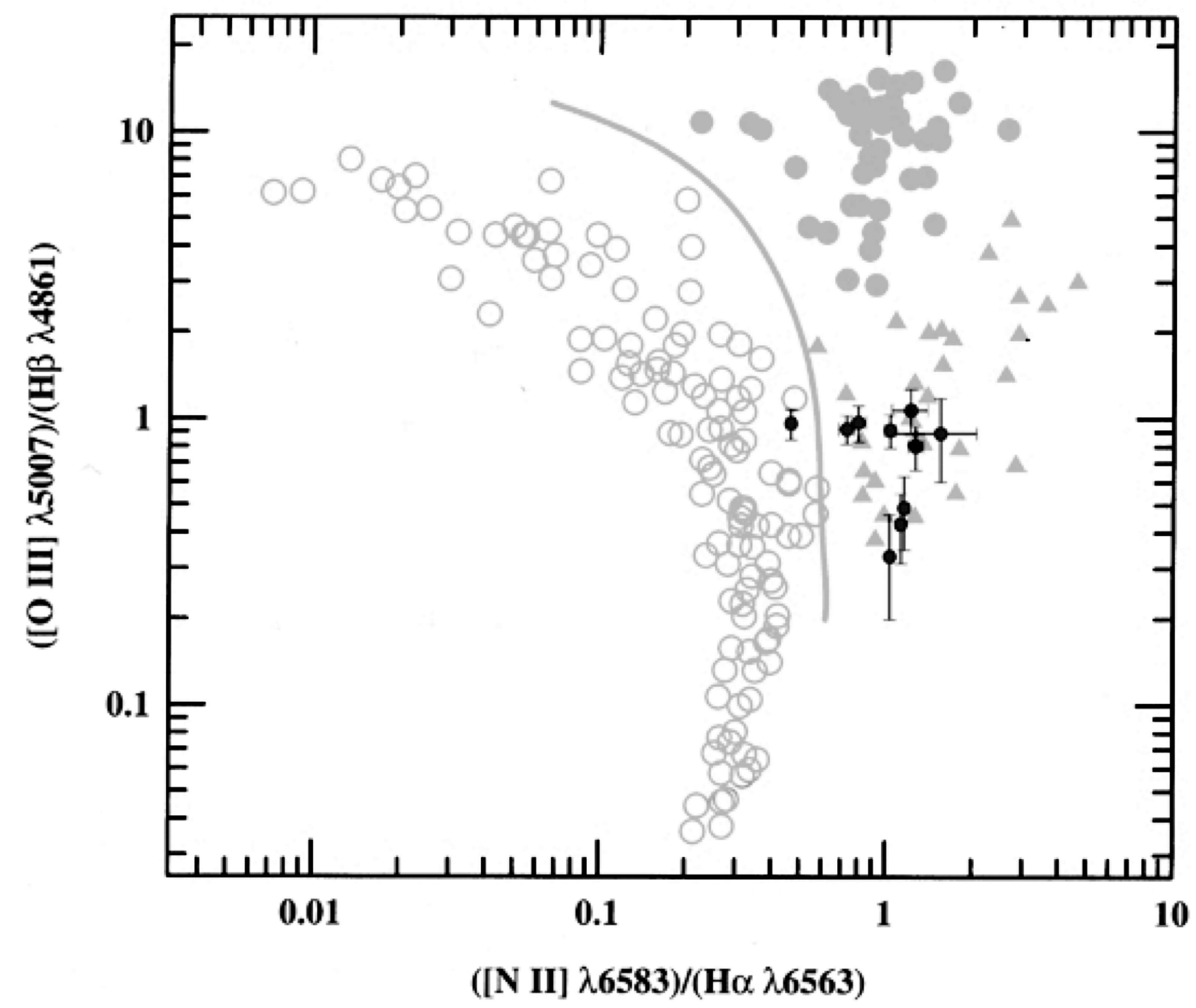}
}
\caption{\label{fig:bpt} Goodman line ratios for ten regions extracted along the brightest portion of the BCG (position angle of 110\degr), shown in the context of the BPT diagram. The curve separates what \cite{1981PASP...93....5B} classify as Seyfert galaxies to the left (open circles) from LINERs to the right (filled triangles) and narrow line AGNs (filled circles). The measurements for RXC~J2014.8-2430 are largely grouped in the LINER portion of the diagram, while the core of the galaxy falls into the region containing Seyfert type galaxies. Figure adapted from \citealt{1997iagn.book.....P} and LEVEL~5.
}
\end{figure}

\begin{figure}
\begin{center}
 \includegraphics[clip,trim=10mm 5mm 10mm 10mm,width=0.5\textwidth]{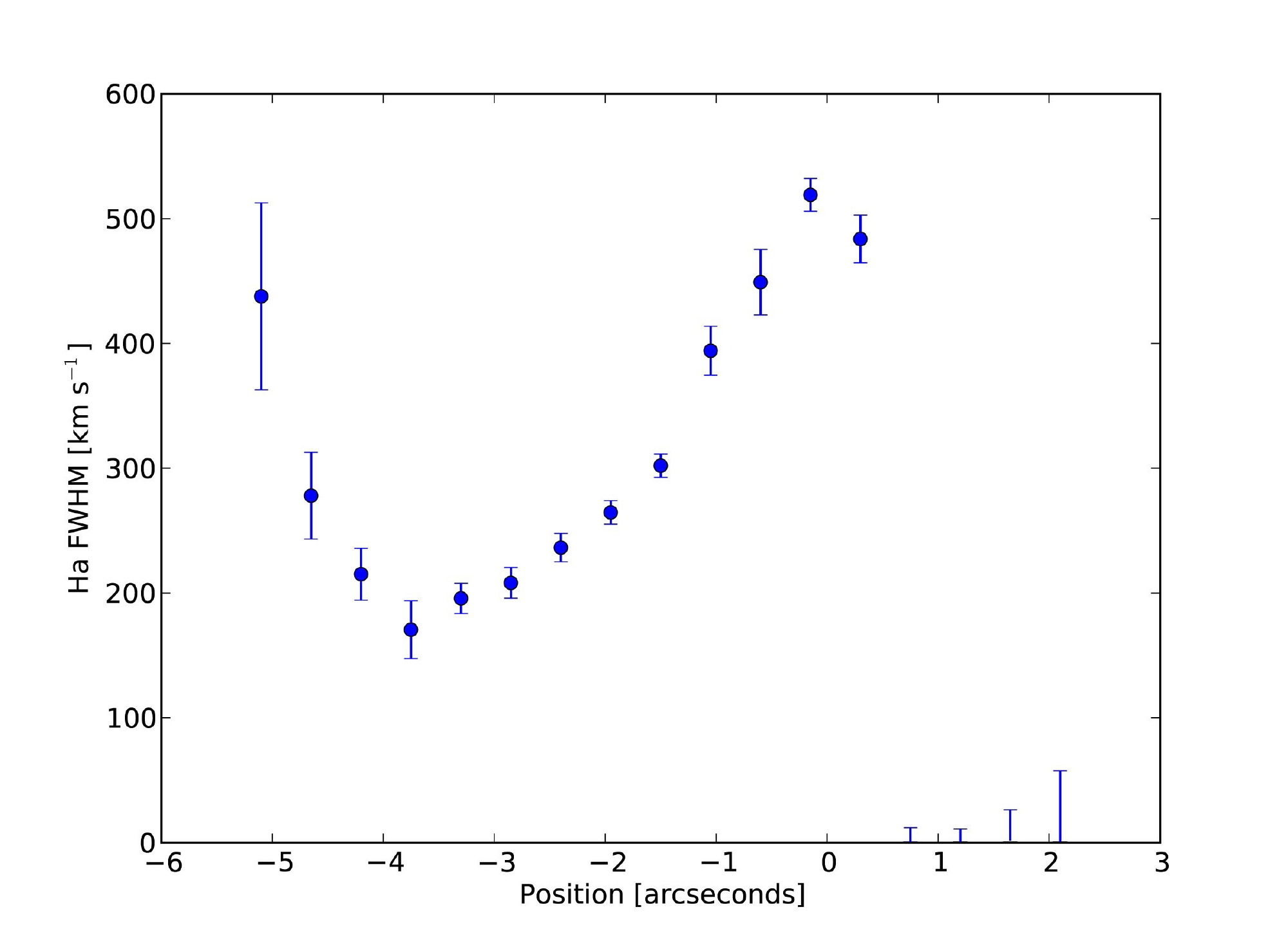}
 \end{center}
\caption{\label{fig:goodmanvelocity} Velocity widths of the H$\alpha$ emission, which have the component of instrumental velocity width ($294$ km~s$^{-1}$) removed in quadrature. The four points on the right have a velocity width less than the instrumental velocity and are unresolved with velocity widths of less than 150 km~s$^{-1}$. Angular distances in arcseconds are relative to the centre of the BCG, taken along position angle of 110\degr, running west to east (i.e. more positive with increasing RA) and centred on the peak (at 0\arcsec), corresponding to the white region displayed in the left-hand panel of Fig.~\ref{fig:core}.}
\end{figure}

We extracted 17 regions of 3 pixels each over the extended H$\alpha$ + [NII] region using the 2D calibrated spectral image discussed earlier. For the H$\alpha$ + [NII] complex, we deblend the features by simultaneously fitting the H$\alpha$ line with the two [NII] lines using {\it deblend} in {\it splot}. However, the [N II] $\lambda$ 6584 is contaminated by a bright sky line and, similar to \citealt{2010ApJ...715..881D}, we used the line measurements for the [NII] $\lambda$ 6548 line and multiplied it by a factor of three (the constant [NII] line ratio fixed by atomic physics) to determine the [NII] $\lambda$ 6584 value for computing abundance ratios. Using these values, the ratios, like ratios of similar regions in nearby clusters of galaxies \citep[e.g.][]{1989ApJ...338...48H}, fall into the lower-right side of the Baldwin, Phillips, and Terlevich (BPT) diagram \citep{1981PASP...93....5B}, which is used to diagnose the difference between ionisation by hot stars and a low ionisation nuclear emission region (LINER). We note, however, that unlike a LINER, which is unresolved and point-like, this emission-line region is extended  and unlikely to be heated by the radiation coming from an AGN, based on arguments similar to those presented in \citealt{1989ApJ...338...48H}: a lack of ionisation gradient that would indicate a central ionisation source, presence of extended emission with nearly constant line ratios, and relatively constant velocity widths of fairly modest width. From a sample of brightest cluster and brightest group galaxies in the Sloan Digital Sky Survey, \citet{2007MNRAS.379..867V} found that most have emission-line ratios that place the galaxies onto the LINER region of the BPT diagram. The [OIII]/H$\beta$ ratio is flat across the cluster, but the [NII]/H$\alpha$ ratio, shown on the BPT diagram adapted from \cite{1997iagn.book.....P} and LEVEL~5\footnote{\url{https://ned.ipac.caltech.edu/level5/}}, in Figure~\ref{fig:bpt}, dips down by a factor of two at the centre of the BCG (leftmost filled circle). The increase in the H$\alpha$/[NII] ratio that we report here could indicate the presence of a weak AGN in the core.  The resolved H$\alpha$ line widths also show a peak at on the brightest emission point in the centre of the BCG (Figure~\ref{fig:goodmanvelocity}), consistent with the position found in the lower-frequency radio and ALMA data. 
Taken together with the steep low-frequency radio spectral index ($\alpha \lesssim -1$) reported in Section~\ref{sec:radio_analysis} and the low X-ray luminosity, we find the AGN in the cluster may be at a unique time in its evolution either due to episodic accretion cycles or the sloshing near the core.
A larger sample of sloshing, cool-core clusters both hosting and lacking bubbles will be required to significantly advance our understanding of the complex interplay of the AGN, the cooler ISM, and the hot ICM.

\begin{figure*}
\centerline{
  \includegraphics[clip,trim=0mm 5mm 0mm 0mm,width=0.95\textwidth]{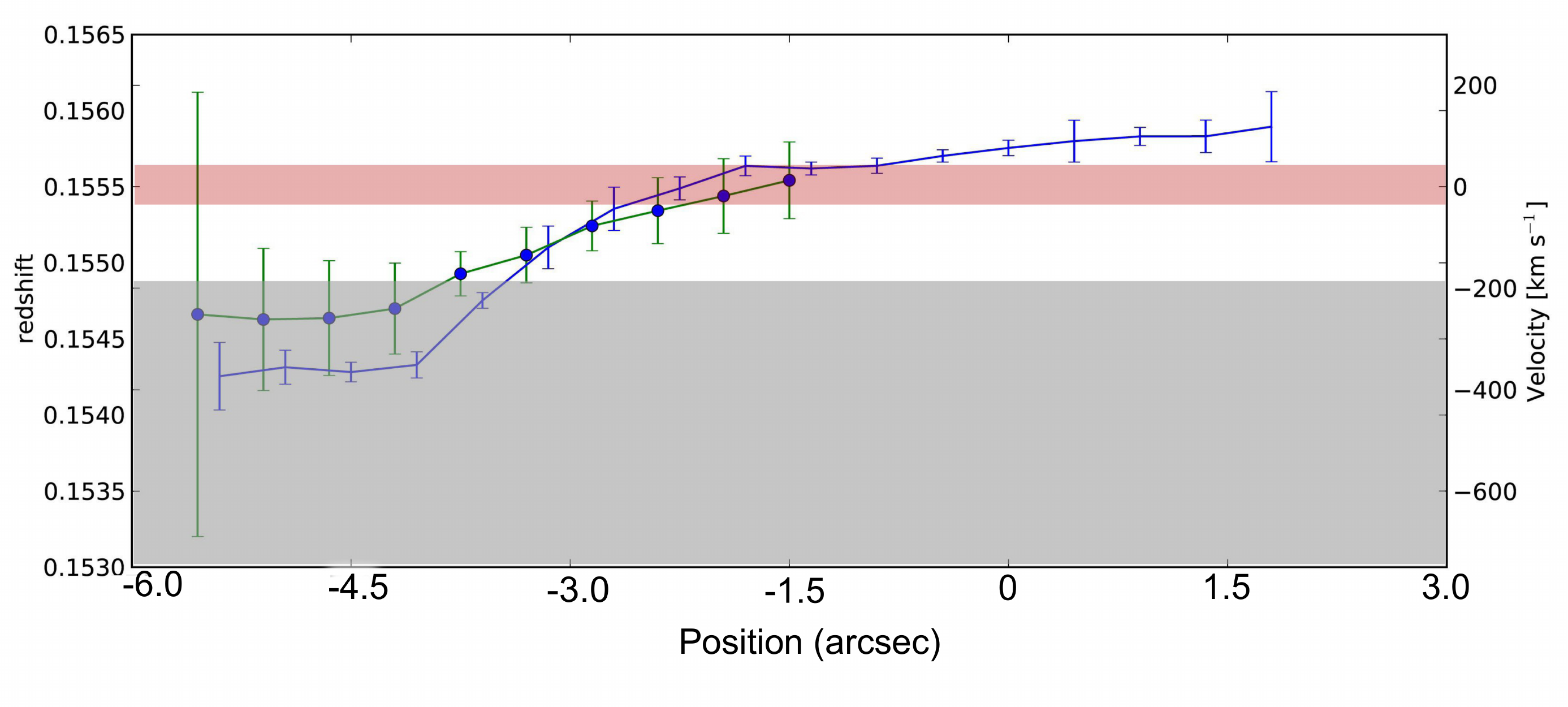}
}
\caption{Redshifts and equivalent line of sight velocities with respect to $z=0.1555$. The data points indicated by blue circles with green error bars are averages from H$\beta$, [OIII] 5007, [NI] 5199, [OI] 6300. The blue line is the average from H$\alpha$ and [NII] 6548. The grey band marks the 90\% confidence range for the \textit{Chandra} redshift fit. The narrow red band marks the cross correlation centre for the stellar absorption spectra. Angular distances in arcseconds are relative to the centre of the BCG, taken along position angle of 110\degr, running west to east (i.e. more positive with increasing RA).} 
\label{fig:goodmanlines}
\end{figure*}

\begin{figure}
 \centering
 \includegraphics[width=90mm]{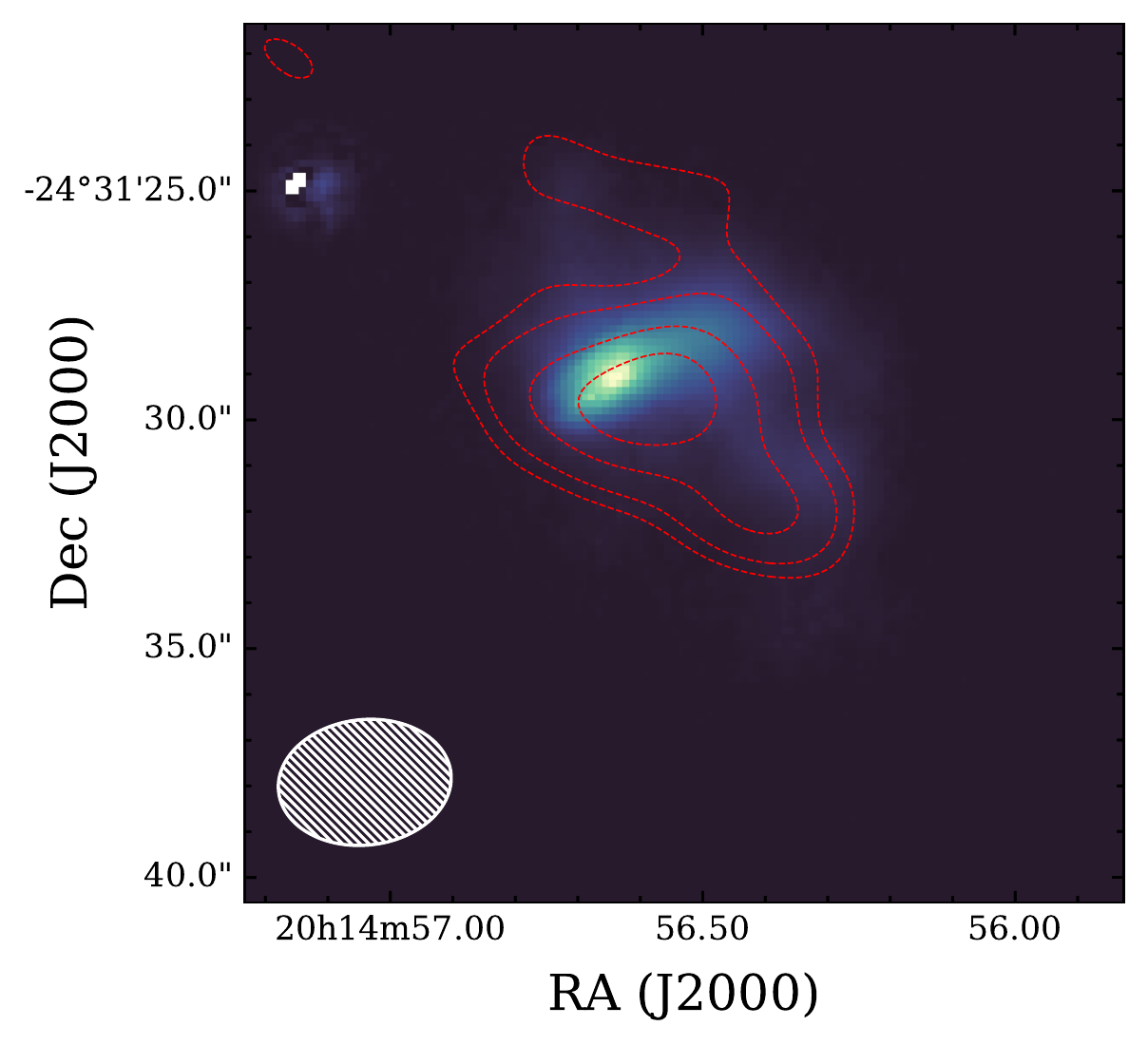}
 \includegraphics[width=90mm]{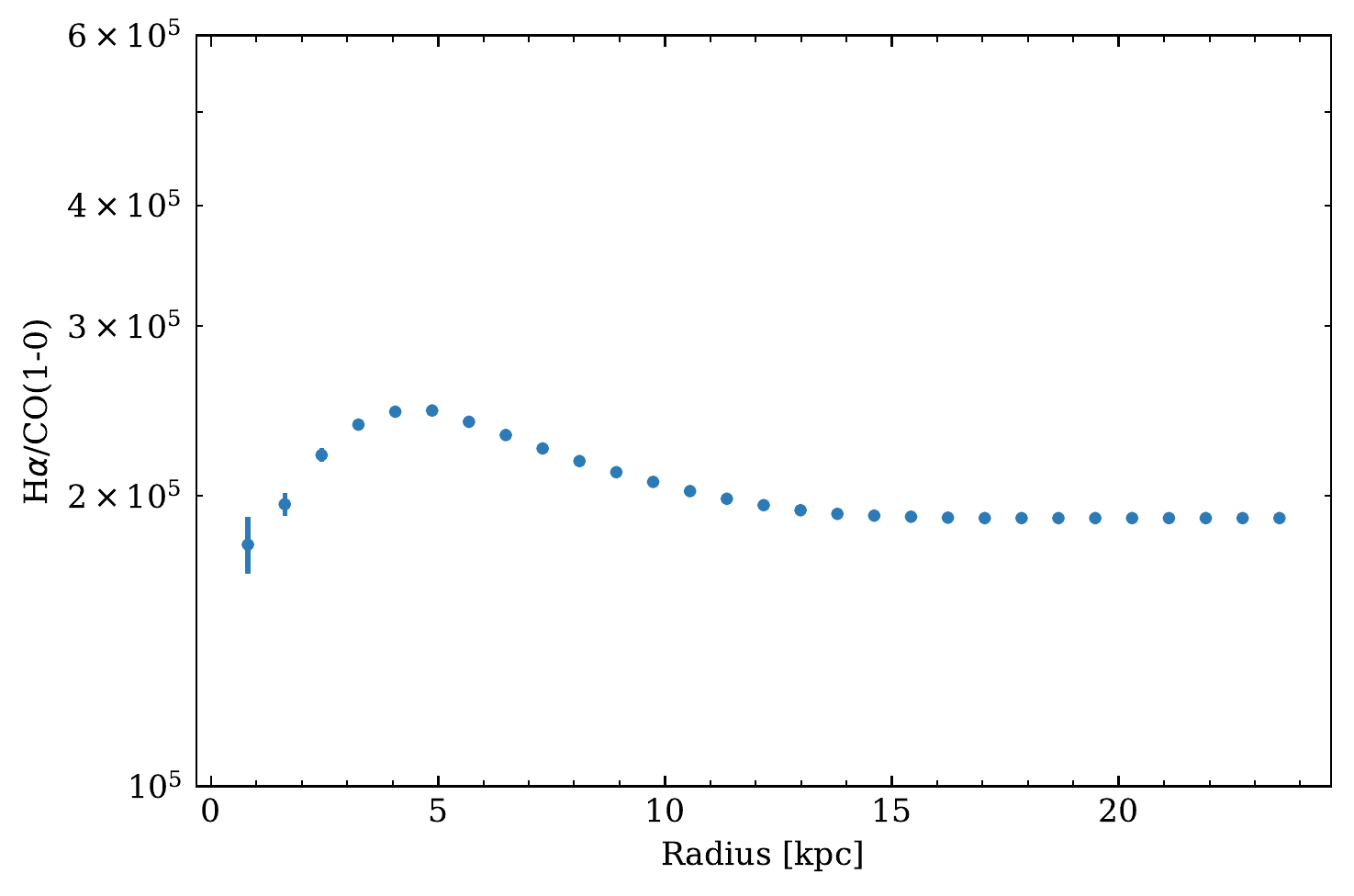}
\caption{\label{fig:Halpha_and_CO}
        Properties of the H$\alpha$ emission seen by SOAR in comparison to the ALMA CO(1-0) measurements. {\bf Upper:}  H$\alpha$ map with contours for the CO(1-0) emission overlaid as red contours (upper panel of Figure \ref{fig:CO_moments}).  This shows the cold molecular gas seen by ALMA to be extended significantly to the south-west, tracing the H$\alpha$ emission seen by SOAR.  Contours start at 2$\sigma$ and are incremented in 2$\sigma$ steps. The synthesised beam for the CO(1-0) observation is depicted in the lower-left corner (hatched ellipse). 
        {\bf Lower:} Ratio of H$\alpha$ to CO(1-0) emission, where the scaling is arbitrary. The plot shows the flux ratio summed in annular apertures centred on the CO(1-0) peak for pixels where the intensity of the CO(1-0) emission is at least 2$\sigma$.}
\end{figure}

\subsection{Velocity structure in the BCG}\label{sec:velocity_struct}

Similar to what has been observed in other clusters \citep[e.g.][]{2004ApJ...607..294S}, the structure in the soft X-ray emission aligns moderately well with the H$\alpha$ structure of an elongated core with perpendicular wings.  Using X-ray spectroscopy, we find that the regions with bright H$\alpha$ emission are slightly cooler and have a higher metallicity than those regions that are not coincident with the H$\alpha$ wings (Table~\ref{tab:xrayhalpha}).  This is also shown clearly in the temperature map in the right panel of Figure~\ref{fig:core}.

We calculated an instrumental velocity width of 294 $\rm{km~s^{-1}}$ based on the width of unsaturated FeAr lines in the calibration spectra. The instrumental velocity we find is similar to the expected spectral resolution for our instrument setup\footnote{\url{https://www.goodman-spectrograph.org/observers.html}}. To estimate the true velocity width, we subtracted the instrumental velocity from the observed velocity width in quadrature. We compared the spatial position of the continuum emission to the spatial position of the H$\alpha$ by extracting 25 \AA~regions around the centre of the H$\alpha$ as well as an equivalent width area of emission bluewards of the H$\alpha$ + [NII] complex. The continuum and H$\alpha$ have a similar peak, within 0$\farcs$3 of 812 pixels, which was the nominal centre pointing. While the emission in the H$\alpha$ is not as spatially symmetric as the continuum light, we do see that the peak of the H$\alpha$ emission and the continuum emission are co-located with the X-ray peak to within 1$\arcsec$.  

\begin{figure*}
 \centering
 \includegraphics[clip,trim=0mm 3mm 0mm 3mm,width=0.95\textwidth]{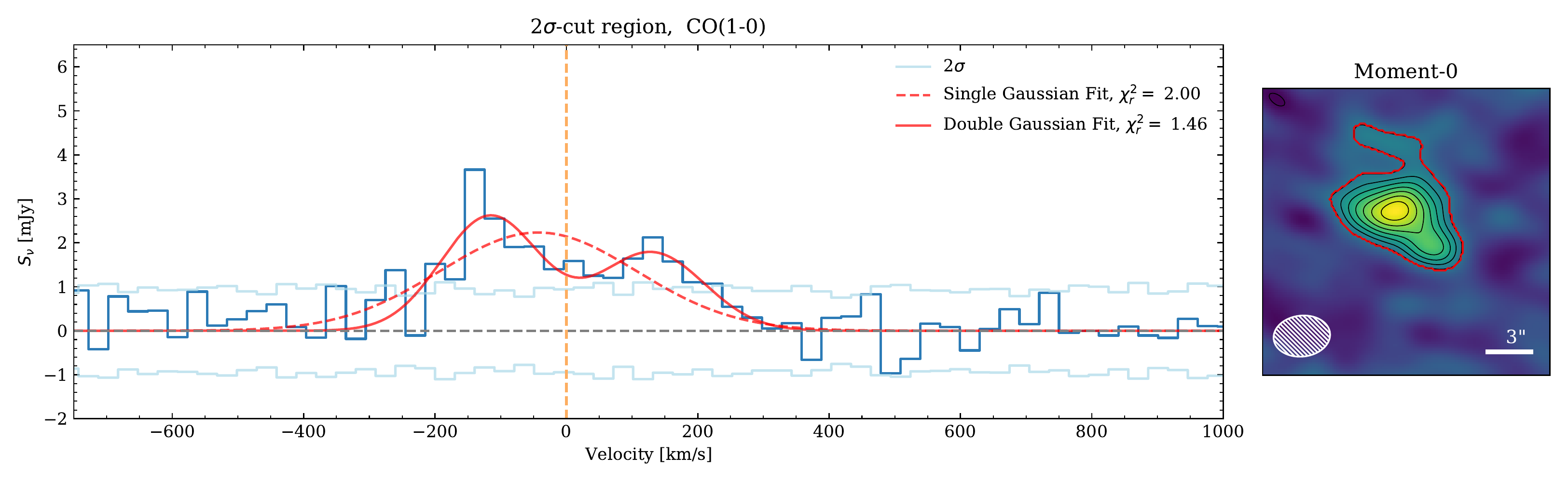}
 \includegraphics[clip,trim=0mm 3mm 0mm 3mm,width=0.95\textwidth]{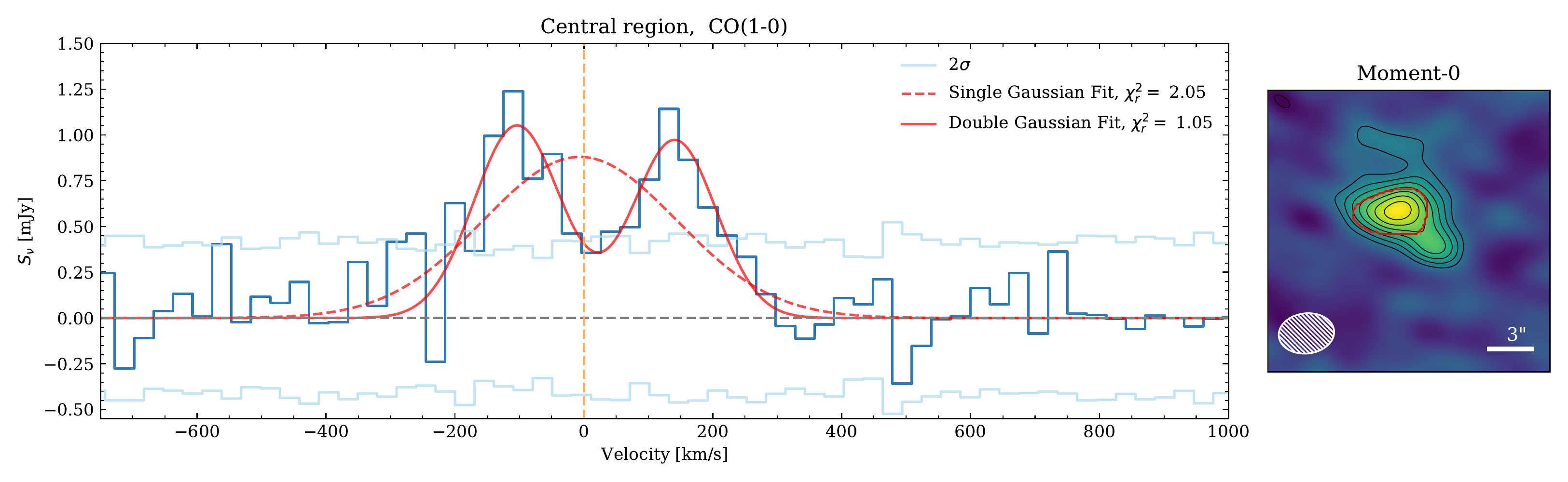}
 \includegraphics[clip,trim=0mm 3mm 0mm 3mm,width=0.95\textwidth]{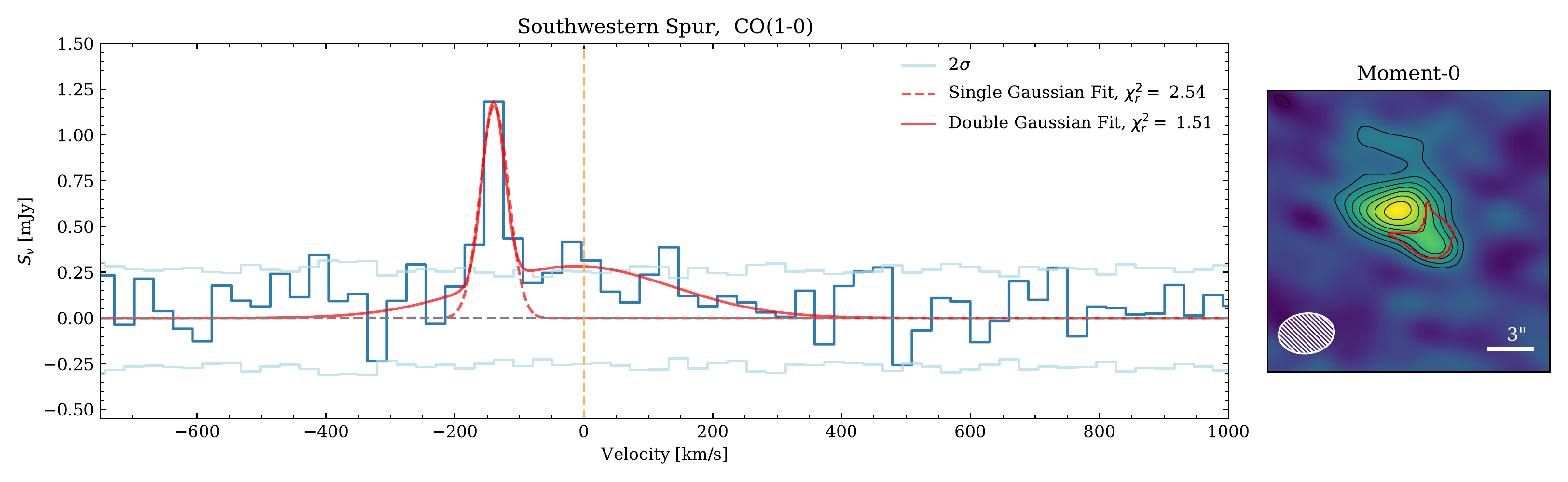}
 \includegraphics[clip,trim=0mm 3mm 0mm 3mm,width=0.95\textwidth]{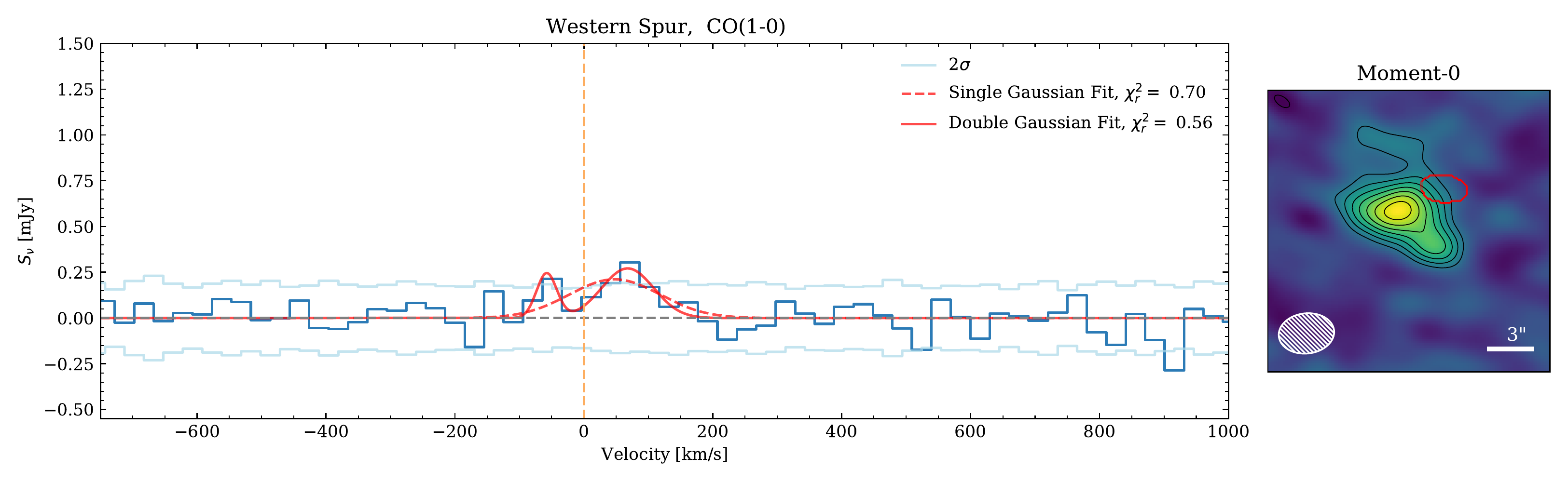}
 \caption{\label{fig:alma_co_regions}
        ALMA spectra for the CO(1-0) emission in select regions using $30~\rm km~s^{-1}$ for the channel width.  The region choices were informed by the channel maps in Fig.~\ref{fig:alma_co_channelmaps}. The lighter blue lines indicate the per channel RMS at 2$\sigma$. The yellow dashed vertical line indicates the expected location of the CO(1-0) line for a redshift $z=0.1555$. For each spectrum, the outset image on the right shows the CO intensity (Fig.~\ref{fig:CO_moments}) with contours in black starting at 2$\sigma$, incremented in 1$\sigma$ intervals. The red outline in each outset panel shows the region for which the spectrum has been extracted.  
        {\bf First row:} Full ALMA spectrum of the CO emission detected at $>2-\sigma$ in total line intensity.  The peak of the line emission is offset in velocity by $\approx -150~\rm km~s^{-1}$.  We note that the intensity scale of the full spectrum differs in this panel.
        {\bf Second row:}  Core region corresponding to the BCG shows a nearly symmetric velocity distribution that can be described by a double Gaussian fit.
        {\bf Third row:}  South-western region, corresponding to the H$\alpha$ spur and the peak of the X-ray emission seen in Fig.~\ref{fig:core}, shows a strong peak offset by $\approx -150~\rm km~s^{-1}$. 
        {\bf Fourth row:}  Western region corresponding to the residual seen in the GMRT radio emission in Fig.~\ref{fig:radio} shows a weak peak at $\approx +70~\rm km~s^{-1}$.
        }
\end{figure*}

In Figure~\ref{fig:goodmanlines} we compare the redshifts and inferred peculiar velocities of different emission and absorption lines. Relative to the BCG centre, the H$\alpha$ shows a velocity gradient with the line centre becoming more relatively blueshifted to the west (more negative position in the Fig.~\ref{fig:goodmanlines}). We compare that to the 90\% confidence interval (0.1531 $\pm$ 0.0017) on a fit for the redshift from the {\it Chandra} spectra using the same XSPEC models but freeing the redshift parameter while requiring it to be the same for all cluster spectra. Given the one-sided nature to the velocity structure, the gas is likely either cooling and infalling onto the BCG or is flowing outwards, entrained by an AGN outburst or by the sloshing ICM surrounding it. The optical velocity structure does not appear to be due to a rotating disk, since it is one-sided and the disk would have to be off-centre from the peak of the emission, even though the emission peaks are aligned. However, we cannot distinguish whether the velocity gradient is indicative of gas infall or outflow. Additionally, the optical emission is inconsistent with a pair of bubbles pushing the emission-line gas in front of each one because that would produce a profile with two peaks if the gas is otherwise optically thin.

\citet{2011ApJ...738L..24V} estimate that the stellar mass loss in BCGs may be as high as 8 M$_\odot$ per year, comparable to the star formation rate in RXC~J2014.8-2430, which \cite{2010ApJ...715..881D} found to be 8-14 M$_\odot$ yr$^{-1}$. This gas is also predicted to remain cool and may be a source of the emission-line gas, such that the emission-line gas seen in the BCG of RXC~J2014.8-2430 could originate from stellar mass loss. While this theory predicts the emission-line gas would have a similar velocity to the stars in the BCG, sloshing in the cluster core could affect the relative motions of the gas and stars in the BCG. 

In the presence of sloshing caused by a minor merger through the centre of the cluster, the ISM of the BCG could be dominated by interactions with the ICM, as discussed in Section~\ref{sec:sloshing}.
Indeed, as seen in the overlays of the H$\alpha$ on the X-ray images (Figure \ref{fig:core}), the H$\alpha$ spur follows the peak of the offset cool core of the ICM.
From the H$\alpha$, one can see that the measured line velocity gradient $\pm 200~\rm{km~s^{-1}}$ is similar to that seen for the CO(1-0), though the values only span the range of the best-fit mean velocities of both the ICM and the BCG stars. Since we only measure radial velocities of the line, the true 3D sloshing speed could be somewhat higher. However, having a centrally peaked metallicity distribution in the core is much more unlikely if the velocity of the sloshing approaches trans-sonic speeds ($\mathcal{M}\sim 1$), while more typical speeds of $\mathcal{M}\sim 0.5$ remain plausible assuming modest inclinations from the plane of the sky.

For comparison with the optical data, we turn to the view ALMA provides of the cold molecular gas, as traced by CO(1-0).
Fig.~\ref{fig:Halpha_and_CO} (upper panel) shows the molecular gas traced by CO emission to be co-spatial with the warmer gas traced by H$\alpha$ emission, consistent with previous studies of the environments surrounding BCGs in cool-core and sloshing clusters \citep[e.g.][]{Vantyghem2016, Tremblay2018, Olivares2019}.
The lower panel of Fig.~\ref{fig:Halpha_and_CO} shows the radial behaviour of the ratio of the H$\alpha$ to CO(1-0) emission.  The observed dip at the smallest radii, followed by a bump at $\sim 4$~kpc, in the H$\alpha$/CO(1-0) ratio can be explained by the fact the profile is centred around the CO(1-0) peak. Compared to \cite{Olivares2019}, this ratio is relatively flat, which could indicate the cold and warm gas are multi-phase and not cooling along a gradient.  More strongly, the flat ratio would support the hypothesis that both the warm and cold gas are being entrained by the surrounding ICM.

Guided by the channel maps tracing the redshifts of the CO(1-0) line emission (see Fig.~\ref{fig:alma_co_channelmaps}), we chose several regions of interest for more detailed spectral analysis: the total CO(1-0) detection, the central region containing the BCG, the south-western spur also traced by H$\alpha$, and the western spur corresponding to the residual radio emission seen in the GMRT data (Fig.~\ref{fig:radio}).
Following, for example, \cite{Narayanan2012} or \cite{Bolatto2013}, we assume a standard CO-to-H$_2$ conversion factor $\alpha_{CO} = M_{\rm mol} / L_{\rm CO} = 3.6~\rm M_{\odot}/L_{\odot}$, finding molecular gas masses of $M_{\rm mol} \approx [3.6, 1.4, 0.63, 0.16]\times 10^9~\rm M_{\odot}$ for the four regions.
The CO(1–0) spectrum taken over the full extent of line velocity, shown in the top row of Fig.~\ref{fig:alma_co_regions}, indicates a strongly bimodal CO(1–0) component roughly centred around the rest frequency, with a significant component redshifted $\approx -150$~km~s$^{-1}$ from that of the optical redshift. 
While the possibility exists that the observed dip could be due to CO(1-0) absorption from foreground molecular gas, we discount this possibility as the emission is well-described by the fitting of two Gaussian profiles, with a reduced $\chi^2=1.05$.  In known cases of CO absorption \citep[e.g.][]{Combes2008, Rose2019}, the dip in emission is generally a few percent of the total, indicating the absorption is intrinsically weak, and is found against a background continuum source.
The second row of Fig.~\ref{fig:alma_co_regions} reveal there is a significant bimodality to the molecular gas velocity distribution for the CO(1-0) emission superposed on the BCG, while the brightest singular offset feature corresponds to the south-west extension tracing the H$\alpha$ spur (third row).  Since one component of the velocity in the core is at the same velocity at the dominant component in the south-western spur, we argue that this supports the notion the molecular gas distribution is being influenced by the sloshing near the core. 

Notably, the CO(1-0) emission to the west (fourth row of Fig.~\ref{fig:alma_co_regions}), where the radio residual seen in the GMRT data is located (see Fig.~\ref{fig:radio}), is much less significant and does not provide any evidence for entrainment by a putative radio bubble.  We therefore favour the interpretation that the extended radio feature in the GMRT data is most likely spurious.

\section{Summary and conclusions}

We have conducted a multi-band analysis -- ranging from low-frequency radio through millimetre-wave, optical, and up to X-ray --  of the cool-core cluster RXC~J2014.8-2430, known colloquially as the `strongest cool core in REXCESS.' Prior observations have shown this cluster is a strong cool core with sloshing on large scales, approaching half the virial radius. The decrements we see in the unsharp masked {\it Chandra} X-ray image indicate that the sloshing structure reaches farther into the core than previously thought.  Surprisingly, the observations also do not reveal the X-ray cavities or radio bubbles that one would expect for such a strong cool-core cluster hosting a powerful radio source. 

Further, we do not find evidence for an X-ray AGN, based on the lack of a point source in the X-ray images consistent with what was found by \cite{2010ApJ...715..881D} in observations taken in the UV portion of the spectrum. Our analysis of the optical line-emission results and the presence of a bright central source in the radio suggest a weak AGN.

Based on models presented by \citet{2010ApJ...717..908Z} and the scale of the sloshing reported in \citet{2014MNRAS.441L..31W}, it is likely that the sloshing near the cluster core producing the inner cold front has only begun recently. We see a very strong metallicity peak in the core of the cluster, which is inconsistent with the suggestion that sloshing will transport metals from the centre of the cluster to the outskirts, effectively flattening the metallicity profile \citep{2010MNRAS.405...91S,2010AA...523A..81D}.  However, it is possible to decrease the impact sloshing would have on the metallicity distribution with additional viscosity in the core, magnetic field lines following the cold fronts that define the sloshing structure, and by including the large gravitational potential of the massive BCG.

We considered the possibility that the sloshing we find in the cluster core may have suppressed any X-ray cavities.  The magneto-hydrodynamic simulations in \citet{2013ApJ...762...78Z} indicate that sloshing may redistribute the particles ejected by feedback.  However, the timescales for this process to quell the bubbles are $\gtrsim0.5$~Gyr, while the radio-loud nature of the central AGN indicates feedback is currently ongoing.  Further, \citet{2013ApJ...762...78Z} find that the redistribution of the relativistic particles from any previous bubbles coupled with turbulence due to sloshing should also result in a radio mini-halo, which is not observed in this system. An earlier work by the same authors \citep{2010ApJ...717..908Z} found that sloshing can prevent the formation of a cool core, raising the entropy and reducing the core luminosity. The lack of a radio mini-halo, the compact distribution of metals in the core, and the strong, luminous, low-entropy cool core all indicate the sloshing near the core is too weak to significantly disrupt any putative bubbles, leading us to consider the possibility that the sloshing has stifled or preempted the accretion process by entraining or otherwise displacing the cool gas that would feed the AGN.

From the cooling inferred from the X-ray luminosity of the core, in the standard picture of episodic feedback, we would expect to detect a pair of radio bubbles at high significance. Since we conclude it is unlikely that sloshing is sufficiently strong to suppress existing bubbles in the core, we hypothesise that this cluster in the unique evolutionary position where the AGN is just turning on and has only started to create X-ray cavities. 
Several other sloshing cool-core clusters have been found to contain X-ray cavities.
An example of a cool-core system with both sloshing and intact X-ray cavities is Abell 2052, reported in \cite{2011ApJ...737...99B}.
Notably, recent simulations from \cite{Fabian2022} have shown that while the Perseus cluster exhibits both sloshing and numerous prominent cavities near the core, cavities farther out could be suppressed by the sloshing.
We also note that RXC~J2014.8-2430 is not the only sloshing cool-core cluster with a radio bright AGN lacking radio bubbles. A2029 \citep{McNamara2016} and RXJ1347.5-1145 \citep{Ueda2018, DiMascolo2019} are also well-studied strongly sloshing systems with no reported bubbles or X-ray cavities.

Examining the optical and millimetre-wave emission, we see that the H$\alpha$ and CO(1-0) emission is elongated in the same direction in which the cluster is thought to be sloshing. 
The east-west sloshing compression is also in the same orientation as the elongation of the central H$\alpha$ emission, and the fact that the north and south wings of the H$\alpha$ are both behind the brightest knot in the centre of the H$\alpha$, near the centroid of the X-rays but ahead of the X-ray peak, suggests either that the X-ray gas is moving past the galaxy or the galaxy is moving through the X-ray gas. There is a significant velocity gradient along the elongated central H$\alpha$ region, indicating this emission-line gas could be gravitationally pulled along by the BCG, or that this gas could be pulled outwards by the ICM sloshing.

And finally, the ALMA CO(1-0) line emission shows a clear velocity separation between the gas near the BCG and that following the south-western H$\alpha$ spur towards the coolest ICM gas seen in the X-ray emission.  We interpret this as a significant mass of cold gas potentially being uplifted by the ICM sloshing, and which could be falling back unto the BCG.
This would indicate that not only do galaxies and AGNs affect the ICM, but the ICM can affect the cold gas supply that feeds AGN feedback.
A forthcoming work will explore the impact of sloshing and the full pressure distribution traced by ALMA, the Atacama Compact Array (ACA), and ACT SZ data, while 170~ksec of additional ACIS-S (P.I.\ S.~Walker) and 4.12~ksec of ACIS-I (P.I.\ G.~Garmire) \textit{Chandra} observations will more definitively answer the question of the existence of X-ray cavities in RXC~J2014.8-2430.  
At the same time, deeper radio observations with improved dynamic range using, for example, MeerKAT \citep{Jonas2016} could answer the question of whether the system contains radio bubbles or a mini-halo.
Looking further ahead, X-ray micro-calorimetric observations with {\it Athena} \citep{Barcons2017} would establish the direction and speed of the ICM surrounding the molecular and atomic gas,  definitively establishing whether it is fully entrained by the surroundings.  This will answer the question of whether the observed velocity structure is dominated by sloshing, uplift by AGN outbursts, or cooling and gravitational accretion of the cold gas onto the BCG.

\begin{acknowledgements}

We thank the referee for the substantive comments that greatly improved the clarity and content of the work presented. 
This work is based on observations obtained at the Southern Astrophysical Research (SOAR) telescope, which is a joint project of the Minist\'{e}rio da Ci\^{e}ncia, Tecnologia, e Inova\c{c}\~{a}o (MCTI) da Rep\'{u}blica Federativa do Brasil, the U.S. National Optical Astronomy Observatory (NOAO), the University of North Carolina at Chapel Hill (UNC), and Michigan State University (MSU).
This paper makes use of the following ALMA data: ADS/JAO.ALMA\#2018.1.00940.S. ALMA is a partnership of ESO (representing its member states), NSF (USA) and NINS (Japan), together with NRC (Canada), MOST and ASIAA (Taiwan), and KASI (Republic of Korea), in cooperation with the Republic of Chile. The Joint ALMA Observatory is operated by ESO, AUI/NRAO and NAOJ. The National Radio Astronomy Observatory is a facility of the National Science Foundation operated under cooperative agreement by Associated Universities, Inc.
This research made use of APLpy, an open-source plotting package for Python \citep{2012ascl.soft08017R}.
Basic research in radio astronomy at the Naval Research Laboratory is funded by 6.1 Base funding. Construction and installation of VLITE were supported by the NRL Sustainment Restoration and Maintenance fund.
LDM is supported by the ERC-StG ``ClustersXCosmo'' grant agreement 716762. GWP acknowledges support from CNES, the French space agency.

\end{acknowledgements}

%
\bibliographystyle{aa} 
\bibliography{RXCJ2014} 
%

\onecolumn

\appendix

\section{Goodman spectral line fits from the SOAR data}\label{sec:appendix:goodman}

Table~\ref{tab:goodmanlines} reports in full the spectral line fits to the SOAR data.
\begin{longtable}[p!]{cccccc}
    \hline    
    Spectral Line & Region Centre & Line Centre & EQW   & Flux                      & FWHM \\ 
    (---)         & (Pixels)      & (\AA)       & (\AA) & (10$^{-16}$ erg/s/cm$^2$) & (\AA)\\
    \hline
    H$\beta$  & 800 & 5616.82 $\pm$ 0.3 &  -24.7 $\pm$ 1.7 &  2.367 $\pm$ 0.16 &  8.882 $\pm$ 0.9 \\
    OIII 5007 &  & 5786.24 $\pm$ 2.28 &  -7.7 $\pm$ 2.9 &  0.7731 $\pm$ 0.3 &  12.44 $\pm$ 5.89 \\
    NI 5199   &  & 6006.91 $\pm$ 1.03 &  -10.9 $\pm$ 2.2 &  0.9945 $\pm$ 0.2 &  9.565 $\pm$ 2.88 \\
    OI 6300   &  & 7279.13 $\pm$ 0.4 &  -22.2 $\pm$ 1.5 &  1.974 $\pm$ 0.14 &  9.464 $\pm$ 0.99 \\
    H$\beta$  & 803 & 5616.55 $\pm$ 0.38 &  -20.0 $\pm$ 1.6 &  2.381 $\pm$ 0.19 &  9.636 $\pm$ 0.88 \\
    OIII 5007 &  & 5784.63 $\pm$ 2.05 &  -8.3 $\pm$ 2.1 &  1.007 $\pm$ 0.26 &  13.68 $\pm$ 4.03 \\
    NI 5199   &  & 6006.66 $\pm$ 1.25 &  -9.0 $\pm$ 1.7 &  1.023 $\pm$ 0.19 &  10.61 $\pm$ 2.61 \\
    OI 6300   &  & 7279.21 $\pm$ 0.58 &  -15.4 $\pm$ 1.5 &  1.735 $\pm$ 0.17 &  11.22 $\pm$ 1.29 \\
    H$\beta$  & 806 & 5616.07 $\pm$ 0.45 &  -13.9 $\pm$ 1.1 &  2.056 $\pm$ 0.16 &  9.529 $\pm$ 0.62 \\
    OIII 5007 &  & 5784.11 $\pm$ 0.94 &  -6.8 $\pm$ 1.8 &  0.9873 $\pm$ 0.27 &  10.95 $\pm$ 3.62 \\
    NI 5199   &  & 6006.35 $\pm$ 1.92 &  -7.4 $\pm$ 1.5 &  0.9811 $\pm$ 0.19 &  12.14 $\pm$ 2.94 \\
    OI 6300   &  & 7278.44 $\pm$ 0.55 &  -14.0 $\pm$ 1.5 &  1.76 $\pm$ 0.18 &  10.98 $\pm$ 1.02 \\
    H$\beta$  & 809 & 5615.08 $\pm$ 0.6 &  -11.6 $\pm$ 0.9 &  1.974 $\pm$ 0.16 &  9.94 $\pm$ 1.08 \\
    OIII 5007 &  & 5783.43 $\pm$ 0.91 &  -9.3 $\pm$ 1.4 &  1.546 $\pm$ 0.24 &  13.76 $\pm$ 2.75 \\
    NI 5199   &  & 6006.43 $\pm$ 1.05 &  -7.2 $\pm$ 1.4 &  1.075 $\pm$ 0.22 &  11.65 $\pm$ 2.22 \\
    OI 6300   &  & 7277.94 $\pm$ 0.83 &  -12.8 $\pm$ 1.3 &  1.83 $\pm$ 0.19 &  12.67 $\pm$ 1.26 \\
    H$\beta$  & 812 & 5614.27 $\pm$ 0.52 &  -12.3 $\pm$ 1.1 &  2.264 $\pm$ 0.21 &  10.11 $\pm$ 0.93 \\
    OIII 5007 &  & 5782.22 $\pm$ 0.73 &  -11.2 $\pm$ 1.0 &  2.014 $\pm$ 0.19 &  13.27 $\pm$ 1.9 \\
    NI 5199   &  & 6005.64 $\pm$ 1.53 &  -5.1 $\pm$ 0.9 &  0.8571 $\pm$ 0.15 &  9.897 $\pm$ 2.94 \\
    OI 6300   &  & 7276.64 $\pm$ 0.75 &  -13.8 $\pm$ 1.2 &  2.104 $\pm$ 0.18 &  13.85 $\pm$ 1.35 \\
    H$\beta$  & 815 & 5613.0 $\pm$ 0.59 &  -12.8 $\pm$ 0.9 &  2.37 $\pm$ 0.16 &  10.53 $\pm$ 1.05 \\
    OIII 5007 &  & 5781.22 $\pm$ 0.49 &  -12.5 $\pm$ 1.2 &  2.228 $\pm$ 0.22 &  12.23 $\pm$ 1.0 \\
    NI 5199   &  & 6005.62 $\pm$ 1.08 &  -5.8 $\pm$ 1.7 &  0.9444 $\pm$ 0.27 &  12.98 $\pm$ 4.4 \\
    OI 6300   &  & 7276.48 $\pm$ 0.81 &  -11.5 $\pm$ 1.4 &  1.714 $\pm$ 0.2 &  13.53 $\pm$ 2.05 \\
    H$\beta$  & 818 & 5612.06 $\pm$ 0.32 &  -14.6 $\pm$ 1.2 &  2.415 $\pm$ 0.21 &  9.398 $\pm$ 0.93 \\
    OIII 5007 &  & 5779.17 $\pm$ 0.37 &  -13.2 $\pm$ 1.0 &  2.168 $\pm$ 0.16 &  8.645 $\pm$ 0.89 \\
    NI 5199   &  & 6004.6 $\pm$ 2.76 &  -5.0 $\pm$ 1.4 &  0.7397 $\pm$ 0.2 &  13.97 $\pm$ 5.03 \\
    OI 6300   &  & 7275.74 $\pm$ 1.63 &  -8.3 $\pm$ 1.7 &  1.148 $\pm$ 0.23 &  13.45 $\pm$ 2.92 \\
    H$\beta$  & 821 & 5611.1 $\pm$ 0.28 &  -13.3 $\pm$ 1.4 &  1.895 $\pm$ 0.21 &  8.033 $\pm$ 1.04 \\
    OIII 5007 &  & 5778.52 $\pm$ 0.33 &  -12.3 $\pm$ 1.1 &  1.791 $\pm$ 0.16 &  6.875 $\pm$ 0.76 \\
    NI 5199   &  & 6005.0 $\pm$ 3.3 &  -4.6 $\pm$ 2.3 &  0.5812 $\pm$ 0.3 &  12.49 $\pm$ 9.32 \\
    OI 6300   &  & 7275.75 $\pm$ 2.5 &  -10.0 $\pm$ 1.8 &  1.056 $\pm$ 0.19 &  18.93 $\pm$ 5.98 \\
    H$\beta$  & 824 & 5611.33 $\pm$ 0.63 &  -10.2 $\pm$ 1.4 &  1.189 $\pm$ 0.17 &  8.039 $\pm$ 1.42 \\
    OIII 5007 &  & 5778.55 $\pm$ 0.45 &  -10.4 $\pm$ 1.1 &  1.239 $\pm$ 0.13 &  6.552 $\pm$ 0.95 \\
    NI 5199   &  & 6005.31 $\pm$ 3.61 &  -6.2 $\pm$ 2.2 &  0.6171 $\pm$ 0.22 &  15.48 $\pm$ 9.44 \\
    OI 6300   &  & 7274.81 $\pm$ 3.82 &  -9.7 $\pm$ 3.1 &  0.7907 $\pm$ 0.25 &  22.96 $\pm$ 13.93 \\
    H$\beta$  & 827 & 5612.19 $\pm$ 1.51 &  -8.9 $\pm$ 2.3 &  0.7779 $\pm$ 0.2 &  11.85 $\pm$ 3.72 \\
    OIII 5007 &  & 5779.16 $\pm$ 1.37 &  -7.2 $\pm$ 1.4 &  0.6745 $\pm$ 0.13 &  8.567 $\pm$ 1.92 \\
    NI 5199   &  & 6006.36 $\pm$ 7.97 &  -2.9 $\pm$ 5.5 &  0.2399 $\pm$ 0.46 &  8.206 $\pm$ 10.71 \\
    OI 6300   &  & 7272.52 $\pm$ 15.44 &  -3.3 $\pm$ 5.4 &  0.2379 $\pm$ 0.38 &  15.7 $\pm$ 27.14 \\
    Deblended H$\alpha$+N II & & & & & \\
    N II 6548 & 778 & 7568.33 $\pm$ 2.01 &  -33.4 $\pm$ 6.7 &  0.6347 $\pm$ 0.13 &  9.687 $\pm$ 3.17 \\
    H$\alpha$ &  & 7586.6 $\pm$ 0.71 &  -111.0 $\pm$ 16.6 &  2.168 $\pm$ 0.33 &  13.43 $\pm$ 1.89 \\
    N II 6548 & 781 & 7568.33 $\pm$ 0.91 &  -27.0 $\pm$ 5.6 &  0.7375 $\pm$ 0.15 &  9.106 $\pm$ 2.52 \\
    H$\alpha$ &  & 7585.78 $\pm$ 0.39 &  -86.9 $\pm$ 9.0 &  2.414 $\pm$ 0.25 &  10.35 $\pm$ 0.88 \\
    N II 6548 & 784 & 7568.54 $\pm$ 0.47 &  -31.7 $\pm$ 5.7 &  1.029 $\pm$ 0.18 &  9.581 $\pm$ 1.94 \\
    H$\alpha$ &  & 7585.56 $\pm$ 0.29 &  -96.4 $\pm$ 5.0 &  3.102 $\pm$ 0.16 &  9.341 $\pm$ 0.53 \\
    N II 6548 & 787 & 7568.5 $\pm$ 1.26 &  -35.2 $\pm$ 5.0 &  1.237 $\pm$ 0.18 &  10.63 $\pm$ 2.66 \\
    H$\alpha$ &  & 7585.2 $\pm$ 0.21 &  -110.4 $\pm$ 6.5 &  3.983 $\pm$ 0.24 &  8.732 $\pm$ 0.59 \\
    N II 6548 & 790 & 7568.14 $\pm$ 0.45 &  -45.8 $\pm$ 3.9 &  1.816 $\pm$ 0.15 &  9.71 $\pm$ 1.12 \\
    H$\alpha$ &  & 7584.96 $\pm$ 0.13 &  -129.2 $\pm$ 3.0 &  5.236 $\pm$ 0.12 &  9.063 $\pm$ 0.31 \\
    N II 6548 & 793 & 7567.73 $\pm$ 0.35 &  -47.6 $\pm$ 3.4 &  2.293 $\pm$ 0.17 &  10.26 $\pm$ 0.75 \\
    H$\alpha$ &  & 7584.68 $\pm$ 0.14 &  -131.5 $\pm$ 3.5 &  6.353 $\pm$ 0.17 &  9.238 $\pm$ 0.31 \\
    N II 6548 & 796 & 7567.25 $\pm$ 0.46 &  -39.3 $\pm$ 3.2 &  2.641 $\pm$ 0.22 &  10.3 $\pm$ 0.86 \\
    H$\alpha$ &  & 7584.32 $\pm$ 0.09 &  -112.1 $\pm$ 2.9 &  7.643 $\pm$ 0.2 &  9.663 $\pm$ 0.29 \\
    N II 6548 & 799 & 7567.18 $\pm$ 0.38 &  -29.9 $\pm$ 2.1 &  2.742 $\pm$ 0.19 &  10.2 $\pm$ 0.63 \\
    H$\alpha$ &     & 7584.15 $\pm$ 0.11 &  -92.1 $\pm$ 2.2 &  8.255 $\pm$ 0.2 &  10.12 $\pm$ 0.24 \\
    N II 6548 & 802 & 7567.38 $\pm$ 0.58 &  -30.5 $\pm$ 2.2 &  3.098 $\pm$ 0.22 &  13.99 $\pm$ 1.43 \\
    H$\alpha$ &     & 7584.17 $\pm$ 0.17 &  -86.0 $\pm$ 2.4 &  8.539 $\pm$ 0.24 &  10.77 $\pm$ 0.24 \\
    N II 6548 & 805 & 7566.03 $\pm$ 0.68 &  -25.0 $\pm$ 3.1 &  3.171 $\pm$ 0.39 &  13.99 $\pm$ 1.37 \\
    H$\alpha$ &     & 7583.59 $\pm$ 0.19 &  -67.8 $\pm$ 2.1 &  8.501 $\pm$ 0.27 &  12.53 $\pm$ 0.5 \\
    N II 6548 & 808 & 7565.3 $\pm$ 1.29 &  -26.7 $\pm$ 4.1 &  3.737 $\pm$ 0.58 &  17.35 $\pm$ 2.15 \\
    H$\alpha$ &     & 7582.56 $\pm$ 0.28 &  -67.0 $\pm$ 3.8 &  9.128 $\pm$ 0.52 &  13.66 $\pm$ 0.66 \\
    N II 6548 & 811 & 7563.2 $\pm$ 1.25 &  -22.8 $\pm$ 2.2 &  3.621 $\pm$ 0.35 &  16.26 $\pm$ 2.14 \\
    H$\alpha$ &     & 7581.34 $\pm$ 0.35 &  -70.2 $\pm$ 3.2 &  10.83 $\pm$ 0.49 &  15.16 $\pm$ 0.33 \\
    N II 6548 & 814 & 7560.82 $\pm$ 0.42 &  -8.7 $\pm$ 2.1 &  1.327 $\pm$ 0.32 &  8.786 $\pm$ 1.3 \\
    H$\alpha$ &     & 7579.15 $\pm$ 0.23 &  -59.6 $\pm$ 2.0 &  8.754 $\pm$ 0.3 &  14.39 $\pm$ 0.48 \\
    N II 6548 & 817 & 7558.29 $\pm$ 0.79 &  -7.8 $\pm$ 2.2 &  1.038 $\pm$ 0.29 &  8.08 $\pm$ 2.12 \\
    H$\alpha$ &     & 7576.12 $\pm$ 0.1 &  -33.8 $\pm$ 2.1 &  4.387 $\pm$ 0.27 &  6.527 $\pm$ 0.3 \\
    N II 6548 & 820 & 7558.12 $\pm$ 0.59 &  -9.6 $\pm$ 2.5 &  1.12 $\pm$ 0.29 &  8.056 $\pm$ 1.92 \\
    H$\alpha$ &     & 7575.67 $\pm$ 0.08 &  -38.2 $\pm$ 2.3 &  4.319 $\pm$ 0.26 &  5.815 $\pm$ 0.28 \\
    N II 6548 & 823 & 7558.58 $\pm$ 1.02 &  -13.4 $\pm$ 4.9 &  1.167 $\pm$ 0.43 &  11.34 $\pm$ 2.98 \\
    H$\alpha$ &     & 7575.63 $\pm$ 0.19 &  -34.5 $\pm$ 4.9 &  2.957 $\pm$ 0.42 &  6.483 $\pm$ 0.67 \\
    N II 6548 & 826 & 7557.72 $\pm$ 2.0 &  -5.6 $\pm$ 1.9 &  0.4371 $\pm$ 0.15 &  8.918 $\pm$ 3.99 \\
    H$\alpha$ &     & 7575.72 $\pm$ 0.5 &  -11.6 $\pm$ 3.9 &  0.8754 $\pm$ 0.29 &  5.318 $\pm$ 1.46 \\
\hline
\caption{Goodman spectral line fits.  The table is organised starting with the westernmost region, and increasing 
 to the east (higher RA), with Pixel \#811-812 located at the peak of the BCG. Each region is 3 pixels, corresponding to 0$\arcsec\!.45$ on the sky.
}\label{tab:goodmanlines}
\end{longtable}

\section{ALMA CO(1-0) channel maps}\label{sec:appendix:channelmaps}

\begin{figure*}
 \centering
 \includegraphics[width=0.99\textwidth]{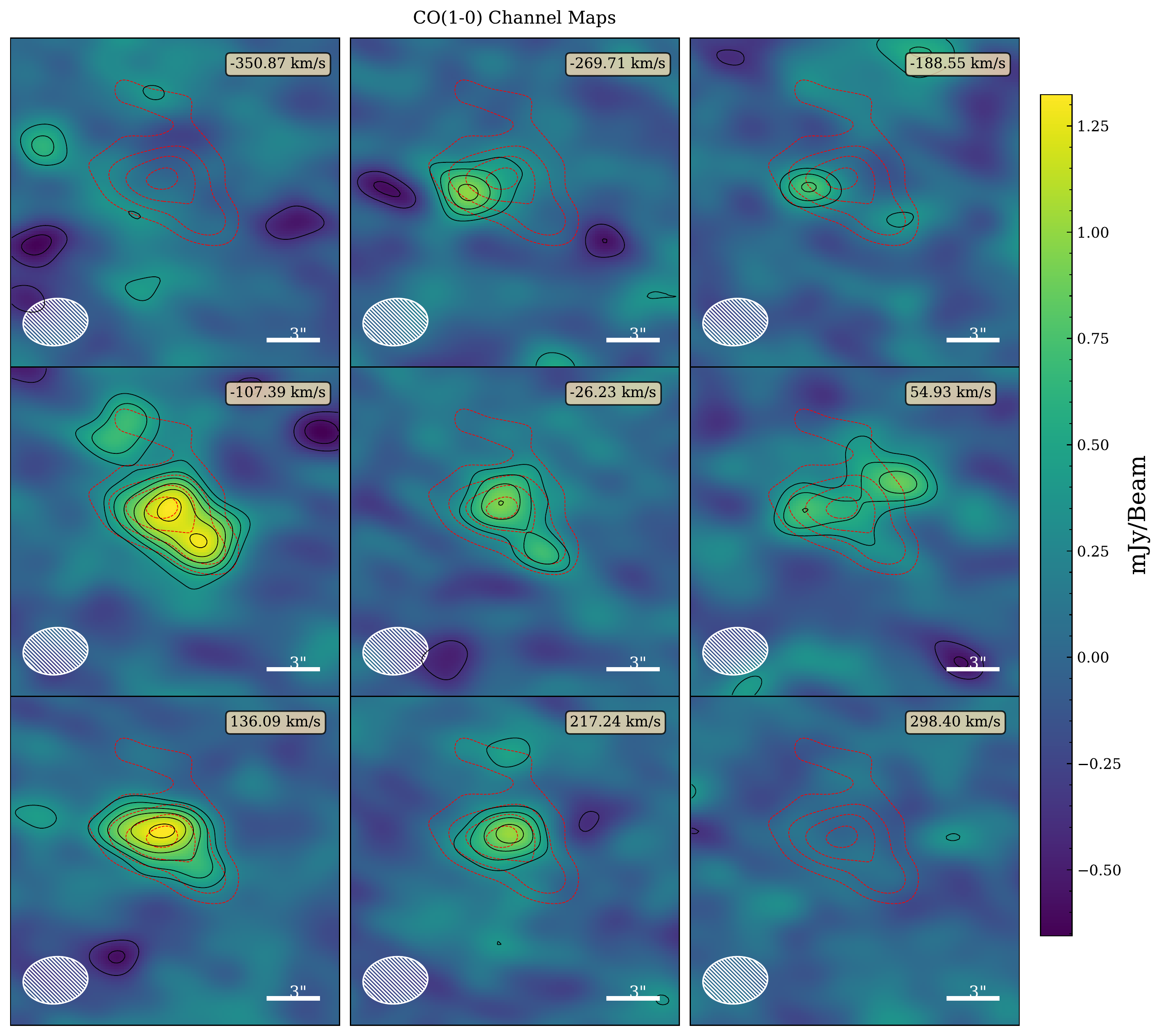}
\caption{\label{fig:alma_co_channelmaps}
        Channel maps binned $80~\rm km~s^{-1}$ determined relative to the redshifted CO(1-0) transition assuming a redshift of $z=0.1555$, consistent with that determined for the BCG in \cite{2010ApJ...715..881D}. The red contours correspond to the $\geq$2$\sigma$ contours in the CO(1-0) moment=0 map shown in the upper panel of Figure \ref{fig:CO_moments}.
    The solid black contours correspond to the signal in each maps and start at 2$\sigma$ significance.
        The channel maps were used to inform the region choices in Fig.~\ref{fig:alma_co_regions}.
        }
\end{figure*}

Figure~\ref{fig:alma_co_channelmaps} provides the CO(1-0) channel maps produced from the ALMA CO(1-0) data.

\end{document}